\documentclass[11pt]{article}
\pdfoutput=1
\usepackage{amsmath,amssymb,epsfig,amsfonts}
\usepackage{graphicx}
\usepackage{subfig}
\usepackage[usenames, dvipsnames]{color}
\usepackage{hyperref}
\usepackage{cite}
\usepackage{multirow}
\usepackage{slashed}
\usepackage{verbatim} 
\usepackage{enumitem}
\usepackage{rotating}
\usepackage{xcolor}
\usepackage{multirow}
\usepackage{bm}
\usepackage[normalem]{ulem}
\usepackage{cleveref}
\usepackage{braket}
\usepackage{diagbox}

\makeatletter

\definecolor{darkgreen}{rgb}{0,0.7,0}

\def\Graph{{\Omega}}
\def\ZGtilde{{Z_{\widetilde{G}}}}
\def\ZPaFu{{\mathcal{Z}}}
\def\tG{{\widetilde G}}
\def\ZtG{Z_{\widetilde{G}}}

\newcommand{\be}{\begin{equation}}
\newcommand{\ee}{\end{equation}}
\newcommand{\ba}{\begin{aligned}}
\newcommand{\ea}{\end{aligned}}
\newcommand{\su}{\mathfrak{su}}

\newcommand{\C}{\mathbb{C}}

\newcommand{\cO}{\mathcal{O}}
\newcommand{\cN}{\mathcal{N}}

\newcommand{\nn}{\nonumber}
\newcommand{\ds}{\displaystyle}
\newcommand{\bea}{\begin{eqnarray}}
\newcommand{\eea}{\end{eqnarray}}

\newcommand{\cA}{\mathcal{A}}

\newcommand{\R}{{\mathbb R}}
\newcommand{\Z}{{\mathbb Z}}

\def\diag{\mathop{\mathrm{diag}}\nolimits}
\def\Im{\mathop{\mathrm{Im}}\nolimits}

\def\Re{\mathop{\mathrm{Re}}\nolimits}

\def\Tr{\mathop{\mathrm{Tr}}\nolimits}

\def\bra#1{{\langle{#1}|}}

\def\ket#1{{|{#1}\rangle}}

\def\half{{\frac{1}{2}}}

\def\p{\partial}

\def\unit{{1\kern-.65ex {\rm l}}}
\def\1{{1\kern-.65ex {\rm l}}}

\def\g{{\frak{g}}}

\def\tk{{\widetilde{k}}}

\def\CA{{\cal A}}

\def\CH{{\cal H}}

\def\CM{{\cal M}}
\def\CN{{\cal N}}

\def\CS{{\cal S}}
\def\CT{{\cal T}}

\def\CV{{\cal V}}
\def\CW{{\cal W}}

\def\CZ{{\cal Z}}

\def\ie{{\it i.e.}}

\def\eg{{\it e.g.}}
\def\Eg{{\it E.g.}}

\newcount\hour \newcount\minute
\hour=\time \divide \hour by 60
\minute=\time
\def\now{%
\ifnum \hour<13
  \ifnum \hour=0 \advance \hour by 12 \number\hour:\else \number\hour:\fi%
     \ifnum \minute<10 0\fi%
     \number\minute%
\ A.M.%
\else \advance \hour by -12 \number\hour:%
  \ifnum \minute<10 0\fi%
  \number\minute%
  \ P.M.%
\fi%
}

\makeatother

\begin{document}

\baselineskip=18pt  
\numberwithin{equation}{section}  
\allowdisplaybreaks  


%
%


\thispagestyle{empty}

\vspace*{-2cm} 
\begin{flushright}
\end{flushright}

\vspace*{0.8cm} 
\begin{center}

{\huge Higher-Form Symmetries, Bethe Vacua, and the \\
\bigskip
3d-3d Correspondence 
 }\\

 \vspace*{1.5cm}
{Julius Eckhard$^1$, Heeyeon Kim$^1$, Sakura Sch\"afer-Nameki$^1$, Brian Willett$^2$}\\

 \vspace*{1.0cm} 
{\it $^1$ Mathematical Institute, University of Oxford \\
Woodstock Road, Oxford, OX2 6GG, United Kingdom}\\
\bigskip

{\it $^2$ Kavli Institute for Theoretical Physics\\
University of California, Santa Barbara, CA 93106, USA}\\
\bigskip

\end{center}

\vspace*{1cm}

\noindent 
By incorporating higher-form symmetries, we propose a refined definition of the theories obtained by compactification of the 6d $(2,0)$ theory on a three-manifold $M_3$. This generalization is applicable to both the 3d $\mathcal{N}=2$ and $\mathcal{N}=1$ supersymmetric reductions.
An observable that is sensitive to the higher-form symmetries is the Witten index, which can be computed by counting solutions to a set of Bethe equations that are determined by $M_3$. This is carried out in detail for $M_3$ a Seifert manifold, where we compute a refined version of the Witten index. In the context of the 3d-3d correspondence, we complement this analysis in the dual topological theory, and determine the refined counting of flat connections on $M_3$, which matches the Witten index computation that takes the higher-form symmetries into account.

\newpage

\tableofcontents


\section{Introduction}

The compactification of higher dimensional Quantum Field Theories has led to a deeper understanding of the physical properties of the lower dimensional theories, especially their dualities and symmetries.   One well-studied example is the compactification of the 6d $\cN=(2,0)$ SCFT with ADE Lie algebra $\mathfrak{g}$ on a three-manifold $M_3$, with a topological twist along $M_3$.  The choice of topological twist determines the amount of preserved supersymmetry in 3d, and we find either a 3d $\mathcal{N}=2$  \cite{Dimofte:2011ju, Terashima:2011qi,Dimofte:2011py,Cecotti:2011iy} or a 3d $\mathcal{N}=1$ theory \cite{Eckhard:2018raj}. These theories are often referred to as $T[M_3]$; more explicitly, we can write $T_{\mathcal{N}=1,2} [M_3,\g]$, when we need to specify the additional data.  These theories depend on the topological manifold $M_3$, and many of their detailed properties can be understood in terms of the topology of $M_3$.

For each of these theories, there exists a 3d-3d correspondence with a `dual' 3d topological theory, which is obtained by considering the 6d theory on the space
\be \label{M3W3}
M_3 \times W_3\,.
\ee
The partition function of this topological theory on $M_3$ is conjectured to compute the partition function of the  supersymmetric theory, $T[M_3]$, on $W_3$. The most prominent choices for $W_3$ for which such 3d-3d duals have been discussed are the squashed three-sphere $S^3_b$ \cite{Dimofte:2011ju,Terashima:2011qi,Alday:2017yxk}, the superconformal index on $S^2 \times S^1$ \cite{Dimofte:2011py}, and the twisted index on $\Sigma_g \times S^1$ \cite{Gukov:2015sna}. The special case $W_3=T^3$ was considered in \cite{Gukov:2015sna, Eckhard:2018raj} and computes the (regularized) Witten index \cite{Witten:1999ds} $I=\Tr(-1)^{\text{F}}$. \footnote{More recently, the partition function on general Seifert manifolds have been considered \cite{Closset:2017zgf,Closset:2018ghr} - for a recent review see \cite{Closset:2019hyt} - although a 3d-3d dual for these has not yet been proposed.}

However, an important characteristic of the theories has so far been largely ignored: their higher-form symmetries \cite{Gaiotto:2014kfa}.  Namely, the theory $T[M_3]$ has a higher-form symmetry, which, as with other properties of $T[M_3]$, is determined by the topology of $M_3$.  In fact, the theory $T[M_3]$ is not fully specified by the manifold, $M_3$, but requires additional topological data.  This is related to the fact that the 6d $\cN=(2,0)$ theory itself is a relative QFT, \ie, it is only well-defined as the boundary of a 7d TQFT \cite{Witten:1998wy,Freed:2012bs}. Equivalently, its observables depend on a choice of {\it polarization}, \ie, a choice of maximal isotropic subgroup of $H^3(M_3,\ZGtilde)$, where $\ZGtilde$ is the center of the simply connected group, $\widetilde{G}$, with Lie algebra $\mathfrak{g}$.
Naturally, we expect that $T[M_3]$ also depends on the polarization, as is the case for 4d theories \cite{Aharony:2013hda,Tachikawa:2013hya}. In fact, we will see that this additional information translates into the residual 0- and 1-form symmetry of the 3d theory. We propose therefore a refined definition of the theories, which specifies this data
\be \label{T[M3,g,H]}
T[M_3,\mathfrak{g},H]\,, \qquad H\leq \widehat{\Upsilon} \equiv H^2(M_3,\ZGtilde)\,.
\ee
This theory has a discrete 0-form (ordinary) symmetry group $H$. Its residual 1-form symmetry is given by the complementary subgroup, $\Upsilon_H$,\footnote{This is defined as the set of elements, $\gamma$, in $\Upsilon \equiv H^1(M_3,\ZtG)$ with $\gamma \cup \omega=0$ for all $\omega \in H$.} inside $H^1(M_3,\ZtG)$. We show that the choice of $H$ can indeed be detected by the Witten index or, more generally, by the partition function on any $W_3$ with non-trivial homology. Thus, the different theories in \eqref{T[M3,g,H]} are indeed physically distinct.

The main interest of this paper is to develop a sound definition of  the theories $T[M_3,\mathfrak{g},H]$ in \eqref{T[M3,g,H]} for $M_3$ a {\it graph manifold} \cite{MR235576}, a class of three-manifolds we review in section \ref{sec:SeifertManifolds}. These manifolds, which also occur as the boundary of plumbed four-manifolds, are sometimes called plumbed three-manifolds, a special case of which are Seifert manifolds. Similar Lagrangians for the 3d  theories associated to these manifolds were studied in \cite{Gadde:2013sca,Gukov:2016gkn,Pei:2016rmn, Gukov:2017kmk,Alday:2017yxk,Dedushenko:2018bpp,Cheng:2018vpl}. 

In the following we point out new features related to the global structure of the gauge groups and higher-form symmetries of these theories, as well as the explicit computation of the Witten index and related observables. 
The approach we take is as follows:
\begin{enumerate}
	\item
	Graph manifolds can be cut along disjoint embedded tori into degree $k$ circle fiber bundles, $M_{g,k}$, over genus $g$ Riemann surfaces with boundary. This allows us to assign to each graph manifold, $M_3$, a graph, $\Graph$, as follows. Each copy of $M_{g,k}$ is represented by a vertex, dressed by the degree $k$ and the genus $g$. These are connected by edges representing gluing along boundary tori by $S$-transformations, which have the effect of exchanging the cycles of the boundary tori.\footnote{Gluing by more general large diffeomorphisms can be achieved by decomposing these into products of $S$ and $T$ generators, as we discuss in section \ref{sec:SeifertManifolds}.}
	From the graph we define theories $\widehat{T}[\Graph,G]$ with the building blocks
	\be\label{RulesThat}
	\ba
	\text{vertex}_{g,k}:& \quad &&\CN=2~ G_k\ \text{CS theory + }\CN=4 ~g~\text{adjoint hypermultiplets}\\
	\text{edge}: &\quad &&T(G)\ \text{$S$-wall theory}\,,
	\ea
	\ee
where $T(G)$ was defined in \cite{Gaiotto:2008ak}.  In addition to \eqref{RulesThat} we need to couple the theory to $\mathcal{N}=2$ adjoint scalar multiplets via holomorphic moment maps for the $G$ symmetries. Note that $T(G)$ has flavor symmetry group $G \times G^\vee$, but here we take all gauge fields in $G$. 
	\item
	The graph $\Graph$ representing $M_3$ is not unique. Instead, it represents a four-manifold $M_4$ with boundary $M_3$. There are various operations that leave $M_3$ topologically invariant but change $\Graph$.
	We can check that $\widehat{T}[\Graph,G]$ is only invariant under these operations up to decoupled topological sectors \cite{Witten:2003ya}.
	
	The physical theory $T[M_3,\mathfrak{g}]$ should only depend on $M_3$ itself.
	Inspired by \cite{Hsin:2018vcg}, we propose that we can decouple the additional topological factors on which the physical 1-form symmetry acts trivially.
	Then, different choices of $\Graph$ correspond to dual descriptions of $T[M_3,\mathfrak{g}]$, confirming various dualities \cite{Jafferis:2011ns,RW2019}.
	
	\item
	We denote the theory obtained in this way by $T[M_3,\mathfrak{g},1]$.  It has no discrete 0-form symmetry and a global anomaly-free 1-form symmetry with group $H^1(M_3,\ZGtilde)$. Any subgroup of this can be gauged and we can always find a subgroup such that the new 0-form symmetry is $H$, giving a theory we denote $T[M_3,\g,H]$.
	
\end{enumerate}

\begin{table}
\centering
	\begin{tabular}{c|c}
		Theory &  Definition  \cr \hline \hline 
		$\widehat{T}[\Graph, G]$ & Quiver theory for Graph manifold $M_3$ associated to a graph $\Graph$. \cr 
		& If $G= G^\vee$, then $\widehat{T}[\Graph, G] = T[M_3, G]$. \cr  \hline 
		$T[M_3, \mathfrak{g}]$ & Theory obtained by decoupling topological sectors from $\widehat{T}[\Graph, G]$   \cr
		&Independent of four-manifold.  \cr \hline 
		$T[M_3, \mathfrak{g}, H]$ & $T[M_3, \mathfrak{g}]$ with a gauged 1-form symmetry, \cr 
		& $H$ is resulting 0-form symmetry \cr 
		& $T[M_3, \mathfrak{g}] = T[M_3, \mathfrak{g}, 1]$ 
	\end{tabular}
	\caption{Summary of the theories defined in this paper. 
		The theories can be defined for both $\mathcal{N}=2$ and $\mathcal{N}=1$ preserving compactifications of the 6d $(2,0)$ theory. 
		\label{tab:SummaryT}
	}
\end{table}

Having established a Lagrangian description for $T[M_3,\mathfrak{g},H]$, we can exploit the Bethe/gauge correspondence \cite{Nekrasov:2009uh,Nekrasov:2014xaa} to compute the Witten index. The twisted superpotential of $\widehat{T}[\Graph,G]$ is obtained from its quiver description, where the building blocks, corresponding to \eqref{RulesThat}, have been determined in \cite{Nekrasov:2014xaa,Gukov:2015sna,Closset:2016arn,Benini:2016hjo,Closset:2017zgf}. From this we can determine the Bethe equations, whose solutions determine the ``Bethe vacua,'' forming a special subsector in the Hilbert space of $\widehat{T}[\Graph,G]$ on $T^2$. 
With this method we can in principle compute them for any choice of $\Graph$ and $G$. We will do this explicitly for $M_3$ a Seifert manifold with $b_1=0$ and $G=SU(2)$. 

As mentioned above, this Hilbert space is unphysical as it depends explicitly on the choice of graph, $\Graph$. However, we outline a procedure to decouple the topological sector by projecting out certain vacua.  To do this, we must understand how the 1-form symmetry of the theory acts on the space of Bethe vacua.  We characterize this action in detail, and in particular derive the factorization of the Hilbert space into two decoupled sectors, one physical and one purely topological, such that the Hilbert space of the $T[M_3,\g] \equiv T[M_3,\g,1]$ is given by the former.  This yields the Witten index of $T[M_3,\g]$.   We may then further refine the Hilbert space into eigenspaces under the action of the 1-form symmetry of $T[M_3,\g]$, and we refer to the trace in the various sectors as the {\it refined Witten index}. We can then compute the Witten index for all choices of $H$ in $T[M_3,\mathfrak{g},H]$ in terms of this refined Witten index.  Already for these simple examples we can confirm that the Witten index detects the choice of $H$ and so the theories in \eqref{T[M3,g,H]} are physically distinct.  

The 3d-3d correspondence is one of the original motivations for this work. Taking the setup in \eqref{M3W3}, we can either reduce along $M_3$ to obtain $T[M_3]$, or along $W_3$, which yields a 3d topological field theory on $M_3$. This TFT depends on the choice of $W_3$ as well as the choice of topological twist \cite{Terashima:2011qi,Dimofte:2011ju,Eckhard:2018raj}. This is summarized in table \ref{tab:3d3d}.  Similar conjectures have also appeared for M5-branes on $\Sigma \times W_4$ \cite{Alday:2009aq} and $M_4 \times W_2$ \cite{Gadde:2013sca,Assel:2016lad,Gukov:2018iiq}.
In our setup, we expect that the Witten index of $T_{\mathcal{N}=2}[M_3,\mathfrak{g},H]$ counts complex flat connections on $M_3$, as discussed also in \cite{Chung:2014qpa}. However, we find the choice of $H$, and more generally the action of the higher-form symmetries, has an important impact on the precise 3d-3d dual observable.  As described in section \ref{sec:3d3dcorr}, we find that the refined Witten index of $T[M_3,\g]$ in a sector of fixed 1-form symmetry charges maps to the number of flat $\tG^\C$ connections on $M_3$ with prescribed values for their second Stiefel-Whitney class and behavior under large gauge transformations.  This can  be understood by reversing the order of compactification and studying the 4d $\cN=4$ Super Yang-Mills theory with gauge group $\tG$.  Then S-duality of this theory maps to modular invariance of the 1-form symmetry charges, and leads to an interesting constraint on the flat connections of $M_3$.

\begin{table}
	\centering
	\begin{tabular}{c||c|c}
		3d SUSY & $W_3= T^3$ & $W_3=L(p,1)$\\
		\hline
		\hline
		$\mathcal{N}=2$ & complex  BF-model  & Complex Chern-Simons (CS) theory at level $p$ \\
		\hline
		$\mathcal{N}=1$ & BFH-model & Real Chern-Simons-Dirac theory at level $p$
	\end{tabular}
	\caption{The $\mathcal{N}=2,1$ 3d-3d Correspondences, with different observables $W_3=T^3$ corresponding to the Witten index, and $W_3=L(p,1)$ to the Lens space partition function. \label{tab:3d3d}}
\end{table}

\begin{figure}
	\centering
	\includegraphics[width=4cm]{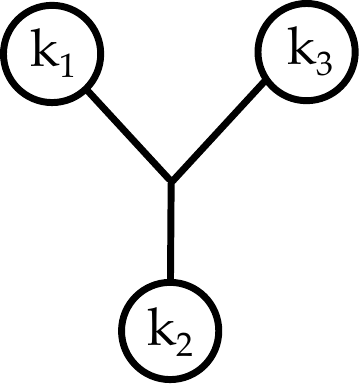}
	\caption{Gauged Trinion: this is the quiver for $T[M_3,\frak{su}(2)]$, where $M_3$ is a Seifert manifold over $S^2$ with special fibers $(k_i,1)$, $i=1,...,3$.  Here the circles denote $SU(2)$ gauge nodes, and the labels are the Chern-Simons levels.}
	\label{fig:gaugedtrinion}
\end{figure}

Let us mention a simple class of examples which illustrates some of these features, and which will be one of the  main examples in the following.  Consider a Seifert manifold, $M_3$, with base $S^2$ and three special fibers of type $(k_i,1)$, $i=1,2,3$ (see section \ref{sec:SeifertManifolds} for more details).  This class of three-manifolds includes the lens spaces, $L(p,q)$, quotients $S^3/\Gamma_{ADE}$, where $\Gamma_{ADE}$ is a finite subgroup of $SU(2)$, and other Brieskorn rational homology spheres.  Then we find a description for $T[M_3,\frak{su}(2)]$ in terms of a {\it gauged trinion}: starting with the $T_2$ trinion theory, given by a trifundamental chiral multiplet of $SU(2)^3$, we gauge the three $SU(2)$  symmetries with CS levels $k_i$, $i=1,2,3$; the quiver is shown in figure \ref{fig:gaugedtrinion}.\footnote{This same description holds also for more general $A_{N-1}$ types, but there one must use the star-shaped quiver \cite{Benini:2010uu} dual description of the trinion in order to write down a Lagrangian for these theories.}  More precisely, depending on 
the $k_i$, this theory may contain decoupled topological sectors which must be projected out.  This simple description passes many non-trivial consistency checks.  For example, it possesses the expected $0$-form and 1-form symmetries.  Moreover, we compute the refined Witten index of this theory and find that it matches the detailed properties of flat $PSL(2,\C)$ and $SL(2,\C)$ connections on $M_3$.

The remainder of this paper is structured as follows. In section \ref{sec:TM3} we review the description of graph manifolds in terms of a graph, $\Graph$.  We prescribe how to translate $\Graph$ into the theory $\widehat{T}[\Graph,G]$, and mention the possible operations on the graph, which also  serves to set our notation. In section \ref{sec:HigherFormSymm} we briefly review some relevant aspects of higher-form symmetries. This is used in section \ref{sec:higherform} to define the theories $T[M_3,\mathfrak{g},H]$. We recall the 6d interpretation of these theories and give an interpretation of the 1-form symmetry in terms of the geometry of $M_3$. This is illustrated in various examples. In section \ref{sec:ResultsU(N)} we explicitly compute the Witten index of $T_{\mathcal{N}=2}[M_3,\mathfrak{g},H]$ for $M_3$ a Seifert manifold and $\mathfrak{g}=\su(2)$ and $\mathfrak{u}(2)$ by solving the corresponding Bethe equations.  In section \ref{sec:3d3dcorr} we discuss the interpretation of these results  in the 3d-3d correspondence, where the Witten index computations can be interpreted as counting flat $G^\C$ connections on $M_3$, refined according to their topological type, which we verify in several examples.
We discuss the $\mathcal{N}=1$ twist in section \ref{sec:3dN=1}. In particular, we use an $\mathcal{N}=2$ enhancement point to propose a Lagrangian description of $T_{\mathcal{N}=1}[L(p,q),\mathfrak{g},H]$ and compute the Witten index.  The appendices contain details on counting solutions to Bethe equations, a discussion of 1-form symmetries in the context of Bethe vacua, and an analysis of the flat connections on $S^3/\Gamma_{ADE}$.  Appendix \ref{app:NotNom} contains a table summarizing our notations.

\section{$T[M_3]$ from Graphs}
\label{sec:TM3}

In this section we begin our study of  the theories $T[M_3]$ obtained by compactification of the 6d $(2,0)$ theory on a Seifert, or more generally, a graph manifold, $M_3$. 
We introduce some of the basic concepts, such as the topological twist, the quiver description of the resulting 3d theories, and dualities.   In particular, as we describe below, we may associate a graph to a decomposition of a three dimensional graph manifold along disjoint embedded tori, and this graph directly determines a quiver gauge theory description.

In this section, we will treat the global structure of the gauge group in a naive way by taking the gauge group to be $G$ for every node in the quiver gauge theories below.  As we will see, this prescription means that the quiver theories we associate to a graph, $\Graph$, are not, in general, the theories $T[M_3]$ obtained by compactification, but rather some closely related theories we define as
\be 
\widehat{T}[\Graph,G]\,.
\ee
When $G$ is equal to its Langlands dual $G^\vee$ (\ie,  the lattice of weights of $G$ is self-dual), then we may define
\be
T[M_3,G] = \widehat{T}[\Graph,G] \,,
\ee
where $\Graph$ is any graph decomposition of $M_3$.  However, in general, two graphs, $\Graph$ and $\Graph'$, which give rise to the same three-manifold do not always give the same 3d theory.  In the following sections we will describe how to pass from the theory $\widehat{T}[\Graph,G]$ to the more physical $T[M_3,G]$ theories, which involves a more careful treatment of the global structure of the gauge groups.

The 6d $\mathcal{N}=(2,0)$ SCFT has 16 supercharges and exists for any $ADE$ Lie algebra. 
Here, we will focus on  $\mathfrak{g}=A_{N-1}$, as this is related to the SCFT that is conjectured to be the effective theory on a stack of $N$ M5-branes, which has gauge algebra $\mathfrak{u}(N)$.  Now consider a three-manifold $M_3$, and compactify the 6d theory on $M_3$ down to 3d. Supersymmetry is preserved for a non-flat $M_3$ only if a suitable R-symmetry background field is turned on, \ie, the theory is partially topologically twisted. 
There are essentially two choices, which result in 3d $\mathcal{N}=2$ and 3d $\mathcal{N}=1$ theories, respectively. The latter will be discussed in section \ref{sec:3dN=1}.
These theories arise by wrapping M5-branes on supersymmetric cycles: either special Lagrangian three-cycles inside Calabi-Yau three-folds or associative three-cycles inside seven-manifolds with holonomy $G_2$, respectively. 

Let us briefly recall how the $\mathcal{N}=2$ twisted version is obtained. 
The local Lorentz group $SO(3)_M$ of $M_3$ gets twisted with an $SU(2)$ subgroup of the $Sp(4)_R$ R-symmetry, $Sp(4)_R\to SU(2)_R \times U(1)_R$. Geometrically, this corresponds  to the action of $SU(2)_R$ on the normal bundle, $N_{M_3}\cong T^*M_3$, of the Lagrangian cycle inside the local Calabi-Yau three-fold. Defining $SU(2)_{\text{twist}}=\diag(SO(3)_M,SU(2)_R)$, the supersymmetry parameters decompose as
\be \label{TwistDGG}
\ba
SO(1,5)_L\times Sp(4)_R  \quad &\to \quad SO(1,2)_L \times SU(2)_{\text{twist}} \times U(1)_R \,,\\
(\bold{4},\bold{4}) \quad &\mapsto \quad \underline{(\bold{2},\bold{1})_{\pm\half}} \oplus (\bold{2},\bold{3})_{\pm\half}\,,
\ea
\ee
preserving four supercharges. We refer to this twist as the {\it{$\mathcal{N}=2$ twist}} and the 3d theory is $T[M_3,G]$ or $T_{\mathcal{N}=2}[M_3,G]$, where here we take $G=U(N)$, or more generally, any self-dual ADE Lie group.
The conjectured 3d-3d correspondence relates this theory to a `dual' 3d topological theory, which depends on the transverse three-dimensional spacetime. In particular if the 6d geometry is $M_3\times T^3$, the partition function of the topological theory that is obtained by reducing first along the $T^3$ is conjectured to compute the Witten index of the theory $T[M_3]$  \cite{Dimofte:2010tz}. 
For the $\mathcal{N}=2$ twist the topological theory is a complex BF-model, and the Witten index of $T[M_3,G]$ is conjectured to be computed by flat $G_{\mathbb{C}}$-connections.   We will return to this interpretation in section \ref{sec:3d3dcorr}.

A useful point of view that was already observed in \cite{Dimofte:2011ju} is to consider the theory on $M_3$ as arising from a 4d $\mathcal{N}=2$ class $\mathcal{S}$ theory $T[\Sigma_{g,n}]$ with boundary conditions. Geometrically this means that locally we embed the curve $\Sigma_{g,n}$ of genus $g$ and $n$ punctures, into $T^* \Sigma_{g,n}$, which is a local K3. The BPS equations for the class $\mathcal{S}$ theory are the Hitchin equations \cite{Gaiotto:2009hg}
\be
F_{z\bar{z}} - [\Phi, \Phi^\dagger] =0 \,,
\ee
where $F_{z{\bar z}}$ is a $(1,1)$-form on $\Sigma$ and $\Phi$ is a section of $K_{\Sigma}\otimes \text{Adj}(G_\mathbb{C})$.
The connection between these equations and the complex flat connections is well-known, by considering $M_3 = \Sigma \times \mathbb{R}$. 

In this paper the main focus will be on a class of three-manifolds, $M_3$, that are naturally associated to graphs, a special case of which are Seifert manifolds. In this section, we define these graph manifolds, and associate quiver gauge theories to them. Geometric identities become dualities in the quivers. We illustrate these in the case of the simplest non-abelian groups, $U(2)$ and $SU(2)$. 

\subsection{Seifert and Graph Manifolds}
\label{sec:SeifertManifolds}

The class of geometries $M_3$ that we will consider here include Seifert manifolds (see \eg,  \cite{MR0426001}), which are 
circle bundles over a Riemann surface $\Sigma_{g,r}$ of genus $g$ with $r$ marked points
\be\label{Seifertfib}
\pi:\ S^1 \hookrightarrow M_3 \to \Sigma_{g,r}\,.
\ee
The fibration is specified by the Seifert data
\be \label{DefSeifert}
M_3\cong[d;g;(p_i,q_i)]\,, \quad i=1,\dots,r\,,
\ee
where $(d,q_i)\in \Z$ and $p_i\in \Z_+$. Furthermore, $p_i$ and $q_i$ are coprime. Away from the punctures, $M_3$ is a smooth degree $d$ circle fibration over $\Sigma_g$. 
The exceptional fibers are located above the marked points. Around each puncture, excise a tubular neighborhood which together with the fiber forms a solid torus, $S^1 \times D^2$.
On each of these boundaries we attach a mapping class torus, \ie, a $T^2$ fibered over an interval  $x_3 \in [0,L]$. We view the torus as a complex manifold with complex structure $\tau$ and $SL(2,\mathbb{Z})$ action $\tau \rightarrow (r \tau+p)/(s \tau +q)$, where\footnote{Here $S$ and $T$ are the generators of $SL(2,\Z)$, for which we take the explicit representations $S=\begin{pmatrix}
	0 & -1 \\ 1 & 0
	\end{pmatrix}$ and $T=\begin{pmatrix}
	1 & 1 \\ 0 & 1
	\end{pmatrix}$.}
\be\label{varphiDef}
\varphi_{p,q}=\begin{pmatrix}
	r & p \\ s & q
\end{pmatrix}=T^{k_1}ST^{k_2}\dots ST^{k_n} \in SL(2, \mathbb{Z})\,,
\ee
and $r,s$ encode the framing, \ie, the choice of trivialization of the tangent bundle.
The integers 
 $k_j$ are given by the continued fraction
\be\label{poverq}
\frac{p}{q}=k_1-\frac{1}{k_2-\frac{1}{\dots-\frac{1}{k_n}}}\equiv [k_1,\dots,k_n]\,.
\ee
This representation ensures that the $(0,1)$-cycle at $x_3=0$ is identified with the $(p,q)$-cycle at $x_3=L$. Here and in the following we denote the $A$- and $B$-cycle of the boundary of a solid torus as the $S^1$ and the $\p D^2$ respectively and denote general cycles in the $(A,B)$-basis. To close the boundary we then glue back in a solid torus.

We will also consider a generalization of Seifert manifolds to so-called {\it graph manifolds}, which are glued together from Seifert manifolds building blocks.  More formally, they can be defined as three-manifolds, which can be cut along tori, where the resulting building blocks are circle-bundles over Riemann surfaces.
There are various equivalent ways to characterize graph manifolds, however here we will make use of the formulation in terms of a {\it plumbing graph}. This construction will first of all produce a four-manifold, whose boundary is the graph manifold. 
The graph will have vertices $v_i$ corresponding to disc-bundles over a Riemann surface,
\be
v_i:\qquad D_i\hookrightarrow M_i \rightarrow \Sigma_i  \qquad \leftrightarrow \qquad   (e(M_i), g(\Sigma_i)) \;,
\ee
where $e(M_i)$ is the Euler number of the disc-bundle and $g$ is the genus of the base curve.   We include edges between $v_i - v_j$ if there is a gluing between two disc bundles as follows: 
consider $D_i \times B_i$, where $B_i \subset \Sigma_i$ is a disc. Then gluing the discs across (\ie, gluing the fibral disc with the disc within the base, and vice versa)
$D_i \times B_i$ with $B_j \times D_j$ corresponds to an edge along which the two vertices $v_i$ and $v_j$ are glued. 
The boundary of such a four-manifold is automatically a graph manifold. 

There are redundancies in the definition of the plumbing data which leave the boundary - \ie, the graph manifold - invariant. For example, there are operations which change the 4-manifold by taking the connected sum with $\mathbb{P}^2$ and with $\overline{\mathbb{P}^2}$, but change the 3-manifold by a connected sum with $S^3$, which is a trivial operation \cite{Gadde:2013sca}.  We will discuss these identities and their implications on the graph shortly.  For a general discussion of graph manifolds see \cite{PedersenThesis}.

\begin{figure}
	\centering
	\includegraphics[width=6cm]{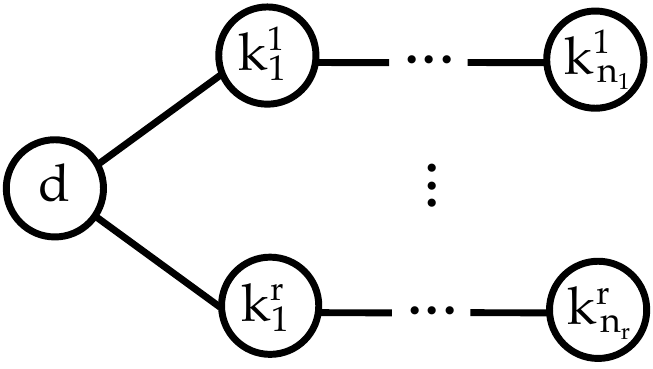}
	\caption{Plumbing graph for Seifert manifolds with $r$ special fibers. For $r=3$ this graph corresponds to the Seifert quivers in the text. We consider here $g=0$ for the central node. This also gives the quiver for the theory $\widehat{T}[\Graph_{[d; 0, [k^i_1,\cdots,k^i_{n_i}]]},G]$.}
	\label{fig:QuiverGeneral}
\end{figure}

The case of Seifert manifolds is a specialization, where the plumbing graph is given by figure \ref{fig:QuiverGeneral}, and the Seifert data is encoded in the continued fractions $p_i/q_i$ given by (\ref{poverq}). In general there is a choice of genus for the central node, which we generally take to be $g=0$ in the following. 

The simplest class of examples are the Lens spaces, $M_3=L(p,q)$, which are obtained by gluing two solid tori, $(S^1\times D^2)_\pm$, identifying the $(0,1)_+$ cycle with the $(p,q)_-$ cycle. This is known as the Heegaard splitting of the Lens space, see \eg, \cite{MR1886684,MR1712769}.
We can easily see that this definition corresponds to Seifert manifolds with $g=0$ and $r\leq2$.
First consider $r=0$, \ie, $M_3$ is a degree $d$ circle fibration over $S^2$. This is a definition of $L(d,1)$ in terms of a generalization of the Hopf fibration. We thus assign to $L(d,1)$ the element $\varphi_{d,1}=T^d$. Now add a puncture with exceptional fiber $(p_1,q_1)$. Excising the tubular neighborhood of the puncture leaves us with a circle-fibration over the disk. We now glue this to the solid torus as explained above. 
After the gluing we obtain
\be \label{varphi1fiber}
\varphi_{[d;0;(p_1,q_1)]}=T^d S \varphi_{p_1,q_1}=\varphi_{d p_1 -q_1,p_1}\quad \Rightarrow \quad
[d;0;(p_1,q_1)]\cong L(d p_1 -q_1,p_1)\,.
\ee
The generalization to two exceptional fibers is obvious. By exchanging the two fibers one can see that $L(p,q)\cong L(p,q^{-1}\mod p)$.

Introducing a third marked point fundamentally changes the geometry and the resulting Seifert manifold is no longer a Lens space (for generic $(p_i,q_i)$). The description in terms of a linear chain of $S$ and $T$ in $SL(2,\Z)$ breaks down but we will present a natural way to couple the central node to three fibers when discussing the quiver description in section \ref{sec:SeifertQuivers}. In a similar fashion we can translate the plumbing graph of a general graph manifold into a web of vertices $M_{g,k}$, degree $k$ circle bundles over $\Sigma_g$, connected by $S$-transformations.  In what follows, we will primarily be concerned with $M_{0,k}$, with $g=0$, and will label the vertices only by $k$.

\begin{figure}
	\centering
	\includegraphics[width=15cm]{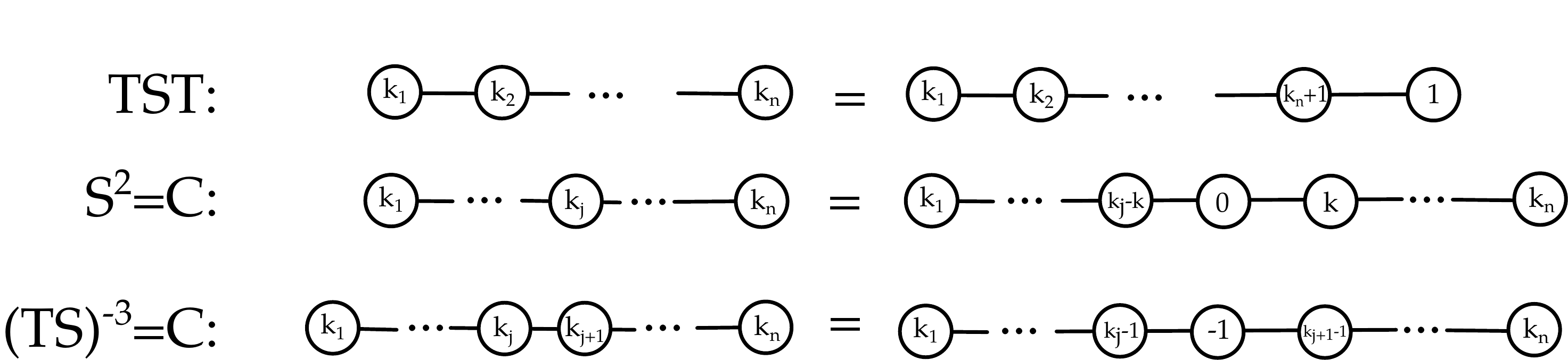}
	\caption{The relations $TST$, $S^2$ and $(TS)^{-3}$ acting on the plumbing data associated to the three-manifold. 
		\label{fig:Plumbs} 
}\end{figure}

The continued fraction presentation of $p/q$ in (\ref{poverq}) is not unique, which implies a non-uniqueness of the Seifert data or, equivalently, the ambient four-manifold. These relations are summarized in figure \ref{fig:Plumbs}, and extend straightforwardly to more general graph manifolds:
\begin{enumerate}[label=\roman*)]
	\item \label{TSTSymm} $[k_1,\dots,k_n]\cong[k_1,\dots,k_n+1,1]$:\\
	This corresponds to adding a factor of $TST$ on the right hand side of $\varphi_{p,q}$. This leaves $(p_i,q_i)$ invariant but affects the framing of the manifold and corresponds to taking the connected sum of $M_3$ with an $S^3$.
	We can also attach this to the central node, adding a new fiber with $p_{r+1},q_{r+1}=(1,1)$, while increasing the degree by one. Thus,
	\be \label{SeifertEquivalencesTST}
	[d;g;(p_i,q_i)]\cong [d+1;g;(p_i,q_i),(1,1)]\,.
	\ee
	\item $[k_1,\dots,k_j,\dots,k_n]\cong[k_1,\dots,k_{j}-k,0,k,\dots,k_n]$ for any $k$: \\
	This corresponds to including a factor of $S^2=C$. This sends $(p,q)\to(-p,-q)$ leaving $p/q$ invariant.
	\item $[k_1,\dots,k_j,k_{j+1},\dots,k_n]\cong[k_1,\dots,k_j-1,-1,k_{j+1}-1,\dots,k_n]$:\\
	This follows from including a factor of $(TS)^{-3}=C$.  	Inserting this element between the central node and a fiber decreases the degree by one and acts on the fiber element as $\varphi_{p,q}\to \varphi_{p,q-p}$. This generalizes to
	\be \label{SeifertEquivalencesST3}
	[d;g;(p_i,q_i)]\cong [d+\sum_{i=1}^r m_i;g;(p_i,q_i+m_ip_i)]\,,
	\ee
	for any choice of $m_i\in \Z$.
\end{enumerate}

Some examples of interesting Seifert manifolds are given in appendix \ref{app:Seife}. 

Before moving on, we note that an efficient way to characterize the graph, $\Omega$, associated to a graph manifold is by  its linking matrix, which we denote by $Q$. Its entries are defined as
\be \label{LinkingMatrix}
Q_{ij}=\left\{
\begin{array}{c l}
	k_i & \text{if }i=j\\
	-1  & \text{if }i \hbox{ connected to } j \text{ by } S\\
	0 & \text{else}
\end{array}
\right.\,.
\ee
Here the choice of taking off-diagonal terms to be $-1$ is a convention, but will be useful in the definition of $T[M_3]$ below.

\subsection{$\widehat{T}[\Graph, G]$}
\label{sec:That}

We now define the theories $\widehat{T}[\Graph, G]$, which will be the building blocks for all the physical theories that we consider in later sections. It is useful to define this theory in terms of boundary conditions on $\cN=2^*$ gauge theory with gauge group  $G$ that preserve a $U(1)_t$ flavor symmetry. 
For the $\mathcal{N}=2$ twist, the construction in terms of boundary conditions  has been extensively studied in \eg,\cite{Gaiotto:2008sd,Gaiotto:2008ak,Gaiotto:2008sa,Dimofte:2011ju,Gukov:2015sna,Gukov:2016gkn,Gukov:2017kmk,Alday:2017yxk}, which we briefly summarize below.
For example, the Lens spaces can be constructed from the Heegaard splitting with one-punctured torus boundaries, which corresponds to the 4d $\cN=2^*$ theories on an interval with half-BPS boundary conditions. In this picture, the puncture on the boundary torus induces a network of line defects in $M_3$, which correspond to the circle actions in the Seifert fibration \eqref{Seifertfib}.

\subsubsection{Boundary Conditions for 4d $\mathcal{N}=4$ SYM}
\label{sec:BounCond}

The basic building blocks can be constructed from the $\half$-BPS equations of the mass deformed 4d $\mathcal{N}=2^*$ $G$ gauge theory, as studied in \cite{Alday:2017yxk}.
The bosonic degrees of freedom of the 4d theory can be decomposed into that of an $\cN=2$ vector multiplet and a hypermultiplet in the adjoint representation,
$(A_{\mu},~\phi)$ and $(X,~Y)$. 
In the $\cN=2$ language, the theory has a $U(1)_t$ global symmetry, under which the complex scalars $(\phi, X,Y)$ have charges $(0,1,-1)$ respectively.

The basic boundary conditions that can be used to build a $\widehat{T}[\Graph,G]$ theory are
\be
\ba
&|N_+\rangle:\quad D_3 X =0\ ,~ Y=0\ ,~N\ ,\qquad && |N_-\rangle: \quad D_3 Y =0\ ,~ X=0\ ,~N\ ,\\
&|{\cal V}_+ \rangle: \quad  D_3 Y=0\ ,~ X=0\ ,~ D({\cal V})\ ,\qquad && |{\cal V}_- \rangle: \quad  D_3 X=0\ ,~ Y=0\ ,~ D({\cal V})\ , \\
\ea
\label{basic boundary}
\ee
where $N$ and $D({\cal V})$ denote the Neumann and Dirichlet boundary conditions for the vector multiplet respectively
\be
\ba
N:& \qquad F^{3\mu}=0\,, \quad D_3\Re\phi=0\,, \quad \Im\phi=0\\
D(\mathcal{V}):& \qquad F^{\mu\nu}=0\,, \quad D_3\Im\phi=0\,, \quad \Re\phi=a\,,
\ea
\ee
where the Dirichlet boundary condition depends on the choice of background 3d $\mathcal{N}=2$ vector multiplet $\mathcal{V}$ whose lowest component is $a$. For the Neumann boundary condition, the vector multiplet ${\cal V}$ remains dynamical.

The overlaps between these boundary conditions can be obtained by studying the low energy limit of the 4d $\cN=2^*$ theory on an interval with two boundary conditions. First of all, one can check
\be
\langle N_\pm | N_\pm\rangle =  \int_{\mathcal V} d{\cal V}~\text{adj}_\pm({\cal V})\ ,
\ee
where we integrate over the dynamical 3d $\cN=2$ vector multiplet ${\cal V}$. In addition to this, we have the $\cN=2$ chiral multiplet valued in the adjoint representation with $U(1)_t$ charge $\pm 1$, which we denote by $\text{adj}_\pm({\cal V})$. If both types of adjoints are present in the theory, the pair can be integrated out in the low-energy effective theory, which leads to the relation $\text{adj}_+({\cal V})\text{adj}_-({\cal V})=1$.
We also need the following relations
\be\label{relations}
\langle {\cal V}_\pm | {\cal V}_\pm\rangle = \text{adj}_\mp ({\cal V})\ ,~~\langle {\cal V}_\pm | {N}_\pm\rangle =1\ ,~~\langle {\cal V}_\pm | N_\mp\rangle = \text{adj}_\mp ({\cal V})\ ,
\ee
which are straightforward to check by studying the propagating degrees of freedom between the overlaps.

For each $T^2$ boundary with a puncture, we specify a choice of polarization, which corresponds to the $A$- and $B$-cycle of the torus. The gauge symmetry  $G$ of the 4d theory is associated to the $A$-cycles of the tori and the holonomy around the $A$- and $B$- cycle correspond to the Wilson and 't Hooft loop expectation values of the gauge theory, respectively. 

A class of boundary conditions for the $\cN=4$ theory are provided by the compactification of the 6d $(2,0)$ theory on a three-manifold with a torus boundary, where the polarization data is specified. The simplest example is the solid torus, $M_3 = D^2\times S^1$, where the $S^1$ is the $A$-cycle. In this case, we claim that the theory corresponds to the 4d theory with the Neumann boundary condition $|N_+\rangle$ assigned \cite{Alday:2017yxk}.

Seifert manifolds with exceptional fibers can be obtained from the surgery procedure discussed in section \ref{sec:SeifertManifolds}. The surgery around each marked point corresponds to an operation inserting a mapping torus between two boundary tori, which implements an $SL(2,\mathbb{Z})$ transformation \eqref{varphiDef}. The puncture associated to the $U(1)_t$ symmetry defines a line defect connecting two torus boundaries. 
When we glue them with another building block of the three-manifold, the polarization data should be identified.
Each mapping torus corresponds to an $SL(2,\mathbb{Z})$ interface in the 4d gauge theory, which defines an operator acting on the space of boundary conditions \eqref{basic boundary}. We now list the operators and the associated boundary conditions for the building blocks:
\begin{enumerate}
	\item The simplest example is the mapping cylinder for the identity element $\varphi_{0,1}=I$. This corresponds to inserting the identity operator in the space of boundary conditions
	\be\label{identity}
	I = \int_{\cal V} ~|{\cal V}_+\rangle \langle {\cal V}_+|~\text{adj}_+({\cal V})\ .
	\ee
	This can be equivalently expressed in terms of the multiplet $\cal V_-$. Note that the measure includes the contribution from the adjoint multiplet that is compatible with the inner product \eqref{relations}.
	
	\item The mapping cylinder implementing the $T$ operation corresponds to adding a background $\cN=2$ Chern-Simons level to the theory
	\be
	T^k = \int_{\cal V} ~|{\cal V}_+\rangle \langle {\cal V}_+|~\text{adj}_+({\cal V})~e^{-k\text{CS}({\cal V})}\ ,
	\ee
	where the last factor denotes an $\cN=2$ CS theory at level $k$.
	\item The $S$ operation corresponds to the $T(G)$ theory defined in \cite{Gaiotto:2008sa}, which has $G\times G^\vee$ flavor symmetry
	\be
	S = \int_{{\cal V},{\cal V}'} ~|{\cal V}_-\rangle~ T(G)({\cal V}_-,{\cal V}'_+)~\langle {\cal V}_+'|\ .
	\ee
	Using the relation $\langle {\cal V}_+'| = \text{adj}({\cal V}')_-\langle {\cal V}_-'|$, we can write the expression more symmetrically in the two background gauge fields as
	\be\label{STrafo}
	\ba
	S &= \int_{{\cal V},{\cal V}'} ~|{\cal V}_-\rangle ~T(G)({\cal V}_-,{\cal V}'_+)~\text{adj}_-({\cal V}')~\langle {\cal V}_-'| \\
	& = \int_{{\cal V},{\cal V}'} ~|{\cal V}_-\rangle ~FT(G)({\cal V}_-,{\cal V}'_+)~\langle {\cal V}_-'|\ ,
	\ea
	\ee
	where $FT(G)$ is the so-called ``flipped" $T(G)$ theory defined by adding an adjoint field in one of the flavor symmetries to the original description of the $T(G)$ theory \cite{Aprile:2018oau}.
\end{enumerate}

Finally, let us consider the building blocks with more than two torus boundaries. Similarly to the identity operator \eqref{identity}, we claim that $\Sigma_{0,r}\times S^1$ with the $A$-cycle along the $S^1$ identifies all the global symmetries associated to the $r$ punctures. For the three-punctured sphere, we have
\be\label{BCTrinion}
\int_{\cal V} ~|{\cal V}_+\rangle|{\cal V}_+\rangle \langle {\cal V}_+|~\text{adj}_+({\cal V})\ , 
\ee
which can be equivalently expressed in terms of the multiplet ${\cal V}_-$.

\subsubsection{Definition of $\widehat{T}[\Graph, G]$}
\label{sec:THatDef}

A description of $\widehat{T}[\Graph,G]$ can be obtained by a combination of the above building blocks according to the surgery procedure discussed in section \ref{sec:SeifertManifolds}.

We start with the plumbing graph $\Omega$ of a graph manifold $M_3$, which corresponds to a web of vertices $M_{g,k}$ connected by $S$-transformations.
From this we construct $\widehat{T}[\Graph,G]$ in the following way:\footnote{Here we specify the action of the $U(1)_t$ symmetry, which is related to the geometrical $U(1)$ isometry in the case where $M_3$ is a Seifert manifold.  This definition can also be applied formally to graph manifolds, but in that case the existence of a $U(1)_t$ symmetry from 6d is not guaranteed.}
\begin{enumerate}
	\item
	To each vertex $M_{g,k}$ we assign an $\mathcal{N}=2$ gauge multiplet at Chern-Simons level $k$ together with an $\mathcal{N}=2$ adjoint chiral multiplet, which we assign $U(1)_t$ charge $+1$.  We also couple this to a set of $g$ $\mathcal{N}=4$ adjoint hypermultiplets \cite{Benini:2010uu,Gukov:2017kmk}, although in this paper we will mostly consider the case $g=0$.
	This $M_{g,k}$ vertex has $r$ boundaries along which it is glued to neighboring vertices. To each of these boundaries we assign a $\bra{\mathcal{V}_+}$.
	\item
	To each edge we assign the $S$-transformation \eqref{STrafo}, \ie, the $T(G)$ theory.  More precisely, it is convenient to use the second line of \eqref{STrafo} to write this more symmetrically using the $FT(G)$ theory, which is the $T(G)$ theory with one flavor symmetry coupled to an adjoint chiral multiplet of negative $U(1)_t$ charge. To the two flavor groups we assign $\ket{\mathcal{V}_-}$ and $\ket{\mathcal{V}'_-}$ respectively.
	\item
	We glue these building blocks together using the overlap $\braket{\mathcal{V}_+|\mathcal{V}'_-}=\delta\left(\mathcal{V},\mathcal{V}'\right)$.
\end{enumerate}
For $\Graph$ the graph associated to a Seifert manifold, we will show that the theories defined in the above way are closely related -- and in the case of self-dual $G$, identical -- to the theories $T[M_3, G]$. 

In this paper we use the following depiction of the quivers for the $\widehat{T}[\Graph, G]$, where we henceforth assume $g=0$ for all vertices:
\be\label{QuivBuild}
\ba
\text{gauge field $\mathcal{V}$ at CS level $k$} \qquad &\longleftrightarrow \qquad &&\text{round node dressed with $k$}\\
\text{flavor group} \qquad &\longleftrightarrow \qquad &&\text{square node}\\
\text{adj}_\pm(\mathcal{V}) \qquad &\longleftrightarrow \qquad &&\text{node dressed with upward/downward arc}\\
\text{$T(G)(\mathcal{V}_-,\mathcal{V}_+')$} \qquad &\longleftrightarrow \qquad &&\text{line with a downward and upward arc adjacent to}\\ &\phantom{\longleftrightarrow} \qquad &&\text{the Higgs and Coulomb symmetries, respectively}\\\
FT(G) \qquad & \longleftrightarrow \qquad &&\text{line with two downward arcs}\,.\\ 
\ea
\ee

\subsubsection{Examples: Seifert Quivers}
\label{sec:SeifertQuivers}

\begin{figure}
	\centering
	\includegraphics[width=13cm]{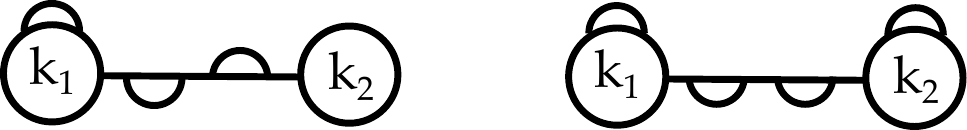}
	
	\caption{Quiver description of Lens space quiver $\widehat{T}[\Graph_{[k_1, k_2]},G]$, where $p/q= k_1 - 1/k_2$ and the dictionary in (\ref{QuivBuild}) is used. On the left we use  the $T(G)$ theory, and on the right the $FT(G)$ theory, which is related by flipping an adjoint field.  This flipping operation can be represented pictorially by contracting an upward arc on a node and an  adjacent downward arc on an edge to leave an upward arc on the edge.  Further equivalent descriptions of this theory are shown in figure \ref{fig:tststduality} below.}
	\label{fig:QuiverTST}
\end{figure}

We now consider some examples of Seifert quivers, \ie, where the graph $\Omega$ describes a  Seifert manifold.  The simplest non-trivial example is the graph $\Omega_{[k_1,k_2]}$ of the Lens space $L(p,q)$ with $\frac{p}{q}=k_1-\frac{1}{k_2}$. It consists of two copies of $M_{0,k_j}$ connected by an edge and therefore
	\be
	\ba
	&\widehat{T}[\Omega_{[k_1,k_2]},G]=\\
	&\int_{\mathcal{V}^{1,2,3,4}} e^{-k_1\text{CS}(\mathcal{V}^{1})}\ \text{adj}_+(\mathcal{V}^{1})\  \underbrace{\braket{\mathcal{V}^{1}_+|\mathcal{V}^2_-}}_{\delta(\mathcal{V}^1,\mathcal{V}^2)}\ T(G)(\mathcal{V}_-^2,\mathcal{V}_+^3)\ \text{adj}_-(\mathcal{V}^3)\ \underbrace{\braket{\mathcal{V}^3_-|\mathcal{V}^4_+}}_{\delta(\mathcal{V}^3,\mathcal{V}^4)}\ \text{adj}_+(\mathcal{V}^{4})\ e^{-k_2\text{CS}(\mathcal{V}^{4})}\,.
	\ea
	\ee
	Using the delta functions we can integrate out $\mathcal{V}^2$ and $\mathcal{V}^3$ after which two adjoints with opposite charge cancel. After these simplifications
	\be
	\widehat{T}[\Omega_{[k_1,k_2]},G]=\int_{\mathcal{V}^{1,4}} e^{-k_1\text{CS}(\mathcal{V}^{1})}\ \text{adj}_+(\mathcal{V}^{1})\ T(G)(\mathcal{V}_-^1,\mathcal{V}_+^4)\ e^{-k_2\text{CS}(\mathcal{V}^{4})}\,.
	\ee
	This theory is depicted in figure \ref{fig:QuiverTST}, illustrating the conventions of \eqref{QuivBuild}.
	More generally, we can write the corresponding quiver for $\Omega_{[k_1,\cdots,k_n]}$ as $n$ $\cN=2$ CS-theories at levels $k_i$ coupled by $T(G)$ theories, with a single adjoint scalar coupled to the left-most node.

For Seifert manifolds with three or more exceptional fibers, \ie,
$M_3\cong[d;0;(p_i,q_i)]$, we label the graph by $\Graph_{[d; 0, [k^i_1,\cdots,k^i_{n_i}]]}$, depicted in figure \ref{fig:QuiverGeneral}.
We identify the degree $d$ with a central $T^d$ coupled to $r$ linear quivers. More generally we understand a node labeled by $\frac{p_i}{q_i}=[k^i_1,\dots,k^i_{n_i}]$ as being the same as the linear quiver representing the Lens space $L(p_i,q_i)$ \cite{Gadde:2013sca}, where we take the left-most node as a $G$ flavor symmetry.  Then, applying the rules above, one finds $\widehat{T}[\Graph_{[d; 0, [k^i_1,\cdots,k^i_{n_i}]]},G]$ is obtained by gauging these common flavor symmetries with level $d$ CS term and a single adjoint chiral multiplet of $U(1)_t$ charge $+1$. 

\subsection{Symmetries and Dualities of $\widehat{T}[\Graph, G]$}
\label{sec:QuiverSymmetries}

\subsubsection{General Gauge Group}

The non-uniqueness of the Seifert data discussed in section \ref{sec:SeifertManifolds}, should map to dualities of the theories $\widehat{T}[\Graph, G]$. However, we will see below that for non-self-dual $G$, the $SL(2,\mathbb{Z})$ actions and equivalence relations of three-manifolds discussed in section \ref{sec:SeifertManifolds} holds only up to a decoupled topological sector at the level of the $\widehat{T}[\Graph,G]$ theory, indicating an undesired dependence on the decomposition graph $\Graph$.  We will discuss how to cure this problem in the following sections. We now summarize the dualities:

\begin{enumerate}[label=\roman*)]
	
	\item\label{dual1} $[k_1,\dots,k_n]\cong[k_1,\dots,k_n+1,1]$:\\
	The equivalence of the quiver under this operation was discussed in \cite{Gadde:2013sca,Gukov:2015sna,Pei:2016rmn}.  In the case of the $G=U(N)$ theory, this is related to the following duality of
	\cite{Kapustin:2011vz}
	\be U(N)_{k=1} \; + \; \text{adjoint chiral} \;\; \leftrightarrow \;\; N \; \text{free chirals}\ . \ee
	We discuss this relation and the role of this duality in more detail in the $N=2$ case below.  For $G=SU(N)$, we use instead the following duality, a special case of which is the duality appetizer of \cite{Jafferis:2011ns} 
	\be \label{sudecoup}
	SU(N)_{k=1} \; + \; \text{adjoint chiral}\;\; \leftrightarrow \;\; (N-1 \; \text{free chirals})\otimes (\text{topological $U(1)_N$ CS theory})\,.
	\ee
	The duality implies the relation \footnote{Up to a relative gravitational CS term which does not affect our discussion.}:
	\be \label{TSTDualitySUN} \text{$[k_1,...,k_n+1,1]$ quiver} \;\;\; \leftrightarrow  \;\;\;( \text{$[k_1,...,k_n]$ quiver} )\otimes (\text{topological $U(1)_N$ CS theory})\,. \ee
	Note the presence of the decoupled topological sector, which is directly related to the fact that $SU(N)$ is not self-dual. This sector will play an important role in what follows.

	\item\label{dual2} $[k_1,\dots,k_i,\dots,k_n]\cong[k_1,\dots,k_{i}-k,0,k,\dots,k_n]$: \\
	This corresponds to two factors of $T(G)$ coupled at a common node and gauged at level zero. However, since we take the gauge field in $G$ (rather than $G^\vee$) this operation adds a topological factor, which for $G=SU(N)$ is $\left(U(1)_N\right)^2$.
	
	\item\label{dual3} $[k_1,\dots,k_i,k_{i+1},\dots,k_n]\cong[k_1,\dots,k_i-1,-1,k_{i+1}-1,\dots,k_n]$: \\
	For $G=SU(N)$, we claim that the $SL(2,\mathbb{Z})$ relation $(TS)^3 = C$ holds only up to the decoupled topological
	$U(1)_N$ CS theory, as in the relation i) discussed above. 
	Let us consider the case $n=2$ and $i=1$ without loss of generality. This implies the duality
	\be
	~[k_1,k_2] \text{ quiver} ~~\leftrightarrow~~ ([k_1-1, -1, k_2-1] \text{ quiver}) \otimes (\text{topological $U(1)_N$ CS theory}) \,.
	\ee
	Using the relation \eqref{TSTDualitySUN} for the theory on the right hand side, we find that this theory precisely corresponds to the trinion theory with the three flavor symmetries gauged at levels $(k_1-1,k_2-1,1)$ respectively.

	\begin{figure}
		
		\begin{centering}
			
			\includegraphics[width=10cm]{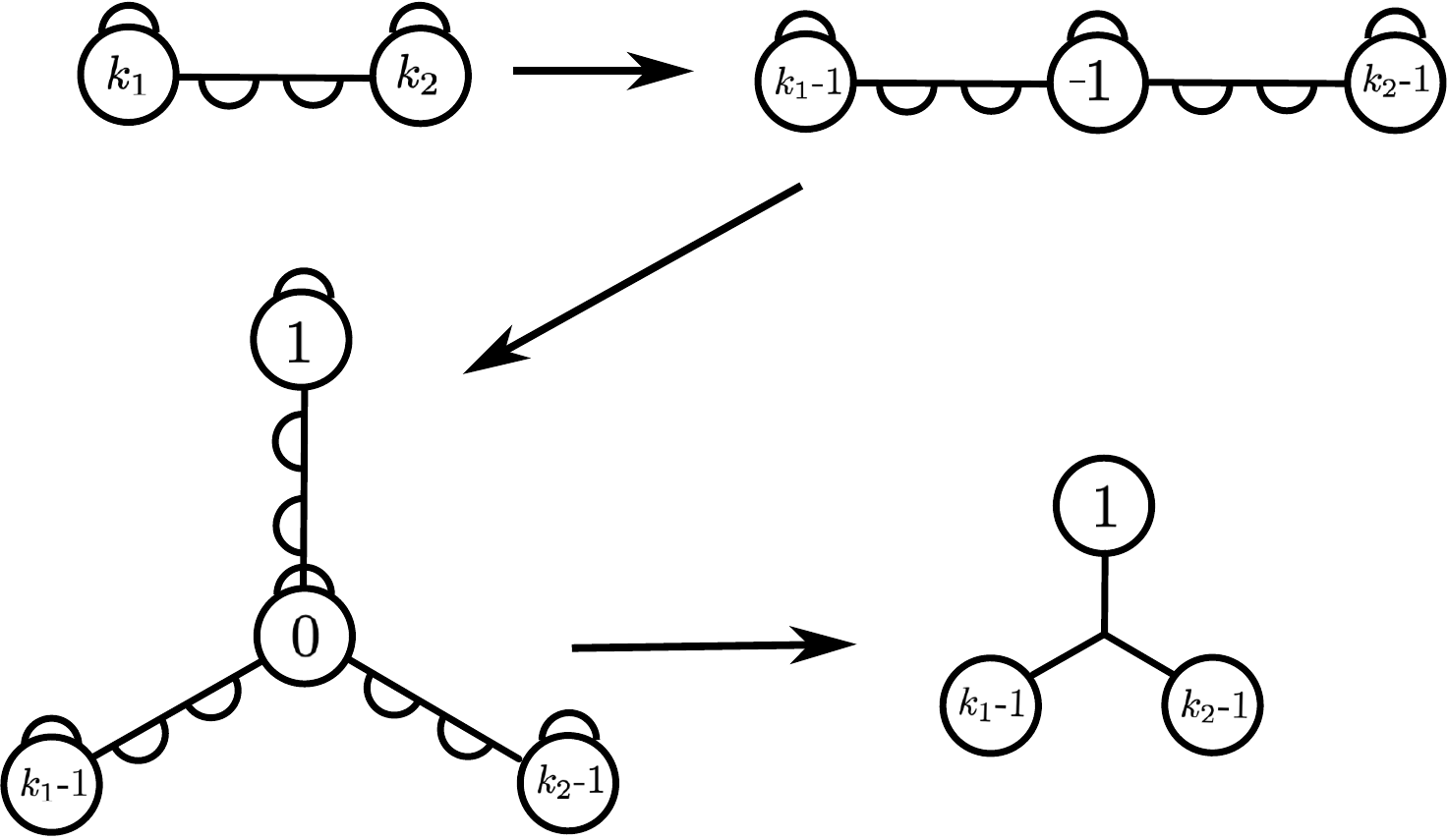}
			
		\end{centering}
		\caption{Sequence of moves leading to the duality between the $FT(G)$ theory and a trinion theory with one $G$ symmetry gauged with level 1 CS term.   In the last step with use the star-shaped quiver description of the trinion of \protect\cite{Benini:2010uu}.}
		\label{fig:tststduality}
	\end{figure}
	
	In fact, we may make a stronger statement.  Rather than taking these nodes with level $k_1$ and $k_2$ CS terms as gauge symmetries, we may retain them as flavor symmetries.  Then, in the former case, we obtain the $T_N$ theory with one of its $G$ flavor symmetries gauged with a level 1 CS term, while in the latter case we obtain the $T(G)$ theory, or more precisely, the flipped theory, $FT(G)$.  Both have $G \times G$ flavor symmetry.  We also notice that the levels of the CS terms are shifted by one in the latter description, which we may attribute to relative contact terms appearing in the duality.  In summary, this yields the following duality
	\bea \label{tgtrinduality} & T_G \;\; \text{trinion theory with one flavor symmetry gauged at level 1} \nonumber\\
	\leftrightarrow &FT(G) \; \text{ with level $-1$ contact terms for $G \times G$ flavor symmetry} \,.
	\eea
	This is illustrated in figure \ref{fig:tststduality}.   This observation was also made in \cite{RW2019} for $G=SU(N)$.  We will discuss this in more detail in the case $G=SU(2)$ below.

\end{enumerate}

\subsubsection{$\widehat{T}[\Graph, SU(2)]$ and $\widehat{T}[\Graph, U(2)]$}
\label{sec:ExSU2}

Let us illustrate some of the properties discussed in the previous subsection in more detail in the special cases $G=SU(2)$ and $G=U(2)$, and also point out some additional subtleties that arise.  

To start, let us briefly recall some relevant features of the $T(SU(2))$ theory.  The usual description of the $T(SU(2))$ theory is as a 3d $\cN=4$ $U(1)$ gauge theory with $N_f=2$ hypermultiplets.  This theory has $SU(2)_F \times SU(2)_J$ flavor symmetry, the former acting manifestly on the hypermultiplets, and the latter being an infrared enhancement of the $U(1)_J$ symmetry visible in the UV gauge theory Lagrangian.  This is the theory living on a domain wall interpolating between $S$-dual instances of the 4d $\cN=4$ $SU(2)$ SYM theory, with each bulk theory coupling to one of the $SU(2)$ flavor symmetries of the domain wall, as discussed in \cite{Gaiotto:2008ak}. 

Now let us consider the relation \ref{dual1} above in the case of $G=SU(2)$.  Consider a linear quiver associated to the RHS of this relation: this ends with an $S$-wall, followed by a $SU(2)$ gauge node with a CS level 1.  Let us take the latter $SU(2)$ symmetry to couple to the $SU(2)_F$ symmetry of the $T(SU(2))$ theory, and let us denote the gauge symmetry of $T(SU(2))$ as $U(1)_g$.  Thus the $SU(2)_F$ node has a single fundamental flavor, also charged under $U(1)_g$, and CS level 1.  Then this theory has a dual description \cite{Giveon:2008zn,Willett:2011gp} as a single chiral multiplet, and importantly, a level 2 contact term for the $U(1)_g$ symmetry.  Thus this part of the quiver is replaced by singlet chiral multiplet along with a $U(1)_g$ level 2 topological CS theory, leading to
\be \label{TSTDualitySU2} G= SU(2): \;\;\;\; \text{$[k_1,...,k_n+1,1]$ quiver} \;\;\; \leftrightarrow  \;\;\;( \text{$[k_1,...,k_n]$ quiver} )\otimes (\text{topological $U(1)_2$ CS theory}) \;,\ee
and so the relation only holds up to a decoupled topological sector, as noted above.  A special case of this, noted in \cite{Pei:2016rmn}, is the case $n=0$, where this relation formally gives the ``appetizer duality'' of \cite{Jafferis:2011ns}, which indeed contains such a decoupled $U(1)_2$ topological sector. 

Next consider $G=U(2)$.  The $T(U(2))$ theory is closely related to the $T(SU(2))$ theory.  As described in \cite{Gaiotto:2013bwa}, this is simply the $T(SU(2))$ theory, whose symmetry we denote by $SU(2)_{F_-} \times SU(2)_{J_-}$, along with a background level 2 FI parameter coupling to an $U(1)_{F_+} \times U(1)_{J_+}$ symmetry.  This exhibits  the flavor symmetry as $SU(2)_{F_-} \times U(1)_{F_+} \times SU(2)_{J_-} \times U(1)_{J_+}$, but one can check that the true symmetry acting on the matter content is
\be 
(SU(2)_{F_-} \times U(1)_{F_+})/\Z_2 \times (SU(2)_{J_-} \times U(1)_{J_+})/\Z_2 \;\; \cong \;\; U(2)_F \times U(2)_J \,.
\ee
As above, this interpolates between $S$-dual instances of the 4d $\cN=4$ $U(2)$ SYM theory.

Now let us consider the relation \ref{dual1} in this case.  The analysis for the $T(SU(2))$ factor goes through as above, but now the final node also contains a $U(1)$ gauge group with level 2 CS term.  Now when we take the $\Z_2$ quotient, these two $U(1)_2$ topological sectors are eliminated,\footnote{Namely, the theory obtained after this quotient can be written as a $U(1)_1 \times U(1)_{-1}$ CS theory, which is a trivial theory.} and so there is no residual topological sector, and we find
\be \label{TSTDualityU2} G= U(2): \;\;\;\; \text{$[k_1,...,k_n+1,1]$ quiver} \;\;\; \leftrightarrow  \;\;\;( \text{$[k_1,...,k_n]$ quiver} )\;. \ee
As above, the $n=0$ case formally leads to the $U(2)$ case of the duality of \cite{Kapustin:2011vz}
\be \label{u2duality} U(2)_1 \; + \; \text{adjoint chiral } \; \leftrightarrow \;\; \text{2 free chirals}\,,\ee
which also holds without any extra decoupled sectors.

Next consider the relation \ref{dual3}.  Recall this follows from the duality of the $T_G$ theory with one $G$ flavor symmetry gauged at  level 1 and the $T(G)$ theory. In general, the $T_G$ theory is non-Lagrangian, and so this duality is not of immediate practical use.  However, for the case of $G=SU(2)$ (the $U(2)$ case is a straightforward extension) the $T_2$ theory is simply a free bifundamental hypermultiplet. Thus, we arrive at the duality\footnote{More precisely, the relation between the two theories involves a ``flipping'' of one of the $SU(2)$ flavor symmetries, and so the theory appearing on the LHS should be thought of as the $FT(SU(2))$ theory of \cite{Aprile:2018oau}.}
\be \label{DualityT(SU2)T2} 
\ba
&3d \; \cN=4 \; U(1) \; \text{with $N_f=2$ hypermultiplets}   \cr 
\leftrightarrow \quad &SU(2)_{k=1} \text{ with $N_f=2$ hypermultiplets with level $-1$ contact terms}\cr 
& \text{for $SU(2) \times SU(2) \cong SO(4)$ symmetry}   \,.
\ea
\ee
This duality was noticed earlier in \cite{Teschner:2012em}, and also discussed in \cite{RW2019}.  Importantly, note that in this dual description, the $SU(2) \times SU(2)$ symmetry is manifest in the Lagrangian, unlike in the usual description of $T(SU(2))$.  This observation makes this description of $T(SU(2))$ more convenient for constructing Lagrangians for general Seifert manifold theories, and it will play an important role in what follows.

\section{Higher-Form Symmetry in QFT}
\label{sec:HigherFormSymm}

In this section we briefly review higher-form, or $q$-form symmetries of QFTs \cite{Gaiotto:2014kfa}.  These turn out to play an important role in understanding the precise definition of the theories, $T[M_3]$, obtained by compactification of the 6d $\cN=(2,0)$ theory on a general three-manifold, $M_3$, which we will discuss in the next section.

\subsection{Higher-Form Symmetries}

A $q$-form symmetry with group $\Gamma$ in a $d$-dimensional QFT can be described by a set of topological {\it charge operators}, $U_\gamma[\sigma]$, which are labeled by an element $\gamma \in \Gamma$ and a $(d-q-1)$-dimensional submanifold $\sigma$ in spacetime, on which the operator is supported.  The fusion of two charge operators obeys the group law, \ie, $U_\gamma[\sigma] U_{\gamma'}[\sigma] \cong U_{\gamma \gamma'}[\sigma]$.  In addition, the operator may have non-trivial commutation relations with a physical $q$-dimensional extended operator, ${\cal O}$, in the theory, which we then say is charged under the $q$-form symmetry.  For example, if we work in the Hilbert space picture on a spacetime $M_{d-1} \times \R$, we have the commutation relations
\be \label{hfcomrel}
\big[ \; U_\gamma[\sigma]\; , \; {\cal O}[\rho] \; \big]  \; = \; R_{\cal O}(\gamma)^{\rho \cap \sigma} \cdot {\cal O}[\rho]  \,,
\ee
where $R_ {\cal O}$ is the representation of $\Gamma$ in which ${\cal O}$ transforms, and $\rho \cap \sigma$ is the intersection number of $\rho$ and $\sigma$ in $M_{d-1}$.  Higher-form symmetries are a generalization of ordinary global symmetries, which correspond to the case $q=0$, where the charged operators are local operators.  For $q>0$, it can be shown that $\Gamma$ must be abelian.

When $q=\frac{d-1}{2}$, the charge operators have the same dimension as the charged operators, and so they may themselves be charged under the symmetry.  When this happens we say the symmetry has an 't Hooft anomaly, and this is an obstruction to gauging it.\footnote{This follows because the charge operators should be trivial in the gauged theory, but this is not compatible with the non-trivial commutation relations in the presence of an anomaly.}  When the symmetry is non-anomalous, the expectation value of a set of charge operators depends only on the cycle $\widetilde{\omega} \in H_{d-q-1}(M_d,\Gamma)$ determined by the choice of $\sigma$ and $\gamma$ of the operators.  By Poincar\'e duality, this is the same as a choice of cocycle, $\omega \in H^{q+1}(M_d,\Gamma)$, and we may equivalently interpret this as computing the partition function in the presence of a background $(q+1)$-form gauge field, labeled by $\omega$, and define
\be 
\ZPaFu_{M_d}[\omega] = \langle U[\widetilde{\omega}] \rangle_{M_d} \,.
\ee
This gives a refinement of the ordinary partition function, which keeps track of the response of the system to sources for the higher-form symmetry.

When a higher-form symmetry is non-anomalous, we may gauge it to obtain a new theory.  This means we promote the background $(q+1)$-form gauge field to be dynamical, and so the partition function of the gauged theory is given by summing over all such background gauge fields.  It is natural to also introduce a coupling to a new background gauge field, and define
\be \widehat{\ZPaFu}_{M_d}[\widehat{\omega}] = \frac{1}{|H^{q+1}(M_d,\Gamma)|^{1/2}} \sum_{\omega \in H^{q+1}(M_d,\Gamma)} e^{i \int_{M_d} \omega \cup \widehat{\omega}} \ZPaFu[\omega]  \,,
\ee
where $\widehat{\omega} \in H^{d-q-1}(M_d,\widehat{\Gamma})$, $\widehat{\Gamma}$ is the Pontryagin dual group, \be
\widehat{\Gamma}=\text{Hom} (\Gamma, U(1))\,,
\ee
and $e^{i \int_{M_d} \omega \cup \widehat{\omega}} \in U(1)$ is the natural pairing.  This may be interpreted as the partition function of a theory with a $(d-q-2)$-form symmetry with group $\widehat{\Gamma}$, which we call the {\it dual symmetry}.  Repeating this gauging operation brings us back to the original theory.  More generally, we may gauge a subgroup of $\Gamma$ or include an additional local action for the background gauge fields, which lead to more general versions of the theory.

An important set of examples that will arise below are the electric and magnetic symmetries of a gauge theory.  Given a theory with gauge group $G$ with center $Z_G$, if a subgroup $\Gamma$ of $Z_G$ does not act on the charged matter of the theory, the theory admits a 1-form $\Gamma$ symmetry, called the {\it electric symmetry}.  The charged operators are Wilson loop operators, which transform according to the action of $\Gamma$ on the representation of the Wilson loop.  The partition function in the background of a 2-form gauge field is given by
\be \label{elecform}
\ZPaFu_G[\omega] = \sum_{P|w_2(P) = \omega} \ZPaFu_P \,, \qquad \omega \in H^2(M_d,\Gamma) \,,
\ee
where $\ZPaFu_P$ is the contribution to the partition function from a principal $G/\Gamma$-bundle, $P$, over spacetime, and we sum over all principal bundles with a specified value for their second Stiefel-Whitney class, $w_2(P) \in H^2(M_d,\Gamma)$, which measures the obstruction to lifting a $G/\Gamma$ bundle to a $G$ bundle.  Then the $G/\Gamma$ theory is simply given by gauging this symmetry, \ie, summing over all $G/\Gamma$ bundles
\be \label{magform} 
\ZPaFu_{G/\Gamma}[\widehat{\omega}] = \frac{1}{|H^2(M_d,\Gamma)|^{1/2}} \sum_{\omega \in H^2(M_d,\Gamma)} e^{i \int_{M_d} \omega \cup \widehat{\omega}} \ZPaFu_G[\omega] \,,\qquad 
{\widehat{\omega} \in H^{d-2}(M_d,\widehat{\Gamma})} \,.
\ee
The  $G/\Gamma$ theory naturally comes with a $d-3$ form $\widehat{\Gamma}$ symmetry which we call the {\it magnetic symmetry}, and whose charged operators are 't Hooft operators, which have dimension $d-3$.  We may also gauge a subgroup of $\Gamma$, or include additional local actions for the background gauge fields, to obtain other versions of the gauge theory.

\subsection{1-Form Symmetries in 3d Gauge Theories}
\label{sec:1formanom}

Let us now review some general features of 1-form electric symmetries in 3d gauge theories, which will arise in the study of $T[M_3]$ below.  

Recall that $d=3$ is the case where a 1-form symmetry may have an 't Hooft anomaly, and such anomalies turn out to arise from CS terms for the gauge groups.  To describe these anomalies more carefully, we follow \cite{Benini:2017dus}.  Consider a 3d gauge theory with gauge and flavor groups, $\widetilde{G}_g$ and $\widetilde{G}_f$, which we take to be the simply connected.  However, there may in general be some subgroup, $\Gamma$, of the center of these groups which does not act on the matter content.  Then in principle, we might try to couple the system to a $(\widetilde{G}_g \times \widetilde{G}_f)/\Gamma$ bundle.  However, anomalies may cause this observable to be ill-defined.

To see this, let us exhibit the 3d spacetime as the boundary of a 4d manifold, and extend the gauge fields into the bulk.  Then the 3d theory may include CS terms for the gauge and flavor symmetries, which we assume are properly quantized for $\widetilde{G}_g \times \widetilde{G}_f$ connections.  However, they may not be well-defined for $(\widetilde{G}_g \times \widetilde{G}_f)/\Gamma$ connections.  To cure this, we include appropriate 4d topological terms, which are also not well-defined on a four-manifold with boundary, but such that the combined 3d-4d system is well-defined.  Thus we obtain an action
\be S_{\text{anom}} = \int_{M_4} c(A)\,, \ee
where $c$ is some characteristic class of a 4d $(\widetilde{G}_g \times \widetilde{G}_f)/\Gamma$ gauge bundle.  We refer to this as the anomaly polynomial.  If we consider two different ways of extending the 3d system into a 4d bulk, the difference is given by the integral of the anomaly polynomial over a closed four-manifold.  This characterizes the extent to which the 3d partition function is not uniquely defined.  While this is not necessarily a problem for the flavor symmetries, the theory must be well-defined with respect to the gauge connection, and so we demand that the anomaly polynomial vanishes for any allowed choice of gauge bundles.

Below we will be mostly interested in theories with $SU(2)^n$ gauge and $SU(2)^m$ flavor symmetry.  The center is ${\Z_2}^n \oplus {\Z_2}^m$, and let us suppose there is a subgroup, $\Gamma$, which acts trivially on the matter.  Let us further include Chern-Simons levels $k_i$, $i=1,...,n$, and $K_j$, $j=1,...,m$, for the gauge fields.  Then for these to be well-defined, we extend the gauge fields into a bulk four-manifold, $M_4$, and include the term
\be S_{\text{anom}} = \pi \int_{M_4}\bigg( \sum_{i=1}^n k_i \frac{{\cal P}(w_2^{(i)})}{2}  +  \sum_{j=1}^m K_j \frac{{\cal P}(w_2^{(j)})}{2} \bigg)\,.
\ee
Here $w_2$ is the second Stiefel-Whitney class, valued in $H^2(M_4,\Z_2)$, and
\be {\cal P}:\quad 
H^2(M_4,\Z_2) \rightarrow H^4(M_4,\Z_4) \ee
is the Pontryagin square operation, which satisfies ${\cal P}(w) = w^2 \; (\text{mod} \; 2)$.  On a spin manifold, where the intersection form is even, ${\cal P}/2$ is well-defined as an element of $H^4(M_4,\Z_2)$.  

For example, consider the trinion theory, given by a $\cN=2$ chiral multiplet transforming in the trifundamental representation of an $SU(2)^3$ flavor symmetry.  Then the subgroup acting trivially on the matter is generated by $(1,1,0)$ and $(1,0,1)$ in ${\Z_2}^3$.  This means the Stiefel-Whitney class of 
the three groups must satisfy
\be \label{w123} w_2^{(1)} + w_2^{(2)} + w_2^{(3)} = 0 \,.\ee
Now we gauge one of the $SU(2)$ symmetries, say the third, with a level $k$ CS term, and we also include level $K_{1,2}$ CS terms for the flavor symmetries.  Note we cannot sum over arbitrary choices of $w_2^{(3)}$, but it is fixed by the classes of the flavor gauge fields by \eqref{w123}.  The anomaly polynomial in this case is
\be S_{\text{anom}} = \pi \int_{M_4} \bigg(K_1 \frac{{\cal P}(w_2^{(1)})}{2} + K_2 \frac{{\cal P}(w_2^{(2)})}{2} + k \frac{{\cal P}(w_2^{(3)})}{2} \bigg) \,.
\ee
Using \eqref{w123}, we may rewrite this independently of the dynamical gauge field as
\be S_{\text{anom}} = \pi \int_{M_4}  \bigg((k+K_1) \frac{{\cal P}(w_2^{(1)})}{2} + (k+K_2) \frac{{\cal P}(w_2^{(2)})}{2} - k w_2^{(1)} \cup w_2^{(2)} \bigg) \,.
\ee
In particular, for $k=-K_1=-K_2 =1$, which gives the $SU(2)_1$ dual description of the $T(SU(2))$ theory, as in \eqref{DualityT(SU2)T2}, we have
\be \label{tsu2anom} S_{\text{anom}} 
= 
-\pi \int_{M_4} w_2^{(1)} \cup w_2^{(2)}\,.
\ee
This result was also found in \cite{Gang:2018wek} using the usual description of $T(SU(2))$ as $U(1)$ with two hypermultiplets, and this gives another check of the duality between these two descriptions.  

More generally, using the fact that the $T_N$ theory has a $\Z_N \times \Z_N$ symmetry, which acts on any two of the three $SU(N)$ flavor nodes, as above, we may use the description of the $T(SU(N))$  theory in section \ref{sec:TM3} to argue analogously that it has the same anomaly polynomial as \eqref{tsu2anom}, where we now interpret the second Stiefel-Whitney classes as elements of $H^2(M_4,\Z_N)$.  Note that this means that we may not gauge both flavor symmetries as $G^\vee = G/\Gamma$ gauge groups, but instead, if we take one as $G^\vee$, the other must be $G$.  This reflects the fact that $T(G)$ exchanges these two versions of the theories.

In general, given a theory with ${\Z_N}^n$ 1-form symmetry, we may encode the anomalies in a symmetric $n \times n$ matrix, $\cA_{ij}$, with entries valued in $\Z_N$, such that the anomaly polynomial is given by
\be 
S_{\text{anom}} = \frac{2\pi}{N} \int_{M_4} \bigg( \sum_{i=1}^n \cA_{ii} \frac{\mathcal{P}(w_2^i)}{2} + 
\sum_{i<j} \cA_{ij} w_2^i \cup w_2^j \bigg) \,.
\ee
Then one finds that non-anomalous symmetries are ones labeled by a vector $\gamma = (\gamma_1,...,\gamma_n) \in {\Z_N}^n$ such that
\be 
\gamma \cA \gamma = 0 \; (\text{mod} \; N)  \,.
\ee
More generally, for a subgroup $\Lambda$ of the set of 1-form symmetries to be non-anomalous, we impose also that all mixed anomalies vanish, \ie,
\be \gamma \cA \gamma' = 0 \; (\text{mod} \; N), \;\;\;\; \gamma,\gamma' \in \Lambda \,.
\ee
Note that when $\cA$ is trivial, every subgroup is  non-anomalous.  In this case we refer to  the group $\Gamma$ as being ``anomaly-free.''

\subsection{Representation of 1-Form Symmetries on $\CH_{T^2}$}
\label{sec:ht2reps} 

One important property of 1-form symmetries we will exploit below is that they act on the Hilbert space of the theory on a non-trivial manifold.  Below we will consider 3d theories with 1-form symmetries on the spatial manifold $T^2$.  Then we may define linear operators on the Hilbert space associated to 1-form generators wrapping the two cycles of the torus
\be 
U_\gamma^{A,B} :\quad  \CH_{T^2} \rightarrow \CH_{T^2} \,.
\ee
Then the Hilbert space must fall into representations of the group of these 1-form generators.  Although the 1-form symmetry group itself is abelian,  in the presence of anomalies this group need not be abelian since we have in general
\be [U^A_{\gamma_1},U^B_{\gamma_2}] = e^{2 \pi i \cA(\gamma_1,\gamma_2)} \,,
\ee
where $\cA:\Gamma \times \Gamma \rightarrow U(1) \cong \R/\Z$ is the anomaly form.  Then the group acting on $\CH_{T^2}$ is the central extension, $E$, of $\Gamma \times \Gamma$ associated to the anomaly form, $\cA$.  It fits into a short exact sequence
\be 
1 \rightarrow U(1) \rightarrow E \rightarrow \Gamma \times \Gamma \rightarrow 1 \,.
\ee

We define the ``center''\footnote{Of course, as a group $\Gamma$ is abelian and so equal to its own center, but the terminology should be clear from context.}  of the 1-form symmetry, $Z \subset \Gamma$, to be the set of elements $z$ such that $\cA(z,\gamma)=0$ for all $\gamma \in \Gamma$.  Then we may simultaneously diagonalize $U^A_z$ and $U^B_{z'}$ for $z,z' \in Z$, and so refine the Hilbert space into characters $\widetilde{\chi}_A,\widetilde{\chi}_B \in \widehat{Z}$, where $\widehat{Z} \equiv \text{Hom}(Z,U(1))$ is the group of characters of $Z$.  In fact, one can show \cite{BWHF} that the irreducible representations, $R_{\widetilde{\chi}_A,\widetilde{\chi}_B}$, of $E$ are labeled by such a pair of characters, and have dimension
\be  \label{dimR1form} \text{dim} \; R_{\widetilde{\chi}_A,\widetilde{\chi}_B} = |\Gamma|/|Z| \,.
\ee
More explicitly, let us suppose that the exact sequence
\be  \label{zses} 1 \rightarrow Z \rightarrow \Gamma \rightarrow \Gamma/Z \rightarrow 1  \ee
splits, so that we may identify
\be \Gamma \cong Z \oplus (\Gamma/Z) \,.
\ee
Then  the representations $R_{\widetilde{\chi}_A,\widetilde{\chi}_B}$ can be written as
\be R_{\widetilde{\chi}_A,\widetilde{\chi}_B} = P_{\widetilde{\chi}_A,\widetilde{\chi}_B} \otimes R_{\Gamma/Z} \,,
\ee
where $P_{\widetilde{\chi}_A,\widetilde{\chi}_B}$ is a 1d space only the $U_{A,B}^z$, $z \in Z$ act on, and $R_{\Gamma/Z}$ is the unique irreducible representation of the quotient 1-form symmetry $\Gamma/Z$, of dimension $|\Gamma/Z|=|\Gamma|/|Z|$.  Thus, in this case, the Hilbert space on $T^2$ can be factorized as
\be  \label{Ht2factor} 
\ba
 \CH_{T^2} 
&=  \bigoplus_{\widetilde{\chi}_A,\widetilde{\chi}_B} \mathcal{H}'_{\widetilde{\chi}_A,\widetilde{\chi}_B}\otimes R_{\widetilde{\chi}_A,\widetilde{\chi}_B} 
= R_{\Gamma/Z} \otimes \left(\bigoplus_{\widetilde{\chi}_A,\widetilde{\chi}_B}  \mathcal{H}'_{\widetilde{\chi}_A,\widetilde{\chi}_B} \otimes P_{\widetilde{\chi}_A,\widetilde{\chi}_B} \right)\cr 
&\equiv R_{\Gamma/Z} \otimes \bigoplus_{\widetilde{\chi}_A,\widetilde{\chi}_B}  \mathcal{H}_{\widetilde{\chi}_A,\widetilde{\chi}_B}\,,
\ea\ee
where $ \mathcal{H}'_{\widetilde{\chi}_A,\widetilde{\chi}_B}$ is not acted on by the 1-form symmetry and 
where all information about the anomaly is encoded in the $R_{\Gamma/Z}$ factor, and the center, $Z$, only acts on the second factor.\footnote{In the more general case where \eqref{zses} does not split, there is still a weaker factorization, where $Z$ still does not act on the first factor, but the full group $\Gamma$ may act on the second factor.}  This factorization property will be important below.

In the case where the 1-form symmetry, $\Gamma$, is anomaly-free, we may also consider ``twisted sectors'' in the Hilbert space.  These are sectors where we let a 1-form charge operator, $U^\gamma$, lie along the time direction, and may be labeled by the element $\gamma \in \Gamma$
\be \label{htwistedsector} \CH^\gamma_{T^2} =  \bigoplus_{{\chi}_A,{\chi}_B}  \mathcal{H}^\gamma_{{\chi}_A,{\chi}_B} , \;\;\;\;\;\; \gamma \in \Gamma \;, \ee
where the untwisted sector corresponds to $\gamma=1$, and we may also simultaneously grade these sectors into characters of the $U_{A,B}$ operators, as above.  These twisted sectors are important when gauging the 1-form symmetry.   Namely, the Hilbert space of the gauged theory is given by
\be \label{hgauged} \widetilde{\CH}^{\text{gauged}}_{T^2} =  \bigoplus_{\gamma \in \Gamma}  \mathcal{H}^\gamma_{{\chi}_A={\chi}_B=1} \;. \ee

\section{$T[M_3,\g]$ and Higher-Form Symmetries}
\label{sec:higherform}

In this section we define the theories obtained by compactification of the 6d $\cN=(2,0)$ theory of type $\g$ on a three-manifold $M_3$. Most of what we discuss in this section is applicable to both $\mathcal{N}=1$ and $\mathcal{N}=2$ versions, so we will drop the subscript in the following. 
What will emerge from this section is that to fully define the theory, we must specify, in addition to the gauge algebra and three-manifold, a subgroup $H$ of the cohomology group $H^2(M_3,\ZGtilde)$, and so we will actually define theories
\be
T[M_3, \mathfrak{g},H]   \,, \quad H \leq H^2 (M_3, \ZGtilde) \,.
\ee
Here $\ZGtilde$ is the center of the simply connected Lie group $\widetilde{G}$ with Lie  algebra $\g$.  For ease of notation, we will often refer to $T[M_3,\g, H= 1]$ simply as $T[M_3,\g]$.

For $M_3$ a graph manifold, the theories ${T}[M_3,\g,H]$ can be naturally defined in terms of  the theories, $\widehat{T}[\Graph,\widetilde{G}]$, defined in the previous section, where $\Graph$ is a graph associated to $M_3$.  To describe the procedure of obtaining the former theory from the latter, it turns out to be natural to use the language of higher-form symmetries, reviewed in the previous section.  In particular, the theories $T[M_3,\g,H]$ in general have several interesting higher-form symmetries, and the theories corresponding to different subgroups, $H$, can be obtained from each other by gauging these symmetries in a suitable way.  This structure can be traced to a higher-form symmetry present already in the 6d theory, and in particular, to its nature as a relative QFT.

\subsection{6d theory as a Relative QFT}
\label{sec:6dO}

Let us first review some properties of the 6d theory relevant to the compactification to 3d, which we consider in the next subsection.  The role of discrete topological data in compactifications of the 6d $\cN=(2,0)$ theory has been discussed extensively in the literature; see, for example, \cite{Witten:1998wy,Witten:2009at,Freed:2012bs,Tachikawa:2013hya,Gang:2018wek,Gukov:2018iiq} for related discussions.  

First we recall that the 6d $\cN=(2,0)$ theory is a {\it relative QFT}, meaning it is not well-defined by itself, but naturally lives on the boundary of a 7d topological theory \cite{Witten:1998wy,Witten:2009at,Freed:2012bs}.  Then the notion of a partition function is replaced by a {\it partition vector}, which is an element in the Hilbert space of this TQFT.  This is analogous to the case of a chiral CFT in 2d, which lives on the boundary of a 3d Chern-Simons theory.  In the 6d case, this 7d TQFT can be described as a {\it Wu-Chern-Simons} (WCS) theory  \cite{Witten:1998wy,Monnier:2017klz}.  

In more detail, let us first consider the $A_{N-1}$ type 6d  $\cN=(2,0)$ SCFT.  Then, roughly speaking, the WCS theory can be thought of as a level $N$ 3-form Chern-Simons theory, and has 3d Wilson surface operators, $\cO_\omega$,  supported on 3-cycles, $\omega \in H_3(M_7,\Z_N)$.  These may be thought of as charge operators for a 3-form symmetry of the Chern-Simons theory, with group $\Z_N$.  This symmetry has an 't Hooft anomaly, which means the charge operators do not commute.  Rather, when we consider the theory on a spacetime $M_6 \times \R$, they obey the appropriate version of \eqref{hfcomrel}, specifically\footnote{As described in \cite{Witten:1998wy}, to define the full algebra, and not just the commutation relations, when $N$ is odd, one needs to choose a spin structure on $M_6$.  In explicit examples below we will mostly take $N=2$, so will not encounter this issue.}
\be \cO_\omega  \cO_{\omega'} =e^{\frac{2 \pi i}{N} \omega \cap \omega'} \cO_{\omega'} \cO_{\omega}, \;\;\;\;\; \omega ,\omega' \in H_3(M_6,\Z_N) \,.\ee
This can also be described by saying the operators generate a Heisenberg group, $W$, which is a central extension of $H_3(M_6,\Z_N)$, and fits into the short exact sequence
\be 1 \rightarrow \Z_N \rightarrow W \rightarrow H_3(M_6,\Z_N) \rightarrow 1 \,.\ee

The Hilbert space of the TQFT on $M_6$ is given by an irreducible representation of $W$, which can be described as follows. First we must choose a {\it polarization} of $H_3(M_6,\Z_N)$, which is a maximal isotropic subgroup, $\Lambda$, \ie, with $\lambda \cap \lambda'=0$ for all $\lambda,\lambda' \in \Lambda$.  Let us assume that we can write
\be H_3(M_6,\Z_N) \cong \Lambda \oplus \widetilde{\Lambda} \,,\ee
where it follows that $\widetilde{\Lambda}$ is also maximal isotropic.  Then we pick a basis in which the operators, $\cO_{\widetilde{\lambda}}$, $\widetilde{\lambda}\in \widetilde{\Lambda}$, are diagonal.  Namely, one can find a unique vector $|0\rangle$, which is fixed by all $\cO_{\widetilde{\lambda}}$, and for ${\lambda} \in {\Lambda}$ we define states
\be |{\lambda}\rangle  \equiv \cO_{{\lambda}} |0\rangle , \;\;\;\; {\lambda} \in {\Lambda} \;. \ee
Then one can check that
\be \cO_{{\lambda}} |{\lambda}' \rangle =  |{\lambda}+{\lambda}' \rangle, \;\;\;\; \cO_{\widetilde{\lambda}} |{\lambda} \rangle = e^{\frac{2 \pi i}{N} \lambda \cap \widetilde{\lambda}} |{\lambda} \rangle \;,  \;\;\; \lambda,\lambda' \in \Lambda, \; \widetilde{\lambda} \in \widetilde{\Lambda} \;,\ee
and so these states form a basis for the irreducible representation of $W$, and so for the Hilbert space of the 7d theory.  The choice of polarization picks out a corresponding basis of the Hilbert space as above.  \Eg, if we had chosen $\widetilde{\Lambda}$ rather than $\Lambda$, the basis states would be  related by a discrete Fourier transform. Note that a choice of polarization will typically break some of the diffeomorphism symmetry of $M_6$.  

For other ADE Lie algebras $\mathfrak{g}$, we expect a similar description, where in general the polarization is determined by a decomposition
\be H_3(M_6,\ZGtilde) \cong \Lambda \oplus \widetilde{\Lambda} \,,
\ee
where $\ZGtilde$ is the center of $\widetilde{G}$, the simply connected Lie group with Lie algebra $\g$.  

After picking the polarization, we may write the partition vector of the 6d $\cN=(2,0)$ theory on $M_6$ in the corresponding basis as
\be 
\left| \ZPaFu^{M_6} \right\rangle = \sum_{{\lambda} \in {\Lambda}} \ZPaFu^{M_6}_{{\lambda}} |{\lambda} \rangle \,.
 \ee
We refer to $\ZPaFu^{M_6}_{{\lambda}}$, ${\lambda} \in {\Lambda}$, as the components of the partition vector.  We emphasize that they depend not only on ${\lambda}$, but also on the choice of polarization, $\Lambda$.

\subsubsection{5d Reduction}
\label{sec:5d}

As an example, let us consider $M_6$ of the form $M_5 \times S^1$.  Then we may write, by K\"unneth,
\be 
H_3(M_6,\ZGtilde) = H_3(M_5,\ZGtilde) \oplus H_2(M_5,\ZGtilde)  \equiv A \oplus B \,.
\ee
Here there are two natural choices of polarization which preserve diffeomorphism invariance on $M_5$, which are to take $\Lambda=A$ or $\Lambda=B$.  In the first case, the components of the partition vector can be written as
\be 
\ZPaFu^{M_5 \times S^1}_\omega, \;\;\; \omega \in A = H_3(M_5,\ZGtilde)   \cong H^2(M_5,\ZGtilde) \,.
\ee
The 6d $\cN=(2,0)$ theory on $M_5 \times S^1$ is believed to be described by  5d $\cN=2$ SYM with Lie algebra $\g$ on $M_5$.  Then $\ZPaFu^{M_5 \times S^1}_\omega$ is the contribution to the path-integral of this theory from principle $\g$ bundles with second Stiefel-Whitney class equal to $\omega$.  In particular, $\ZPaFu^{M_5 \times S^1}_0$ is the partition function of the theory with simply connected gauge group, $\widetilde{G}$, and more generally, $\ZPaFu^{M_5 \times S^1}_\omega$ is  the refinement by the electric 1-form symmetry of this theory, as in \eqref{elecform}.

On the other hand, in the $B$-polarization the states are labeled by $H_2(M_5,\ZGtilde) \cong H^3(M_5,\ZGtilde)$, and this corresponds to the $\widetilde{G}/\ZGtilde$ theory, refined by its 2-form magnetic symmetry.  Indeed, this change in polarization is implemented on the representation by a discrete Fourier transform, and we obtain
\be 
\widetilde{\ZPaFu}^{M_5 \times S^1}_{\widetilde{\omega}} 
= \frac{1}{|H^2(M_5,\ZGtilde)|^{1/2}} \sum_{\omega \in B} e^{i \omega \cup \widetilde{\omega}} \ZPaFu^{M_5 \times S^1}_\omega , \;\;\; \widetilde{\omega} \in H_2(M_5,\ZGtilde)  \cong H^3(M_5,\ZGtilde) \,,
\ee
which agrees with \eqref{magform}.  Note these two choices preserve diffeomorphism invariance on $M_5$, which is natural in the context of describing a 5d QFT.  Other such choices of polarization give rise to the different global forms of the group with Lie algebra $\g$.  Thus we see the different choices of polarization gives rise to different 5d theories.  

\subsubsection{Abelian case}

In addition to the simple Lie  algebras considered above, we may define the 6d $\cN=(2,0)$ theory in the case $\g=\frak{u}(1)$, where it corresponds to a free $\cN=(2,0)$ tensor multiplet.  Although this theory is free, there are still several subtleties that arise, as discussed, \eg, in \cite{Witten:1996hc,Witten:2009at}.

A useful analogy to this theory is the chiral periodic scalar theory in 2d.  This theory depends on the period $R$, of the scalar, where  the  self-dual radius $R=1$ corresponds to a free chiral fermion. For all other choices of $R$, the theory is not modular invariant by itself, but  instead, the partition function is mapped by modular transformations among a space of  {\it conformal blocks}.  If we focus on $R =N \in \Z_{>0}$ and spacetime $\Sigma$, these blocks are labeled by elements of an isotropic subgroup
$\Lambda \subset H_1(\Sigma,\Z_N)$.
For example, for $\Sigma=T^2$, $|\Lambda|=N$, and there are $N$ conformal blocks on the torus.  As modular transformations act on $H_1$ and so on the choice of $\Lambda$, they also act on the space of conformal blocks.  The space of conformal blocks may also be naturally interpreted as the Hilbert space of 3d $U(1)$ level $N$ Chern-Simons theory on $\Sigma$.  Indeed, the chiral CFTs are not defined on their own, but must live on the boundary of this 3d TQFT, and so their partition  functions on $\Sigma$ are naturally elements in the Hilbert space of these TQFTs.

The self-dual 2-form in 6d is completely analogous.  Here we also must specify a periodicity, $N \in \Z_{>0}$, and rather than a partition function, the observables on a spacetime $M_6$ are given by conformal blocks or a partition vector, as we saw in the ADE case above, namely 
\be 
\ZPaFu_{\lambda} \,,\qquad  \lambda \in \Lambda \subset H_3(M_6,\Z_N) \,.
\ee
These may also be understood as vectors in the Hilbert space of a 7d TQFT on $M_6$, a 3-form version of the Chern-Simons theory.  In particular, for $N>1$, the theory is not an ordinary QFT, but a relative QFT, as we saw in the ADE case above.

We note for later reference that the self-dual 2-form with periodicity $N$ has a $U(1)$ 2-form symmetry \cite{Gaiotto:2014kfa}, whose charge operators are
\be \label{u12form} U_\alpha[\Sigma_3] = e^{2 \pi i \alpha \int_{\Sigma_3} H} \,,
\ee
where $H=dB$ is the self-dual 3-form field strength.  This symmetry has an 't Hooft anomaly, which implies that when we place the $U_\alpha$ operator on a manifold with boundary, $\partial \Sigma_3=\Sigma_2$, this configuration has a non-trivial 2-form charge supported on $\Sigma_2$.  We will return to the consequences of this statement below.

\subsubsection{Self-dual $G$}
\label{sec:6dselfdual}

Let $\g$ be a semi-simple, simply-laced Lie algebra, and suppose we can find a subgroup, $L$, of $\ZGtilde$, such that $L \cong \ZGtilde/L$.  In this case, the group $G \equiv \widetilde{G}/L$, is isomorphic to its Langlands dual group, and we call such groups ``self-dual.''  Some examples are $SO(2N)=\text{Spin}(2N)/\Z_2$, $(SU(N) \times SU(N))/{\Z_N}$, and $SU(N^2)/\Z_N$.  For such $G$,  there is a corresponding choice of polarization of the 6d theory, which is\footnote{For convenience, we sometimes specify the polarization in terms of the cohomology, rather than homology, which is equivalent by Poincar\'e duality.}
\be \Lambda_G \equiv H^3(M_6,L) \subset H^3(M_6,\ZGtilde)\,.
 \ee
One can check that this is a maximal isotropic subgroup. Note that this choice does not break any diffeomorphism symmetry of $M_6$, and so we may naturally interpret these observables as partition functions of an ordinary QFT.  Specifically, since the partition function is labeled by an element in $H^3(M_6,L)$, we see this theory has a 2-form symmetry with group $L$.  We may naturally refer to the 6d theory of type $\g$ with the polarization $\Lambda_G$ as the ``6d $\cN=(2,0)$ theory of type $G$,'' which is an ordinary rather than relative QFT.  We emphasize this terminology only makes sense for $G$ self-dual.

Similar considerations apply in the non-semi-simple case.  For example, $G=U(1)$ is self-dual, and we saw above that the $N=1$ case of this theory is indeed an ordinary QFT.  More generally, we may define the ``6d $\cN=(2,0)$ theory of type $U(N)$'' as follows.  We start with the tensor product of the theory of type $\g=\frak{su}(N)$, along with the $\frak{u}(1)$ theory with periodicity $N$.  These are separately relative QFTs, and so we must choose a  maximal isotropic subgroup of
\be H^3(M_6,\Z_N) \oplus H^3(M_6,\Z_N) \cong H^3(M_6,\Z_N \oplus \Z_N)\,. 
\ee
However, as above, a natural choice of polarization that preserves  diffeomorphism symmetry is given by
\be \Lambda \cong H^3(M_6,\Z_N^{\text{diag}}) \subset H^3(M_6,\Z_N \oplus \Z_N) \,.
\ee
This gives rise to an ordinary 6d QFT, which is acted on by a $\Z_N$ 2-form symmetry.  In fact, this $\Z_N$ sits inside a larger $U(1)$ 2-form symmetry coming from the $\frak{u}(1)$ factor, as in \eqref{u12form}.  This theory corresponds to the low energy theory of $N$ M5-branes, without decoupling the center of mass motion.

In the above cases, the reduction to 5d gives rise to the 5d $\cN=2$ SYM theory with gauge group $G$, which has both 1-form and 2-form symmetries with group $Z_G$.  

\subsection{Compactification on $M_3$}
\label{sec:tm3ghdef}

In this subsection we describe some general features of the compactification of the 6d theory on $M_3$, which follow from properties of the 6d theory described above.\footnote{Similar global considerations in the 3d-3d correspondence were discussed in \cite{Gang:2018wek}.  See also \cite{Gukov:2018iiq} for related discussions  in the 4d-2d correspondence.}   In this context, we are interested in the case where $M_6$ is a product manifold of the form
\be M_6 = M_3 \times W_3\,, 
\ee
which describes a 3d theory, $T[M_3,\g]$, on the spacetime $W_3$.  As we will see in a moment, additional data is necessary to fully specify the theory.

For $M_6=M_3 \times W_3$, the universal coefficient version of the K\"unneth theorem states that for $\ZGtilde$-valued cohomology
\be \label{Kuenneth}
\ba
H^3 (M_3 \times W_3, \ZGtilde) 
\cong \ &  H^3 (M_3,\ZGtilde) \oplus H^3 (W_3, \ZGtilde)  \cr 
&\oplus  H^2 (M_3,\ZGtilde) \otimes_{\ZGtilde} H^1 (W_3,\ZGtilde)
\oplus H^1 (M_3\ZGtilde )\otimes_{\ZGtilde} H^2 (W_3,\ZGtilde ) \cr 
& \oplus \hbox{Tor}_{\ZGtilde} (H^3 (M_3, \ZGtilde), H^1 (W_3, \ZGtilde)) \cr 
&\oplus \hbox{Tor}_{\ZGtilde} (H^2 (M_3, \ZGtilde), H^2 (W_3, \ZGtilde))\cr 
&\oplus \hbox{Tor}_{\ZGtilde} (H^1 (M_3, \ZGtilde), H^3 (W_3, \ZGtilde)) \,.
\ea
\ee
Here the $\hbox{Tor}_{\ZGtilde} (X, Y)$ are the torsion groups over the rings $\ZGtilde$. 

Below we will mostly be interested in the case $W_3=T^3$, and assume from now on that $W_3$ has no torsion.\footnote{However, examples of $W_3$ with torsion, such as the Lens spaces, naturally arise in the 3d-3d correspondence, and it would be interesting to explore the new features that arise here.}  This implies that all the $\hbox{Tor}$ groups vanish.   Under this assumption, we may decompose the third cohomology group as
\be   \label{lldecompm3w3}
H^3(M_3 \times W_3 ,\ZGtilde)   \cong \ZGtilde  \oplus A \oplus B \oplus \ZGtilde \,,
\ee
with the notation
\be   \ba \label{abdef}
A &=  \left(H^1(M_3,\ZGtilde) \otimes_{\ZGtilde} H^2(W_3,\ZGtilde)\right) \cong H^2(W_3,{\Upsilon}) \cr
B  &=  \left(H^2(M_3,\ZGtilde) \otimes_{\ZGtilde} H^1(W_3,\ZGtilde)\right) \cong H^1(W_3,\widehat{\Upsilon})  \,,
\ea \ee
where we have used the universal coefficient theorem to rewrite these groups (using the assumption that $W_3$ is torsionless), and have also defined
\be \Upsilon \equiv H^1(M_3,\ZGtilde), \;\;\;\; \widehat{\Upsilon} \equiv H^2(M_3,\ZGtilde)\,, \ee
which are naturally Pontryagin dual.

We now have to make a choice of polarization.\footnote{We may always choose the polarization to include one of the $\ZGtilde$ factors in \eqref{lldecompm3w3}, and ignore these factors from now on, as they do not depend on the topology in an interesting way.}  One possibility is to pick the $A$-polarization.  In this case, the observables of the theory will be labeled by a choice of element in $H^2(W_3,\Upsilon)$.  This naturally suggests that this 3d theory has a 1-form symmetry with group $\Upsilon$.  Another choice is the $B$-polarization, in which case the observables are labeled by an element in $ H^1(W_3,\widehat{\Upsilon})$, which gives a theory with $0$-form $\widehat{\Upsilon}$ symmetry.  These choices are analogous to the choices which led to the 5d $\widetilde{G}$ and $\widetilde{G}/\ZGtilde$ theories, and we should interpret these as leading to different versions of the $T[M_3,\g]$ theories.

More generally, we may choose a subgroup
\be H \leq \widehat{\Upsilon}\,, \ee
and correspondingly define
\be \Upsilon_H = \{ x \in \Upsilon \; | \; \chi(x) = 1, \; \; \forall \chi \in H \} \subset \Upsilon \,.\ee
Then we may take our polarization, $\Lambda$, to correspond to a subgroup
\be \Lambda =  H^1(W_3,H) \oplus H^2(W_3,\Upsilon_H) \,.
\ee
Here the $A$ and $B$ polarizations above correspond to $H=1$ and $H=\widehat{\Upsilon}$, respectively.  These give rise to a general class of polarizations preserving the diffeomorphism symmetry of $W_3$, which is natural in the context of defining a 3d QFT.

Thus we find there is not a unique theory $T[M_3,\g]$, but rather, different choices corresponding to the choice of $H$ above.  We define the theory
\be T[M_3,\g,H] \ee
to be the 3d theory corresponding to the compactification of the type $\g$ 6d $\cN=(2,0)$ theory with the polarization labeled by $H$ above.  We see that the observables are labeled by
\be H^1(W_3,H) \oplus H^2(W_3,\Upsilon_H) \,,
\ee
which we may interpret as corresponding to symmetries
\be 
\ba
\text{$0$-form symmetry:} &\qquad \Gamma_0 = H \cr 
 \text{1-form symmetry:} &\qquad  \Gamma_1 = \Upsilon_H \nonumber \,.
\ea \ee

\subsection{$T[M_3,\g, H]$ for $M_3$ a Graph Manifold}
\label{sec:Tm3gH}

Let us turn to an explicit description of the $T[M_3,\g,H]$ in the case of $M_3$ a  graph manifold.  Here we will use the theories $\widehat{T}[\Graph,G]$ defined in the previous  section, but we will see that, in general, some additional operations are necessary to produce the $T[M_3,\g,H]$ theory.   In this section we will mostly consider the $\cN=2$ twist, and return to the $\cN=1$ twist in section \ref{sec:3dN=1}.

\subsubsection{$U(1)$ Case}

\label{sec:DefinitionU1}

Let us start in the abelian case, and consider the reduction of the 6d theory of type $\frak{u}(1)$.  Recall we must also choose the periodicity, $N$, and we first consider the case $N=1$ which gives rise to an ordinary QFT in 6d, which we refer to as the $U(1)$ theory.  Then we denote the 3d theory obtained by compactification of this theory on $M_3$, with the appropriate twist, as
\be T[M_3,U(1)] \,.
\ee

We focus on the case of $M_3$ a graph manifold, for which we take a particular graph decomposition, $\Graph$.  This graph gives a prescription for forming $M_3$ by gluing simple pieces along boundary tori by $S$ and $T$ transformations.  Then, as discussed in \cite{Gadde:2013sca}, the theory may be built by iteratively applying the $S$ and $T$ operations of Witten \cite{Witten:2003ya}, or more precisely, their supersymmetric completions.  Namely, we assign a $U(1)$ gauge multiplet to each node in the graph associated to $M_3$, a level $k_i$ CS term for the gauge group in the $i$th node, where $k_i$ is the label of  the node, and an off-diagonal CS term to two nodes connected by an $S$-gluing.  In summary, this results in
\be S = \frac{1}{4 \pi} \int Q_{ij} A_i \wedge dA_j + \cdots \,,
\ee
where $Q_{ij}$ is the linking matrix, defined in \eqref{LinkingMatrix}, and the dots denote additional fields in the ($\cN=1$ or $\cN=2$) vector multiplet, which we suppress.  This is precisely the description of $\widehat{T}[\Graph,U(1)]$ given in the previous section, where we recall the duality wall theory, $T(U(1))$, is simply a mixed BF-term for a $U(1) \times U(1)$ flavor symmetry. Thus we have, in this case
\be T[M_3,U(1)] \equiv \widehat{T}[\Graph,U(1)] \,.
\ee
As a consistency check, we must verify that this theory is independent of the choice of graph decomposition $\Graph$ of $M_3$.  This is almost true, but there is a slight subtlety, which is that the $SL(2,\Z)$ operations used above do not quite close onto a representation of $SL(2,\Z)$, but rather form a projective representation, as noted in \cite{Witten:2003ya,Alday:2017yxk}.  For example, applying the $(ST)^3$ operation does not completely leave the theory invariant, but introduces a tensor factor with an invertible TQFT, the $U(1)$ level 1 CS theory, and correspondingly the partition function is not invariant, but incurs a phase.  This can be attributed to a gravitational anomaly of the 6d theory.  We will work modulo such invertible TQFTs in what follows.

More generally, we may consider the theory,
\be T[M_3,\frak{u}(1)_N] \;, \ee
obtained by the compactification of the free tensor multiplet with periodicity $N$.  The case above corresponds to $N=1$, and more generally, we claim we should now consider the action
\be S_{\widehat{T}[\Graph,U(1)_N]}= \frac{1}{4 \pi} \int N \; Q_{ij} A_i \wedge dA_j + \cdots \ee
Let us denote this theory by $\widehat{T}[\Graph,U(1)_N]$, as we have not yet checked independence on $\Graph$.  Indeed, once again we find that the $SL(2,\Z)$ relations are not satisfied on the nose, but now the discrepancy is more severe.  For example, the $(ST)^3$ relation now introduces a tensor factor of the $U(1)$ level $N$ CS theory, which is not invertible.  This means the partition function is multiplied by a more complicated factor than the phase above, and even the Hilbert space of the theory will in general be modified.  Thus the theory $\widehat{T}[\Graph,U(1)_N]$ is not independent of the graph decomposition of $M_3$, and so is not yet equal to the theory $T[M_3,\frak{u}(1)_N]$.  We will see below that a similar issue arise in the ADE case, and will describe the resolution in that context.

\subsubsection*{1-form Symmetries}

Let us restrict our attention to the case with periodicity $N=1$.  As mentioned above, the 6d self-dual tensor theory has a 2-form $U(1)$ symmetry, and upon compactification, this descends to a higher-form symmetry of the $T[M_3,U(1)]$ theory.  Namely, we find
\be \text{1-form symmetry:} \qquad \Gamma_1 = H^1(M_3,U(1)) \cong H^2(M_3,\Z) \,.
\ee
This holds for general $M_3$, but in the case of $M_3$ a graph manifold, we may see this 1-form symmetry explicitly in the Lagrangian, as follows.  To each $U(1)$ factor in the gauge group, there is naively a $U(1) \cong \R/\Z$ 1-form symmetry.  However, the Chern-Simons terms break this symmetry to the subgroup
\be \Gamma_1 = \text{ker}(Q: \R^n/\Z^n \rightarrow \R^n/\Z^n)\,, \ee
where $Q$ is the $n \times n$ linking matrix.  Then one may check that the above group is indeed isomorphic to $H^2(M_3,\Z)$.  

Moreover, as a result of the anomaly of the 6d 2-form symmetry, one finds that this 1-form symmetry can have an anomaly.  Similarly to the discussion in \cite{Freed:2006ya,Gaiotto:2014kfa}, suppose we wrap a charge operator of the 6d theory on a $\Z_k$ torsion cycle in $M_3$.  Topologically, this cycle may have a non-trivial boundary, but one which is a multiple of $k$, so that it vanishes with $\Z_k$ coefficients.  However, due to the anomaly, this boundary may render the charge operator itself to be charged under the higher-form symmetry.  As a result, the 1-form symmetry in $T[M_3,U(1)]$ has an anomaly associated to the torsion subgroup, $TH^2(M_3)$, of $H^2(M_3)$, and the anomaly  form gives a topological invariant of $M_3$, which is the ``linking form''
\be \ell: \ TH^2(M_3) \times TH^2(M_3) \rightarrow \mathbb{Q}/\Z \,.
\ee
We may see this explicitly  in the Lagrangian for $T[M_3,U(1)]$ above, as one computes that the anomaly form of the Chern-Simons theory is
\be \cA(\gamma_1,\gamma_2) = e^{2 \pi i \gamma_1^T Q \gamma_2} \,, \ee
which indeed reproduces the linking form on $M_3$.

\subsubsection{Self-dual $G$}
\label{sec:selfdualtm3}
The case $G=U(1)$ above is a special case of the more general situation where $G$ is self dual.  Then, as discussed in section \ref{sec:6dselfdual}, there is a natural choice of polarization which preserves all diffeomorphism symmetry of the spacetime, $M_6$, and may naturally be interpreted as an ordinary QFT, which we call the 6d theory with group $G$.  Then we claim that, in this case, compactification on $M_3$ gives a theory
\be 
T[M_3,G] \equiv \widehat{T}[\Graph,G] \,.
\ee
In particular, this is independent of the choice of graph decomposition of $M_3$. Indeed, one can check that, in the self-dual case, the $SL(2,\Z)$ relations discussed in section \ref{sec:QuiverSymmetries} are satisfied on the nose (or at least up to invertible TQFTs).  We saw this in section \ref{sec:ExSU2} for the case $G=U(2)$.

The higher-form symmetry of this 3d theory may be inferred by dimensional reduction of the 2-form $Z_G$-symmetry of the 6d theory, and the theory has
\be \ba \label{sdhfs}
\text{$0$-form symmetry:} &\;\;\;\;\; \Gamma_0 = H^2(M_3,Z_G) \cr 
 \text{1-form symmetry:}& \;\;\;\;\; \Gamma_1 = H^1(M_3,Z_G) \,.
 \ea
 \ee
 
In the case of $G=U(N)$, the 1-form symmetry has a contribution from the $U(1)$ sector, and is given as above by $\Gamma_1=H^2(M_3,\Z)$.  However, due to the periodicity of the $U(1)$ theory being $N$, the anomaly form is now $N$ times the linking form.  Then one can check that the non-anomalous subgroup is given as in \eqref{sdhfs}, but with $Z_G \rightarrow \Z_N$.

In the language of section \ref{sec:tm3ghdef}, this theory corresponds to
\be T[M_3,\g,H] \;\;\; \text{with} \;\;\; H = H^2(M_3,L) \leq H^2(M_3,\ZGtilde) \,,
\ee
where we recall $L$ is defined by $G=\widetilde{G}/L$.  We describe the construction of general $T[M_3,\g,H]$ next.

\subsubsection{General $\mathfrak{g}$}
\label{sec:tm3decoupling}

Let us now consider the general case.  As described in section \ref{sec:tm3ghdef}, on general grounds, we expect the theories, $T[M_3,\g,H]$, obtained by compactification to be labeled by a subgroup
\be H \leq  \widehat{\Upsilon} \equiv H^2(M_3,\ZGtilde) \ee
and to have $0$-form symmetry $H$ and 1-form symmetry $\Upsilon_H$.

Let us first describe the case $T[M_3,\g,1]$, which we denote by $T[M_3,\g]$ for short.  Here our naive guess might be
\be T[M_3,\g,1]  \overset{?}{=} \widehat{T}[\Graph,\widetilde{G}]\,, 
\ee
where we recall $\widetilde{G}$ is the simply connected group with Lie algebra $\g$, and $\Graph$ is some graph decomposition of $M_3$.  However, one basic problem we run into is that this theory is not independent of the graph.  Indeed, as observed in section \ref{sec:ExSU2}, for $\g=\frak{su}(2)$, the theory can pick up decoupled tensor factors of topological theories upon operations on $\Graph$ which preserve the topology of $M_3$.  In order to find a definition of $T[M_3,\g]$ which depends only on $M_3$, we must somehow eliminate these decoupled topological sectors.  

First, let us review the result of \cite{Hsin:2018vcg}, which states that, for a 3d TQFT ${\cal T}$ with anomalous 1-form symmetry, $\Gamma$, one finds that the theory factorizes
\be {\cal T} = {\cal T}' \otimes  \CT_{\Gamma,\CA}  \,,
\ee
where
\be \label{ctgzdef} \CT_{\Gamma,\CA} \equiv \text{minimal TQFT with $\Gamma$ symmetry with anomaly $\CA$}
\,. \ee
One way to isolate the sector ${\cal T}'$ is by coupling to another topological theory, $\CT_{\Gamma,-\CA}$, which has the opposite anomaly, and gauging the non-anomalous diagonal subgroup
\be \label{TpfromT} ({\cal T} \otimes \CT_{\Gamma,-\CA})/\Gamma  \; = \; {\cal T}' \otimes (\CT_{\Gamma,\CA} \otimes \CT_{\Gamma,-\CA})/\Gamma \;  \cong  \; {\cal T}'\,. \ee
We conjecture that a similar decoupling takes place in the theories $\widehat{T}[\Graph,\widetilde{G}]$.  Namely, we conjecture that we may write
\be \label{decoupconj} \widehat{T}[\Graph,\widetilde{G}] \cong T[M_3,\g] \otimes  \CT_{\Gamma/Z,\CA}\,. \ee
Here $\Gamma$ is the 1-form symmetry of $\widehat{T}[\Graph,\tG]$, with center $Z$, so that $\Gamma/Z$ is the anomalous part.  We conjecture that this anomalous symmetry acts only on a decoupled topological sector, and the physical theory, $T[M_3,\g]$, appears as a subsector.  Then one way to isolate the latter is as in \eqref{TpfromT}, namely
\be \label{TM3decoupdef} 
T[M_3,\g] \cong   (\widehat{T}[\Graph,\widetilde{G}] \otimes  \CT_{\Gamma/Z,-\CA})/(\Gamma/Z) \,.
\ee
However, any procedure which eliminates this decoupled sector will lead to an equivalent result.  For example, this can sometimes be achieved by gauging a suitable subgroup of the symmetry $\Gamma$ in the $\widehat{T}$ theory, as we will see in examples below.

It would of course be desirable to find a first principles derivation of the conjecture \eqref{decoupconj}.  However, let us mention several pieces of evidence in favor of it.  First note that with this assumption, the dependence on the graph, $\Graph$, due to the appearance of decoupled sectors, as discussed in section \ref{sec:QuiverSymmetries}, is cured.  Namely, as we will describe in more detail below, such sectors contribute only to $\CT_{\Gamma/Z,\CA}$, and so will be removed in \eqref{TM3decoupdef}.

One tempting interpretation of this decoupling is as follows.  Recall from section \ref{sec:SeifertManifolds} that the graph, $\Graph$, can naturally be interpreted as describing a plumbed four-manifold, $M_4$, bounded by $M_3$.  Then there are operations which preserve $M_3$, but modify $M_4$ by taking a connected sum with $\mathbb{P}^2$.  Now, as we saw above, the 6d theory naturally lives on the boundary of a 7d TQFT, and so the compactification of the 6d theory on $M_3$ can be interpreted as being accompanied by a compactification of the 7d theory on $M_4$.  Then the compactification on the 7d theory on these connected $\mathbb{P}^2$ summands will lead to tensor factors with decoupled topological sectors, precisely as we observed above.  The decoupling procedure can then be understood as removing the dependence on the auxiliary four-manifold, and obtaining a theory, $T[M_3]$, which depends intrinsically on $M_3$.

Next, from \eqref{Ht2factor} we see that the anomalous part of the 1-form symmetry can be  taken to act only on a decoupled tensor factor of the Hilbert space of the theory on $T^2$.  Then we claim that the Hilbert space of the theory $T[M_3,\g]$ is given by the second tensor factor, which has an anomaly-free 1-form symmetry, and the procedure above is defined so as to isolate this factor.  We will see this explicitly when we study the $T^2$ vacua of these theories in the next section.

Finally, we will see in section \ref{3d3dflat} that this decoupling arises naturally when we study the topological side of the 3d-3d correspondence, and is necessary to achieve a precise matching to the counting of flat connections. 

An important feature of this procedure is that it removes the anomalous 1-form symmetry, but does not modify the central subgroup, $Z$.  In particular, we claim
\be
	\ba \label{1formident}  
	\text{1-form symmetry of $T[M_3,\g]$}  & = \text{center of 1-form symmetry of $\widehat{T}[\Graph,\widetilde{G}]$}  \cr 
	& \cong \Upsilon =  H^1(M_3,\ZGtilde) \,,
	\ea
	\ee
which is the expected 1-form symmetry for this theory from the discussion in section \ref{sec:tm3ghdef}.  We will see this explicitly in examples below.

Finally, to define the more general theories $T[M_3,\g,H]$, we must gauge a subgroup of the $\Upsilon$ 1-form symmetry isomorphic to $H$
\be \label{TM3Hdef} 
T[M_3,\g,H] \equiv T[M_3,\g,1]/H \,.
\ee
This will leave a theory with 1-form  symmetry $\Upsilon_H$ and a new $0$-form symmetry $H$, as expected.  As in the 5d example above, this gauging of 1-form symmetry is precisely the same as changing the polarization of the 6d theory. 

\subsection{Examples}

\label{sec:Examples1form}

Let us consider some examples of how this 1-form symmetry structure arises in $T[M_3,\su(N)]$.  While this is applicable to general $M_3$, here and below we focus on the case of $M_3$ a Seifert manifold, and moreover a rational homology sphere, and we use the Lagrangians discussed in section \ref{sec:That}.

\subsubsection{Lens Space Quiver}
\label{sec:ZNAnomalies}

We start with the case of the Lens space, $M_3=L(p,q)$. Then,
\be \label{h12lpq}
H^1(L(p,q),\Z_N) \cong H^2(L(p,q),\Z_N) \cong  \Z_{(N,p)}\,,
\ee
where $(N,p)$ denotes the greatest common divisor. Thus, we expect that if $(N,p)\neq 1$ there are multiple versions of the theory, \eg, one with 1-form $\Z_{(N,p)}$ symmetry and no $0$-form symmetry, and one with $0$-form $\Z_{(N,p)}$ symmetry and no 1-form symmetry, and that we may pass between these two theories by gauging these symmetries.  On the other hand, when $(N,p)=1$, there is no such symmetry, and correspondingly only one choice of theory.  

\subsubsection*{$T[L(p,1)],\frak{su}(N)]$}

Let us first consider the case of $L(p,1)$.  Then we may take the decomposition graph, $\Graph_{[p]}$, to have a single node with label $p$, and so $\widehat{T}[\Graph_{[p]},SU(N)]$ is given by
\be  \widehat{T}[\Graph_{[p]},SU(N)] =  SU(N) \; \text{with level $p$ CS term and an adjoint scalar}  \,.
\ee
This theory has a $\Z_N$ 1-form symmetry, however, due to the Chern-Simons term, it has an 't Hooft anomaly with coefficient $p$, and so only a $Z = \Z_{(N,p)}$ subgroup is non-anomalous.  Comparing to \eqref{h12lpq}, we see we may identify this with
\be Z=  H^1(L(p,1),\Z_N) \cong \Z_{(N,p)}\,, \ee
as expected from \eqref{1formident}.

In the case where $p$ is a multiple of $N$, there is no remaining anomalous symmetry, and we have simply
\be  N|p \; : \;\;\;\; {T}[L(p,1),\frak{su}(N)] = SU(N) \; \text{with level $p$ CS term and an adjoint scalar}\,. \ee
More generally, we must also decouple the anomalous part of the symmetry, which is contained in the quotient, $\Gamma/Z= \Z_N/\Z_{(p,N)} \cong \Z_d$, where $d \equiv N/(N,p)$.  For example, consider the case where $(N,p)=1$, so that $d=N$.  Then, to accomplish this, we must couple to the theory $\CT_{\Z_N}$ with a $\Z_N$ 1-form symmetry of maximal anomaly, which in the present case may be taken as
\be \CT_{\Z_N}  = U(1)_N \; \text{Chern-Simons theory}\,. \ee
Thus, the prescription in \eqref{TM3decoupdef} tells us we should define
\be\ba
T[L(p,1),\frak{su}(N)] \equiv  & \bigg((SU(N) \; \text{with level $p$ CS term and an adjoint scalar}) \cr 
& \quad   \otimes (U(1)_{-N} \; \text{Chern-Simons theory}) \bigg)/\Z_N   \,.
\ea
\ee

This defines the theory corresponding to the subgroup $H=1$.  When $(N,p)=1$, this is the only choice, but more generally, let us choose some integer $k$ that divides $(N,p)$. Then the theory has an anomaly-free 1-form $\Z_k$ symmetry which we may gauge.  This then defines the theory
	\be \label{TLp1SUNZk}
	T[L(p,1),\su(N),\Z_k]\cong T[L(p,1),\frak{su}(N)]/\Z_k\,,
	\ee
	and this theory has a $0$-form $\Z_k$ symmetry. 
	We may gauge any of the non-anomalous $\Z_k$ symmetries so there might be various choices of global structure.
	The simplest example is $N=2$. Then, if $p$ is even we may gauge a 1-form $\Z_2$ symmetry passing over to gauge group $SO(3)$. If $p$ is odd this is not possible and only the $SU(2)$ version exists.
	
	An important special case is the theory $T[S^3,\frak{su}(N)]$. Recall that $S^3=L(1,1)$ so in this case, we have
	\be \widehat{T}[\Graph_{[1]},SU(N)] = SU(N) \; \text{with level 1 CS term and an adjoint scalar} \,.\ee
	From \eqref{sudecoup}, this theory is dual to $N-1$ free chiral multiplets tensored with a $U(1)_N$ topological sector.  This is precisely the decoupling behavior conjectured in \eqref{decoupconj}, providing an explicit example where this conjecture holds.  Then applying the decoupling procedure above, we see that, simply
	\be T[S^3,\frak{su}(N)] \cong \text{ $N-1$ free chiral multiplets} \,.\ee
	It is important that this theory has only a single BPS state in order for the number of such states to be a topological invariant of general $M_3$.  Namely, for any such $M_3$ we can take a connected sum with $S^3$, which tensors $T[M_3]$ with a copy of $T[S^3]$, and for this to leave the number of BPS states invariant, the latter should have only one BPS state.  Note  this is only true after performing the decoupling procedure above.

\subsubsection*{$T[L(p,q)],\frak{su}(N)]$}

For more general $L(p,q)$, we use the linear graph, $\Graph_{[k_1,\cdots,k_n]}$, associated to the continued fraction expansion, $[k_1,...,k_n]$, of $\frac{p}{q}$, as described in section \ref{sec:TM3}.  Then $\widehat{T}[\Graph_{[k_1,\cdots,k_n]},SU(N)]$ is given by a sequence of $SU(N)$ gauge groups with level $k_i$ Chern-Simons terms, and connected by copies of the $T(SU(N))$ theory.  However, as noted above, this linear quiver theory is not quite the theory $T[L(p,q),\frak{su}(N)]$ itself, as it contains an additional decoupled topological sector that must be removed.

From the general considerations above, we expect that the $T[L(p,q),\frak{su}(N)]$ theory has a 1-form symmetry group isomorphic to
\be \label{lpq1form} \Upsilon = H^{1}(L(p,q),\Z_N) = \Z_{(N,p)}\,. \ee
To identify this symmetry, first, using \eqref{tsu2anom}, we find that the anomaly matrix agrees with the linking matrix \eqref{LinkingMatrix}, \ie,
\be \label{AnomalyMatrix}
\mathcal{A}= \begin{pmatrix}
	k_1 & -1 & 0 & \cdots & & 0 \\
	-1 & k_2 & -1 &  & &  \\
	0 &  -1 & k_3 &  &&  \\
	\vdots & & &\ddots &  & \vdots \\
	& & &  & k_{n-1} & -1 \\
	0 & & & \cdots &-1 & k_n 
\end{pmatrix} \mod N \,.
\ee
Thus this theory has a $\Gamma = {\Z_N}^n$ 1-form symmetry.  However, many of these symmetries are anomalous, and  we are interested in the non-anomalous subgroup, $Z$.  This can be identified as the kernel of $\cA$ acting on $\Gamma$.  One may compute that
\be \label{detA}
\det \mathcal{A}= p \mod N\,.
\ee
This implies the anomaly matrix has a (unique, up to scalar multiplication) null-vector $\delta$ precisely when the RHS of  \eqref{lpq1form} is non-trivial.  We call $\delta$ the central element, and in general it generates the central subgroup $Z \cong \Z_{(N,p)}$.  Thus we have identified the expected 1-form symmetry in \eqref{lpq1form}.

Next let us discuss the remaining, anomalous 1-form symmetry.  For simplicity, we consider the case of $N$ prime.  Then, since $\Gamma \cong {\Z_N}^n$ is freely generated, we find that it splits into a direct sum of the center, $Z$, and the remaining, anomalous part, $\Gamma/Z$
\be \label{G=Z+G/Z} \Gamma = {\Z_N}^n \cong Z \oplus (\Gamma/Z)\,, \ee
where
\be Z = \begin{cases} \Z_N & N | p \\1  & \text{else} \end{cases} \,, \qquad \Gamma/Z = \begin{cases} {\Z_N}^{n-1} & N | p \\ {\Z_N}^n  & \text{else} \end{cases} \,.
\ee
Then to decouple the anomalous $(\Gamma/Z)$ 1-form symmetry and obtain the $T[L(p,q),\frak{su}(N)]$ theory, following  \eqref{TM3decoupdef} we define
\be \label{DefTLpq} T[L(p,q),\frak{su}(N)] \equiv  (\widehat{T}[\Graph,SU(N)] \otimes \CT_{\Gamma/Z,-\cA} )/(\Gamma/Z)\,. \ee
For example, if $\cA$ can be diagonalized over $\Z_N$, then we may take $\CT_{\Gamma/Z,-\cA}$ to equal $\text{rank}(\Gamma/Z)$  copies of the $U(1)_{-N}$ CS theory.

The theory obtained after this procedure retains the expected anomaly-free $Z \cong \Upsilon=H^1(L(p,q),\Z_N)$ 1-form symmetry, and we claim this is the theory obtained by compactification of the 6d theory on $L(p,q)$.  One may then obtain $T[L(p,q),\frak{su}(N),H]$ for $H \leq \Upsilon$ by gauging non-anomalous subgroups, as above.

\subsubsection*{Symmetries of the Quiver}

In section \ref{sec:QuiverSymmetries}, we observed that the theories $\widehat{T}[\Graph,\widetilde{G}]$ are not yet topological invariants, as they are not invariant under operations on the graph, $\Graph$, that preserved the underlying three-manifold, $M_3$.  Let us now verify that the theories $T[M_3,\g]$ do not suffer from this problem.  We will focus on the case of $M_3=L(p,q)$ and $\g=\frak{su}(N)$, but the arguments generalize straightforwardly.  

First we consider the operation of appending a $TST$ transformation at one end of the linear quiver.  As we saw in \eqref{sudecoup}, the only effect of this operation for $G=SU(N)$ is to introduce a factor of the topological $U(1)_N$ theory in the quiver.  But such a decoupled topological sector is expected to contribute only to the second factor in \eqref{decoupconj}, and not to the physical theory, $T[L(p,q),\frak{su}(N)]$.  Indeed, this decoupled sector comes with an anomalous $\Z_N$ 1-form symmetry, which appears as a factor in $\Gamma/Z$, and correspondingly there is a factor of $U(1)_{-N}$ in $\CT_{\Gamma/Z,-\cA}$.  When we  perform the decoupling procedure, these two factors of $U(1)_N$ are eliminated by the diagonal $\Z_N$ gauging.  Thus the theory is precisely the same as the one before applying this $TST$ transformation.  Similar comments apply to the insertions of $(ST)^3$ and $S^2$, which also add $U(1)_N$ topological sectors.  Namely, these are also removed by the decoupling procedure above, and so $T[L(p,q),\frak{su}(N)]$ is independent of such operations on the quiver.

Finally, let us consider the operations leading to the dual description of $T(G)$ as a trinion with one flavor symmetry gauged at level 1, as in \eqref{tgtrinduality}.  As above we focus on the case of $G=SU(N)$, but expect this to hold for more general $G$.  This involves applications of both the $(ST)^3$ relation, and the $TST$ relation, and each of these introduce a decoupled $U(1)_N$ factor.  On the one hand, this may be removed by the decoupling procedure above.  An alternative and equivalent way to remove these sectors and isolate the physical theory is to simply gauge the diagonal $\Z_N$ of these two $U(1)_N$ factors.  But one can check that, in the star-shaped quiver description, this $\Z_N$ gauging maps precisely to gauging the 1-form symmetry of the central $SU(N)$ node, leaving the gauge group as $SU(N)/\Z_N$.  This is in fact the expected result, as the duality of the star-shaped quiver of the trinion holds only when the central node is taken as $SU(N)/\Z_N$ \cite{Razamat:2014pta}. Similarly, $S^2$ becomes trivial after gauging the $\Z_N$ at the middle node because $T(G)$ squares to the identity \cite{Gaiotto:2008ak}, when the global structure is taken into account.

\subsubsection{Seifert Quivers}
\label{sec:1-FormGeneral}

Next we look at more general quivers associated to Seifert manifolds. For simplicity, we focus on the gauge algebra $\mathfrak{g}=\su(2)$, which already reveals much of the interesting structure.
We also restrict ourselves to graphs $\Omega_{(k_i)}$ with intersection matrix
\be
Q= \begin{pmatrix}
	0 & 1 & 1 & 1 \\
	1 & k_1 & 0 & 0 \\
	1 & 0 & k_2 & 0 \\
	1 & 0 & 0 & k_3 
\end{pmatrix}\,,
\ee
with the $k_i$ chosen to correspond to the manifolds $S^3/{\Gamma_{ADE}}$ in \eqref{S3/Gamma}. If any $k_i=1$, $M_3$ is a Lens space, which were discussed in the previous section, so we will consider $k_i=(2,-2,n-2)$ for $D_n$ and $k_i=(3,-2,m-3)$ for $E_m$.
The theory $\widehat{T}[\Omega_{(k_i)},SU(2)]$ has a $\Gamma={\Z_2}^4$ 1-form symmetry, which in the notation of \eqref{G=Z+G/Z} splits as
\be
\Gamma\cong Z\oplus (\Gamma/Z)\,, \qquad Z=H^1(M_3,\Z_2)\,,
\ee
depending on the choice of $M_3$. Concretely,
\be \label{H1Z2Coeff}
H^1(S^3/\Gamma_{D_n},\Z_2)=\left\{
\begin{array}{l l}
	\Z_2 \times \Z_2 &n \text{ even}\\
	\Z_2 & n\text{ odd}
\end{array}\right. \,,
\qquad 
H^1(S^3/\Gamma_{E_m},\Z_2)=\left\{
\begin{array}{l l}
	1 & m\text{ even}\\
	\Z_2 & m\text{ odd}
\end{array}\right. \,.
\ee
We then define $T[S^3/\Gamma_{DE},\mathfrak{g}]$ in analogy to \eqref{DefTLpq}.

Let us now focus on the $D_n$ case.
The kernel of the anomaly matrix $\mathcal{A}_n$\footnote{Recall that this is given by $\mathcal{A}=Q\mod 2$.} has dimension two if $n$ is even and dimension one if $n$ is odd, in agreement with \eqref{H1Z2Coeff}. For $n$ odd the 1-form symmetry of $T[S^3/\Gamma_{D_n},\frak{su}(2)]$ is generated by $\delta_1=(0,0,1,1)$. Gauging this symmetry gives us the theory $T[S^3/\Gamma_{D_n},\su(2),\Z_2^{(1)}]$.  For $n$ even there is an additional $\Z_2$ 1-form symmetry generated by $\delta_2=(0,1,0,1)$. Thus there are multiple choices for $H$, namely
\be \label{DnChoices}
\ba
&T[S^3/\Gamma_{D_n},\su(2),\Z_2^{(i)}]\,, \quad i=1,2,3\\
&T[S^3/\Gamma_{D_n},\su(2),\Z_2 \times \Z_2]\,,
\ea
\ee
where $\Z_2^{(3)}$ is the diagonal subgroup generated by $\delta_3=\delta_1+\delta_2$. The $E_m$ case does not unveil any additional structure with $E_7$ being equivalent to $D_n$ with $n$ odd and for $E_m$, $m$ even, there is no 1-form symmetry.

\subsection{Geometric Interpretation of Global Structure}

\begin{figure}
	
	\begin{centering}
		
		\includegraphics[width=5cm]{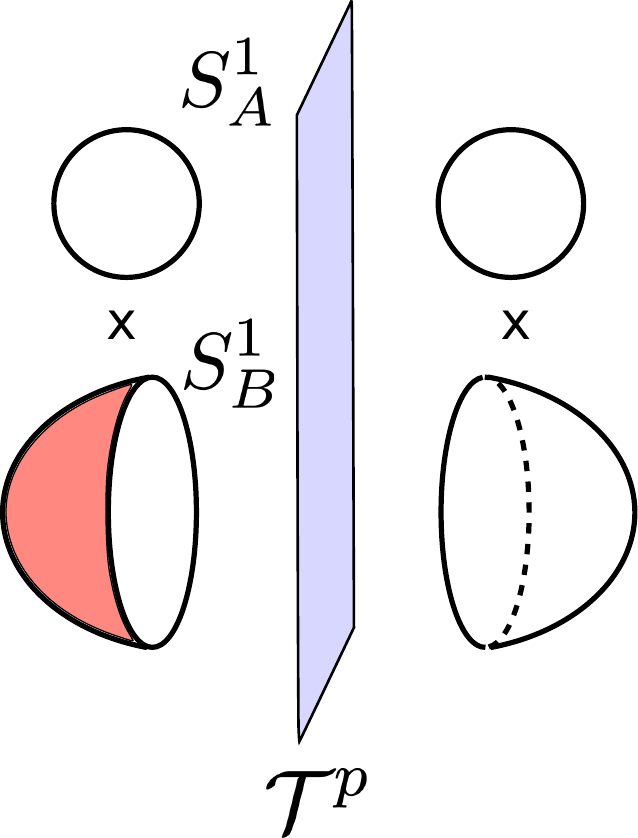}
		
	\end{centering}
	\caption{Heegaard decomposition of $L(p,1)$, where the two solid tori are glued with a $T^p$ transformation.  In red is shown the 2 cycle with $\Z_N$ coefficients.  For this to be consistently extend to the right side, we must have $p=0 \mod N$, which reflects the fact that the homology group is non-trivial only in this case.}
	\label{fig:tk}
\end{figure}

Here we comment on a geometric picture for the 1-form symmetry above in terms of compactification of the 6d theory.  Let us first focus on the case $M_3=L(p,1)$ for simplicity, where we found a non-anomalous $\Z_N$ 1-form symmetry generated by $\delta$ if $p=0 \mod N$.  Recall the possible 1-form symmetry group operators on $W_3$ are constructed by first picking a 1-cocycle on $M_3$, or equivalently, a 2-cycle, with $\Z_N$ coefficients.  We may describe this explicitly using the Heegaard decomposition of $L(p,1)$, as shown in figure \ref{fig:tk}. Here we show the 2-cycle in red on the left solid torus, which we take to generate a $\Z_N$ 2-chain on the LHS.  At the boundary, we have labeled the two cycles of the torus by $S^1_A$ and $S^1_B$.  We see the 1-cycle wrapped by the 2-cycle as it crosses the boundary is $(0,1)$ in the $(A,B)$ basis.  Upon passing through the $T^p$ transformation, this will now wrap the cycle
\be \begin{pmatrix} 1 & p \\ 0 & 1 \end{pmatrix}  \begin{pmatrix} 0 \\  1 \end{pmatrix} = \begin{pmatrix} p \\  1 \end{pmatrix}  \,.\ee
Then for this to be contractible on the right side, so  that we can consistently close the 2-cycle, we see that we must impose $p=0 \mod N$ (recalling we  are working with $\Z_N$ coefficients).  This is the geometric origin of the fact that the $\Z_N$ homology is equal to $\Z_N$ only in this case.  Then this maps precisely to the generator of the $\Z_N$ 1-form symmetry that is present in $T[L(p,1),\frak{su}(N)]$ in this case.

\begin{figure}
	
	\begin{centering}
		
		\includegraphics[width=11cm]{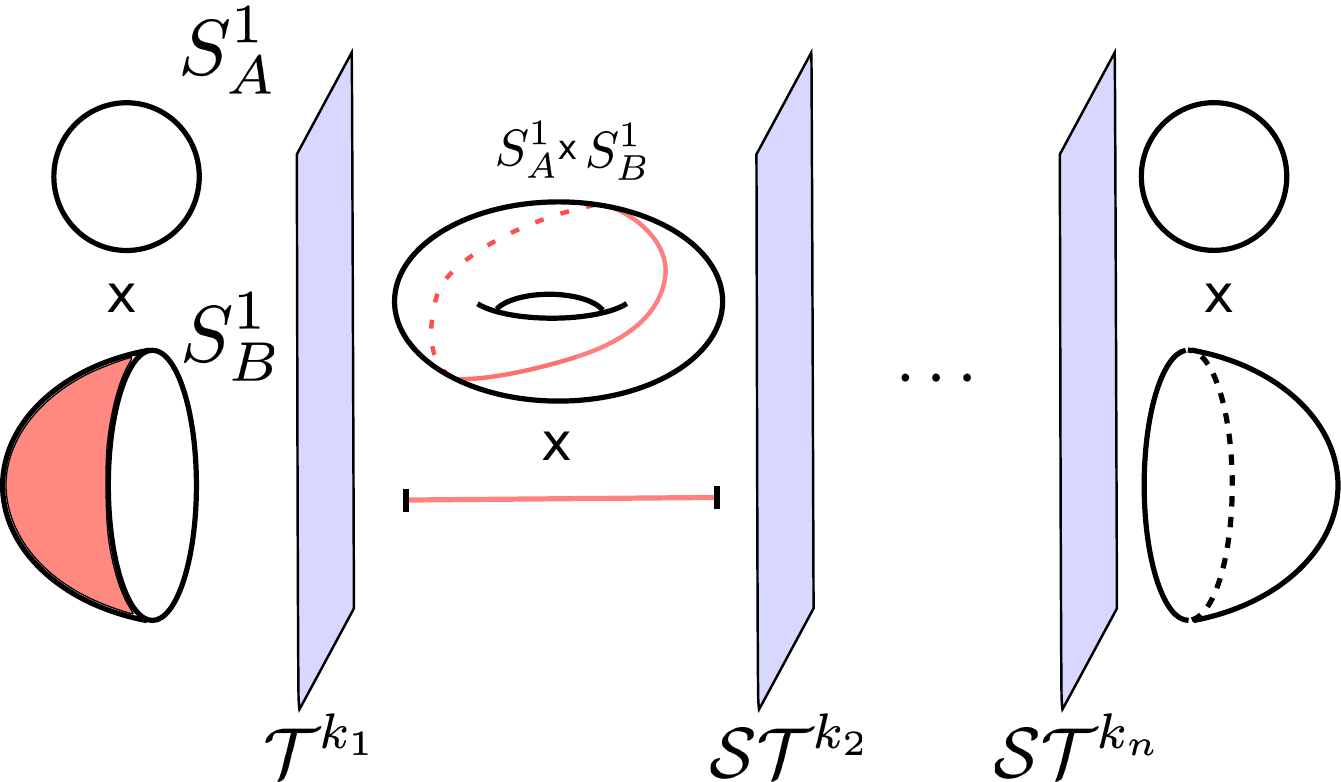}
		
	\end{centering}
	\caption{Heegaard decomposition of $L(p,q)$ for linear quiver, where the two solid tori are glued with an $SL(2,\Z)$ transformation.}
	\label{fig:genvarphi}
\end{figure}

Let us now consider a more general Lens space. Recall from the previous subsection that, in this case, the 1-form symmetry is $\Z_N$ when $p$ is a multiple of $N$, and is generated by an element $\delta$ inside the ${\Z_N}^n$ 1-form symmetry of the linear quiver theory, $\widehat{T}[\Graph_{[k_1,\cdots,k_n]},SU(N)]$.  Explicitly, the components of $\delta$ are
\be \label{deltaN} \delta= (p_1,...,p_n) \;, \ee 
where we have defined
\be p_j = \det( \cA_j ), \;\;\; j=0,...,n , \;\;\;\; \cA_j = j \times j \; \text{upper left submatrix of $\cA$}\,. \ee
We have
\be \label{pidef} p_0 = 1, \;\;\; p_1 = k_1, \;\;\; p_2 = k_1 k_2 -1,  \;\;\; \cdots, \;\;\; p_{n-1} = q, \;\;\; p_n = p\,. \ee
In figure \ref{fig:genvarphi} we have drawn the analogue of figure \ref{fig:tk} for a more general Lens space corresponding to gluing by an $SL(2,\Z)$ element
\be \varphi_{p,q} = T^{k_1} S T^{k_2} .... S T^{k_n} \,.
\ee
As above, we start with the 2-cycle wrapping the disk inside the solid torus on the left.  We call this the generating 2-cycle. The intermediate segments in the diagram have the form of an interval times a torus, which we label as $S^1_A \times S^1_B$ as above, and  the 2-cycle will lie along this interval, and wrap some cycle of the torus.  In fact the cycle wrapped after the $i$th wall is given in the $(A,B)$ basis by
\be \begin{pmatrix} p_i \\ q_i \end{pmatrix} \equiv  \begin{pmatrix} p_i \\ p_{i-1 } \end{pmatrix} \,,
\ee
where the $p_i$ are defined in \eqref{pidef} above.  Here $L(p_i,q_i)$ is the space we would obtain if we were to cap off the space after the $i$th step.

Note that in each segment, we are compactifying the 6d theory on a torus times an interval, which leads to the 4d $\frak{su}(N)$ $\cN=4$ theory on an interval.  Moreover, the cycles of the torus Poincar\'e dual to the polarization, $\Lambda$, determine the global form of the ${\frak su}(N)$ theory we obtain in 4d \cite{Tachikawa:2013hya,Aharony:2013hda}.
Let us give a quick review. Given a Lie algebra $\su(N)$ we can characterize the global structure of the gauge group in a phase of the 4d $\mathcal{N}=4$ by the allowed line operators. They are characterized by two integers
\be
(z_e,z_m)\in \Z_N \times \Z_N\,,
\ee
where with $(1,0)$ and $(0,1)$ we label the minimal Wilson and 't Hooft lines respectively. Then, the charge lattice of a gauge group $G$ is spanned by the allowed line operator charges.
For $N$ prime the structure is particularly simple. For $G=SU(N)$ only the Wilson lines are allowed, \ie, the charge lattice is generated by $(1,0)$. On the other hand, there are $N$ different ways to define the $\Z_N$ quotients. The charge lattice of $G=PSU(N)_n$ is generated by the dyonic line $(n,1)$. Here, $n=0,\dots,N-1$ denotes the different choices of gauge groups. An $SL(2,\Z)$ transformation acts on the line operators as
\be \label{TSChargeLattice}
T:\ (z_e,z_m)\to(z_e+z_m,z_m)\,, \qquad
S:\ (z_e,z_m)\to(z_m,-z_e)\,.
\ee
From this prescription we can determine how the $T$ and $S$ transformations change the gauge group in the related phases. If $N$ is prime they simply act as
\be \label{GtoGvee}
\ba
T:\quad &SU(N) \to SU(N) \qquad &&PSU(N)_n\to PSU(N)_{n+1}\\
S:\quad  &SU(N) \leftrightarrow PSU(N)_0 && PSU(N)_n \leftrightarrow PSU(N)_{1/n}\,.
\ea
\ee
For $N$ not prime this is more involved as we have to also take into account the phases where a subgroup of $\Z_N$ is gauged. The precise prescription can be found in \cite{Aharony:2013hda}.

If we look at the transformation behavior of \eqref{TSChargeLattice} we see that it agrees with the geometric behavior of $(p_i,q_i)$, \ie,
\be 
T:\ \frac{p}{q}\to \frac{p+q}{q}\,, \qquad
S:\ \frac{p}{q}\to \frac{q}{-p}\,.
\ee
It is thus intuitive to identify the values $(p_i,q_i)$, labeling the generating 2-cycle at some point in the Heegaard splitting, with the gauge group of the 4d $\mathcal{N}=4$ theory on the respective interval. In order to make sense of this we have to assign the value $(0,1)$, \ie, gauge group $PSU(N)_0$, to the generating 2-cycle on the left cap of the Heegaard splitting. The condition $p=0$ ensures that the gauge group on the right cap is also $PSU(N)_0$, \ie, the cycle is contractible. Let us for example consider $\mathfrak{g}=\su(2)$. Then, the three choices of gauge groups are
\be (p_i,q_i) = \left\{ \begin{array}{ccl} 
	(1,0) & \rightarrow &  SU(2) \\
	(0,1) & \rightarrow &  SO(3)_+ \\
	(1,1) & \rightarrow &  SO(3)_- \end{array} \right. \,. \ee
In order to obtain the $H=\Z_2$ version of the theory we now have to gauge all the nodes where the gauge group is $SO(3)$. In order for this to be globally consistent, we have to end up with an $SO(3)_+$ gauge group on the right cap, \ie, $p=0$.	We can easily generalize this to any prime $N$.

More generally, for $N$ not prime, we can define any other $H=\Z_k$ version, where $k$ divides $N$, provided $k$ is a multiple of $(N,p)$. To this end we assign the charge vector $(0,\frac{N}{k})$ to $(SU(N)/\Z_k)_0$, \ie, we let the generating 2-cycle wrap the disk $\frac{N}{k}$ times. Then, we can go through the Heegaard splitting as before, but at each step the charges are multiples of $\frac{N}{k}$ and thus non-anomalous.


\section{Bethe Vacua for $T[M_3,\mathfrak{g}]$}

\label{sec:ResultsU(N)}

In this section we study the set of supersymmetric vacua of the theory $T[M_3,\mathfrak{g}]$ for $M_3$ a Seifert manifold with $g=0$. This provides a detailed test of several of the general features described above.  For example, we will see that the set of 1-form symmetries of $T[M_3,\mathfrak{g}]$ act on the  supersymmetric vacua, and constrain the states that can appear. Furthermore, in the next section, we compare the number of vacua to the expectation from the 3d-3d correspondence, which maps this observable to the counting of certain flat connections on $M_3$, and find agreement.
Applications to the $\mathcal{N}=1$ twist are considered in section \ref{sec:3dN=1}.

We will focus on the cases of $\mathfrak{g}= \mathfrak{u}(N)$ and $\su(N)$. The first case that will be considered is $\mathfrak{u}(1)$, where the theory is free.  In the other cases, and specializing to the $\cN=2$ twist, we will use the Bethe ansatz equation approach to studying supersymmetric vacua of 3d $\cN=2$ gauge theories of \cite{Nekrasov:2014xaa}, which we briefly review in section \ref{app:TwistedSupo}.

\subsection{Abelian Case}
\label{sec:Abelian}

The theory with $G =U(1)$ is free and has featured in \cite{Gadde:2013sca}. As discussed in section \ref{sec:DefinitionU1} the theory is independent of the specific choice of graph $\Graph$ for $M_3$, so we can immediately define $T[M_3,U(1)]$. The bosonic Lagrangian is determined by the linking matrix of $M_3$
\be \label{LCSU1}
\mathcal{L}_{\text{CS}}=\frac{1}{4\pi} \sum_{i,j=1}^n\ Q_{ij}\ A_i \wedge dA_j\,,
\ee
as well as a set of adjoint scalars. Since the center of mass field is massless there is a flat direction and the index is ill-defined. This can be remedied by removing this mode and defining the index of $T[M_3,U(1)]$ as the number of vacua of \eqref{LCSU1}.
This modified index is computed by consecutively integrating out the gauge fields and equals
\be
I(T[M_3,U(1)])=|\det(Q_{ij})|\,.
\ee
For example for the Lens spaces $L(p,q)$ this is 
\be \label{U1IndexLens}
I(T[L(p,q),U(1)])=p\,.
\ee
Note that this number agrees, as expected, with the number of flat abelian connections on $L(p,q)$ given by $|H_1(L(p,q),\Z)|=|\Z_p|$.
For the general case of a Seifert manifold $M_3=[d;0;(p_i,q_i)]$, the number of vacua is 
\be
I(T[M_3,U(1)])=\left|\left(\sum_{i=1}^r \frac{q_i}{p_i}-d\right) \prod_{i=1}^r p_i\right| \,.
\ee
Let us look more closely at the manifolds discussed in appendix \ref{app:Seife}. For the Prism manifolds $M_{p_3,q_3}^{\text{Prism}}=[0;0;(2,-1),(2,1),(p_3,q_3)]$ we obtain
\be \label{IU1Prism}
I(T[M_{p_3,q_3}^{\text{Prism}},U(1)])=4|q_3|\,,
\ee
with the special case
\be
I(T[S^3/\Gamma_{D_n},U(1)])=4\,.
\ee
Similarly, we find
\be
I(T[S^3/\Gamma_{E_m},U(1)])=(9-m)\,.
\ee
All of these results match with the respective orders of the first holonomy groups.
For a general Brieskorn manifold $M^{\text{Brieskorn}}_{p_1,p_2,p_3}=[1;0;(p_1,1),(p_2,1),(p_3,1)]$ the index is
\be
I(T[M^{\text{Brieskorn}}_{p_1,p_2,p_3},U(1)])=p_1p_2p_3\left|\frac{1}{p_1}+\frac{1}{p_2}+\frac{1}{p_3}-1\right|\,.
\ee

Recall from section \ref{sec:DefinitionU1} that the 1-form symmetry group, $\Gamma$, of this theory expected from the 6d reduction is $H^1(M_3,U(1)) \cong H^2(M_3,\Z)$, and can be identified in the CS theory as the electric 1-form symmetry left unbroken by the CS terms.
Explicitly, in the case that $M_3$ is a rational homology sphere we find 
\be  \Gamma = \text{ker} ( Q : (\R/\Z)^n \rightarrow (\R/\Z)^n ) \cong H^2(M_3,\Z) \,.
\ee
Moreover, the bilinear form $\cA$ determining the anomaly is the linking form of $M_3$.  As discussed in section \ref{sec:ht2reps}, the space of vacua on $T^2$ must fall into representations of the 1-form symmetry generators, $U^A$ and $U^B$, wrapping the two cycles of the torus.  In the present case, one can check that the center $Z\subset \Gamma$  is trivial for $M_3$ a rational homology sphere, and so the size of the unique irreducible representation is
\be \text{dim} \; R = |\Gamma| = |H^2(M_3,\Z)| \,.\ee
But this is precisely the dimension of the space of vacua, showing that the Hilbert space itself is in an irreducible representation of the 1-form symmetry generators.  This is a special feature of $U(1)$ Chern-Simons theory, and in general we will find the Hilbert space is given by a direct sum of irreducible representations.

\subsection{Twisted Superpotentials and Coulomb Branch Vacua}
\label{app:TwistedSupo}

To compute the vacua for non-abelian gauge groups, we will use the framework developed in  \cite{Nekrasov:2014xaa}, \ie, we compute the vacua using the twisted superpotential and counting the solutions to the resulting Bethe equations. 

Let us consider a 3d $\CN=2$ gauge theory of gauge group $G$ on $\mathbb{R}^2\times S^1$. 
The dynamics of the Coulomb branch is governed by the twisted superpotential,
$\CW(u,m)$, where $u$ is the Coulomb branch variable
\be
u = i\beta (\sigma + ia_0) \in {\frak t}^{\mathbb{C}}/\Lambda_{\text{co-char}}\ ,
\ee
where $\frak{t}^\mathbb{C}$ is the complexified Cartan subalgebra, and $\beta$ is the radius of the $S^1$.
Here $a_0$ is the holonomy of the abelianized gauge field along the $S^1$
\be
\frac{1}{2\pi}\oint_{S^1} A = a_0\ ,
\ee
and $\sigma$ is the abelianized constant real scalar field in the vector multiplet. Similarly, $m = i\beta(\sigma_F + ia_{0,F})$ is a flavor group variable.
Let us define $T$ as the maximal torus of $G$ and $\frak t$ the Cartan subalgebra. Then the lattice $\Lambda_{\text{co-char}}$ is the co-character lattice of $G$ defined as the kernel of the exponential map
\be
\text{exp}: {\frak t} \rightarrow T\ .
\ee
The large gauge transformation along $S^1$ is defined by
\be
u\rightarrow u+\lambda\ ,
\ee
where $\lambda$ is an element in the lattice $\Lambda_{\text{co-char}}$.

The twisted superpotential receives a contribution from the classical CS level as well as the chiral multiplets of the theory. The $\CN=2$ CS action contributes
\be\label{CS term}
\CW_{\text{CS}}(u) = \frac12 \sum_{a,b=1}^{\text{rk}(G)}k_{ab}u_a u_b + \sum_{\substack{a> b\\  \text{abelian}}} K_{ab} u_a u_b \ ,
\ee
up to the gravitational CS level, which we ignore\footnote{In what follows we will ignore all the numerical constants in $\CW$, which are irrelevant for the vacuum count.}. For each simple factor $G_{s}$ of the gauge group, it is understood that $k_{ab}= h_{ab}k_{s}$ where $h_{ab}$ is the Killing form for $G_s$. The second term in \eqref{CS term} is the contribution from the mixed CS levels, $K_{ab}$, between two abelian factors of the gauge group. A chiral multiplet $\Phi_R$ in the representation $R$ of $G$ contributes
\be \label{wphi}
\CW_{\Phi_R}(u) = \sum_{\rho\in R} \frac{1}{(2\pi i)^2} \text{Li}_2(x^\rho)\, ,
\ee
where $x = e^{2\pi i u}$. We adopt the $U(1)_{-1/2}$ quantization of the CS level as discussed in \cite{Closset:2017zgf}. For example, the chiral multiplet with charge $\pm 1$ under the Cartan $U(1)$ contribute as follows
\be\ba
\CW_{\Phi}(u) &=\frac{1}{(2\pi i)^2}~ \text{Li}_2(x) \cr 
\CW_{\widetilde\Phi}(u) &= \frac{1}{(2\pi i)^2}~\text{Li}_2(x^{-1}) + \frac12 u(u+1)\ .
\ea\ee
We will use the notation
\be
x = e^{2\pi i u},~~ z = e^{2\pi i \zeta}\ ,~~ y = e^{2\pi i m}\ ,~~t = e^{2\pi i \tau}\ ,
\ee
where $t$ is the fugacity for the distinguished flavor symmetry in $\CN=4$ theories, which corresponds to the $\CN=2^*$ mass. If $G$ contains a $U(1)$ factor, we turn on the FI parameter $\zeta$ for the associated topological symmetry.

For convenience, let us define the flux operators
\be \label{FluxOperator}
\Pi_a =\exp \left(2\pi i \frac{\partial \CW(u,m)}{\partial u_a}\right)\ ,\qquad 
\Pi_\alpha =\exp \left(2\pi i \frac{\partial \CW(u,m)}{\partial m_\alpha}\right)\ ,
\ee
for the gauge and flavor symmetry, respectively. The Coulomb branch vacua are described by the solutions to the Bethe equations  \cite{Nekrasov:2014xaa}
\be \label{BetheVacua}
\CS_{\text{BE}} = \left\{u_a~|~ \Pi_a=1\ ,\forall a=1,\cdots \text{rank}(G),~~w(u) \neq u ,\forall w \in W_G \right\}/W_{G}\ ,
\ee
where $W_G$ is the Weyl group. Note that we removed the solutions which are fixed under the Weyl group action. We call such solutions degenerate vacua.  We provide further details on the twisted superpotential and Bethe equations in appendix \ref{app:BetheEquations}.

\subsection{Bethe Vacua for $\widehat{T}[\Graph,SU(2)]$ and $T[M_3,U(2)]$}
\label{sec:THatCount}

Before constructing the theories $T[M_3, \mathfrak{g},H]$ in general, we study the unphysical theory $\widehat{T}[\Graph,G]$, defined in section \ref{sec:TM3}, focusing on $\Graph$ the graph of a Seifert manifold $M_3$ and $G=SU(2)$.
This theory is defined by taking the Seifert quiver of section \ref{sec:SeifertQuivers}, connecting $SU(2)$ gauge nodes with level $k_i$ CS-terms and $T(SU(2))$ $S$-walls and taking all gauge groups as $SU(2)$.
A refinement including other choices of gauged 1-form symmetries will be the topic of section \ref{sec:BVSU2SO3}.

The $\widehat{T}[\Graph,SU(2)]$ theories are also used to construct the theories $T[M_3,U(2)]$.
As discussed in section \ref{sec:6dselfdual}, we can define $T[M_3,U(2)]$ as a tensor product of the two decoupled theories
\be \label{tu2decoup}
\widehat{T}[\Graph,SU(2)] \quad  \text{and} \quad  \widehat{T}[\Graph,U(1)_2]\,.
\ee
For each gauge node, we identify a non-anomalous $\Z_2$ subgroup of the 1-form symmetry, and gauge it to replace the gauge group at this node by $(U(1) \times SU(2))/\Z_2 \cong U(2)$.\footnote{Note this is the same description one obtains from the quiver $\widehat{T}[\Omega,U(2)]$, as follows from the more general fact that
\be
\label{unsun} 
U(N)_k = { SU(N)_k \times U(1)_{Nk} \over \Z_N }\,.
\ee
Then, $T(U(N))$ can essentially be regarded as a tensor product of $T(SU(N))$ and $T(U(1))$ with level $N$ BF coupling, with an additional operation to account for the $\Z_N$ quotient.}

\subsubsection{The $\Omega_{[k]}$ Theory}

As a simple example, let us first consider $\widehat{T}[\Omega_{[k]},SU(2)]$, which can be described as the $\CN=2^*$ Chern-Simons theory at level $k$. The twisted superpotential for the $\CN=2^*$ CS theory is given by
\be
\CW_{T^k}^{\CN=2^*} = \CW_{\text{CS}}(u) + \CW_{\text{adj}}(u,\tau)\ ,
\ee
where
\be
\CW_{\text{CS}}(u)  =k u^2
\ee
is the contribution from the $\CN=2$ CS action and
\be
 \CW_{\text{adj}}(u,\tau)= \CW_\Phi(-2\tau \pm 2u) + \CW_\Phi(-2 \tau) \ee
is the contribution from the $\CN=2$ adjoint multiplet, where here and below  we use the convention that each  choice of sign in `$\pm$' is  summed over.  More explicitly, this is given by
 \be \CW_{\text{adj}}(u,\tau) =  \frac{1}{(2\pi i)^2}\left[\text{Li}_2(x^2 t^{-2}) + \text{Li}_2(x^{-2}t^{-2}) + \text{Li}_2(t^{-2}) \right] +  u(u+1) \,.
\ee
The Coulomb branch vacua are then determined as solutions to the Bethe equations
\be
\Pi \equiv \exp \bigg( 2\pi i \frac{\partial \CW_{T^k}^{\CN=2^*}}{\partial u} \bigg) =  x^{2k}\bigg( \frac{x^2-t^{-2}}{1-x^2 t^{-2}}\bigg)^2 = 1 \,.
\ee
It is convenient to split this into two polynomial equations
\be\label{bethe su2}
\Pi^{1/2} = x^{k}\frac{x^2-t^{-2}}{x^2 t^{-2}-1}= \pm 1\,.
\ee
Since the Weyl symmetry acts as
\be
x\rightarrow x^{-1} \;,
\ee
the solutions at $x=\pm 1$ are degenerate vacua. When $k$ is even, we have $k+2$ solutions from each sign of the equation \eqref{bethe su2}. The degenerate vacua are contained in the solution for the bottom sign. Therefore we have $2k+2$ regular solutions paired by the Weyl symmetry. The number of   Coulomb branch vacua are $k+1$. When $k$ is odd, we have $k+1$ regular vacua from each sign, paired by the Weyl symmetry.\footnote{A similar calculation was considered in \cite{Chung:2014qpa}; see footnote \ref{foot:chung} below for the relation between these computations.}



\subsubsection{The Linear Quivers $\Omega_{[k_1,\cdots,k_n]}$}

\begin{figure}
\centering		
\includegraphics[width=12cm]{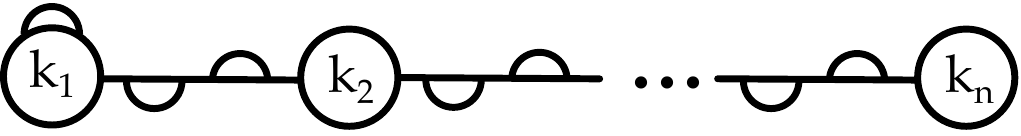}
\caption{Quiver corresponding to $\Graph_{[k_1,\cdots,k_n]}$ for the $\mathcal{N}=2$ twist.
		\label{fig:tststdgg}}
\end{figure}

The graph for a general Lens space $L(p,q)$ is linear and we denote it by $\Omega_{[k_1,\cdots,k_n]}$. The corresponding quiver is shown in figure \ref{fig:tststdgg}.
To simplify the computation, we recall the duality \eqref{DualityT(SU2)T2} which relates the $T(SU(2))$ theory to the trinion $T_2$.\footnote{We should note that this also allows us to write a standard Lagrangian for these theories, which is not true using the usual $T(SU(2))$, as the $SU(2)$ is not manifest.} The effect of this duality on the quiver is shown in figure \ref{fig:tststduality}. In principle this duality could be used for any $SU(N)$, however for $N>2$ the $T_N$ theory is more complicated than $T(SU(N))$.

In appendix \ref{app:ExplicitComp} we compute the number of vacua for a single trinion, with the three nodes gauged at CS-levels $\widetilde{k}_i$ using the Bethe equations. The number of vacua for $\widetilde{k}_i=(k_1-1,k_2-1,1)$ is 
\be \label{NvacTST}
I({\widehat{T}[\Graph_{[k_1,k_2]},SU(2)]}) = 2 \left(|k_1 k_2-1|+1\right) = 2(p+1)\,.
\ee

One can similarly study longer linear quivers.  This quickly leads to a large number of coupled polynomial equations, which  can be solved using Gr\"obner basis techniques.  By studying several examples up to length $n=4$, we have found evidence for the following conjectural number of vacua for a general linear quiver 
\be \label{NvacTSTST}
I({\widehat{T}[\Graph_{[k_1,\cdots,k_n]},SU(2)]}) = 2^{n-1} (p+1)\,.
\ee
It would be interesting to derive this formula analytically, but we leave this to future work.

Since $\widehat{T}[\Graph,U(1)_2]$ is the abelian Chern-Simons theory of section \ref{sec:Abelian} with each Chern-Simons level multiplied by 2, the number of vacua is $|\det\left(2Q_{ij}\right)|=2^n p$.
Putting this together, the tensor product in \eqref{tu2decoup} has
\be 
I(\widehat{T}[\Graph,SU(2)] \ \otimes \   \widehat{T}[\Graph,U(1)_2]) = 2^{2n-1} p(p+ 1)\,.
 \ee
Finally, we must implement the 1-form gauging. As we argue in appendix \ref{app:SU(2)Anomalies:u2gauging}, each $\Z_2$ gauging reduces the number of vacua by a factor of 4, and so we find
\be I(T[L(p,q),U(2)])=\frac{1}{2}p(p+1)\,.
\ee
Note this result no longer depends on the choice of $\Graph$, in agreement with the logic in section \ref{sec:6dselfdual}.

\subsubsection{Seifert Quivers}
\label{sec:GeneralQuivers}

\begin{figure}
\centering
\includegraphics*[width=12cm]{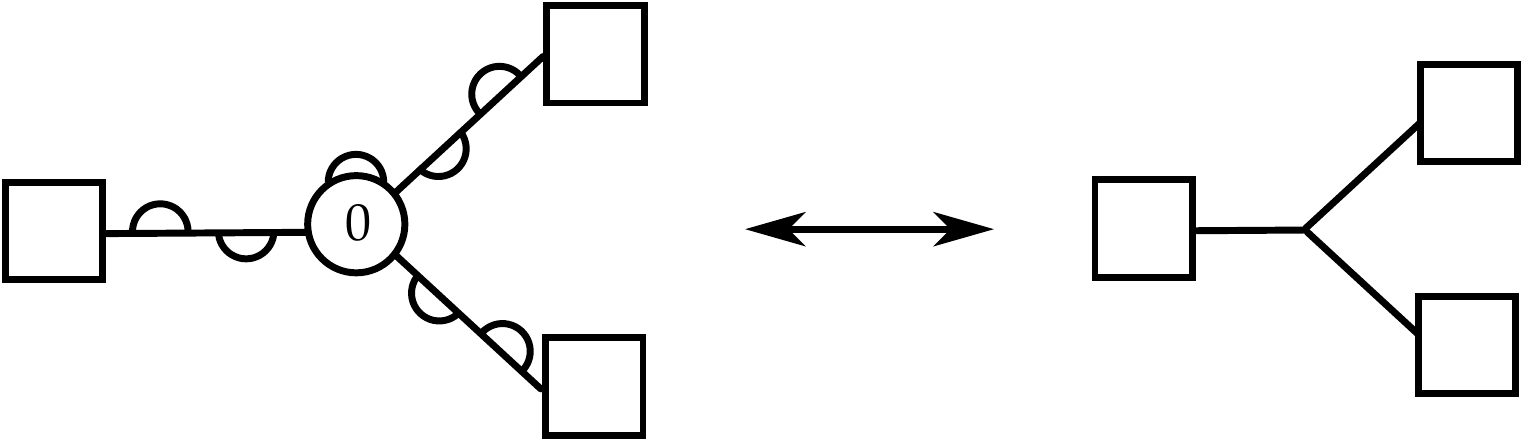}
\caption{Duality to trinion theory: Three copies of $T(SU(N)) \cong T_N$.}
\label{fig:Trinion}
\end{figure}

\begin{figure}
\centering		
\includegraphics[width=6cm]{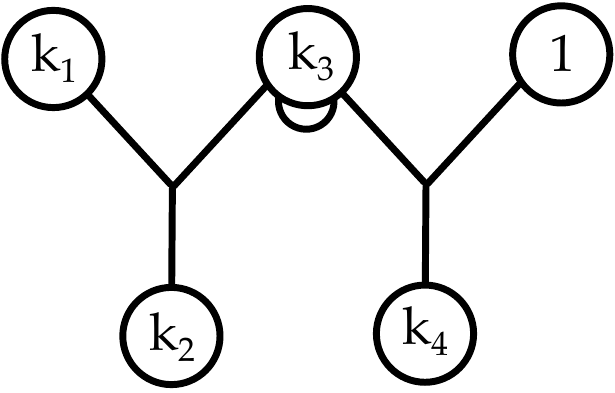}
\caption{Quiver for two coupled trinions and an adjoint at the central node indicated by the arc. We denote this quiver by $\Omega_{(k_1, k_2, k_3, k_4,1)}$.}
\label{fig:TwoTrinions}
\end{figure}

For general Seifert manifolds the counting of vacua becomes computationally difficult. However for a subclass we are able to perform the analysis explictly. These are Seifert manifolds, which have base $S^2$ and three special fibers, and are discussed in appendix \ref{app:Seife} and summarized in table \ref{tab:ResultsTrinions}.
We can simplify the quiver using the trinion theory $T_N$,  using the following duality: couple three copies of $T(SU(N))$ by identifying one of their flavor groups each and gauging it at Chern-Simons level zero, together with an adjoint scalar. This theory is dual to the trinion theory \cite{Benini:2010uu}, see also \eqref{BCTrinion}. This means that if the degree in the Seifert data vanishes, we can write the quiver using the trinion description, as in figure \ref{fig:Trinion}. The vanishing of the degree can always be assumed using \eqref{SeifertEquivalencesTST} and \eqref{SeifertEquivalencesST3}. Then, the graph $\Graph_{(k_i)}$\footnote{For linear quivers we used the subscript $[k_1, \cdots, k_n]$. For the nonlinear quivers we usually indicate the CS-levels, depending on the associated graph.} for the Seifert manifolds that we are interested in is given by figure \ref{fig:TwoTrinions}. In order to guarantee that this corresponds to a Seifert manifold we have to impose $k_5=1$.

First, consider the Prism manifolds $M^{\text{Prism}}_{p,q}\cong[0;0;(2,-1),(2,1),(p,q)]$, assuming that $\frac{p}{q}=\widetilde{k}_1-\frac{1}{\widetilde{k}_2}$ and $q>0$ for simplicity. 
The graphs for Prism manifolds are characterized by
\be
(k_i)^{\text{Prism}}=(-2,2,\widetilde{k}_1-1,\widetilde{k}_2-1,1)\,.
\ee
We find that the number of vacua of the quiver is given by
\be
I({\widehat{T}[\Graph_{(k_i)^{\text{Prism}}},SU(2)])}=2(p+1+8q)\,.
\ee
Now, in order to determine the Witten index we have to take the $U(1)$ index and the $\Z_2$ gauging into account, leading to an extra factor of $\frac{q}{2}$, see \eqref{IU1Prism}. Thus,
\be
I(T[M^{\text{Prism}}_{p,q},U(2)])=q(p+1+8q)\,.
\ee
Note that for $q=1$ this reduces to
\be
\label{u2s3dn}
I(T[S^3/\Gamma_{D_n},U(2)])=n+7\,.
\ee
Similarly, we can compute that the number of vacua of $\widehat{T}$ for the $E$-series is 24, independent of $m$, where we use the graph with two trinions in figure \ref{fig:TwoTrinions}. Including an additional factor of $\frac{9-m}{8}$ from the $U(1)$ sector and gauging, this yields
\be
I(T[S^3/\Gamma_{E_m},U(2)])=3(9-m)\,.
\ee

Finally, let us look at the general Brieskorn manifolds $M^{\text{Brieskorn}}_{p_1,p_2,p_3}=[0;0;(p_1,1),(p_2,1),(p_3,1-p_3)]$. We can always write them with two trinions using that $\frac{p_3}{1-p_3}=[-1,p_3-1]$. The levels are given by
\be
(k_i)^{\text{Brieskorn}}=(p_1,p_2,-2,p_3-2,1)\,,
\ee
which is invariant under relabeling of the $p_i$. We find that the number of vacua of $\widehat{T}$ is
\be
I({\widehat{T}[\Omega_{\text{Brieskorn}}, SU(2)]})=6p_1p_2p_3-6(p_1p_2+p_2p_3+p_3p_1)+2(p_1+p_2+p_3)+2\,,
\ee
as long as the manifold has negative curvature, \ie, $\sum_{i=1}^3 p_i^{-1}<1$. Together with the index of the abelian theory and the $\Z_2$ gauging we find
\be \label{IndexBrieskorn2qr}
\ba
I(T[M^{\text{Brieskorn}}_{p_1,p_2,p_3},U(2)])=\frac{1}{4}&\left(3p_1p_2p_3-3(p_1p_2+p_2p_3+p_3p_1)+(p_1+p_2+p_3)+1\right)\\&
\times \left(p_1 p_2 p_3-(p_1p_2+p_2p_3+p_3p_1)\right)\,.
\ea
\ee
For convenience we summarize the results in table \ref{tab:ResultsTrinions}.
\begin{table}
\begin{center}
\begin{tabular}{c|c|c}
$M_3$ & $k_i$  & $I(T[M_3,U(2)])$\\
\hline\hline
$L(p,q)\,,\ \frac{p}{q}=[\widetilde{k}_1,\widetilde{k}_2,\widetilde{k}_3]$ & $\left(1,\widetilde{k}_1-1,\widetilde{k}_2-1,\widetilde{k}_3-1,1\right)$  & ${{p+1} \choose {2}}$\\
$M^{\text{Prism}}_{p,q}\,,\ \frac{p}{q}=[\widetilde{k}_1,\widetilde{k}_2]$ & $\left(-2,2,\widetilde{k}_1-1,\widetilde{k}_2-1,1\right)$  & $q(p+1+8q)$\\
$S^3/\Gamma_{E_m}$ & $\left(-2,3,m-3,0,1\right)$ & $3(9-m)$\\
$M^{\text{Brieskorn}}_{p_1,p_2,p_3}\,,\ \mathcal{R}<0$ & $\left(p_1,p_2,-2,p_3-2,1\right)$  & \eqref{IndexBrieskorn2qr}
\end{tabular}
\end{center}
\caption{Three-manifold, quiver data and vacuum count  in terms of the Witten index $I$ for various Seifert manifolds $M_3$ with $G=U(2)$.}
\label{tab:ResultsTrinions}
\end{table}

Let us briefly comment on the 1-form symmetry structure of these theories. After the ${\Z_2}^n$ gauging of the tensor product appearing in \eqref{tu2decoup}, one finds that the remaining 1-form symmetry present in $T[M_3,U(2)]$ is the same as that in $T[M_3,U(1)]$.  However, due to the different normalization of the Chern-Simons levels, the anomaly coefficient, $\cA$, is now multiplied by 2.  In general, for $T[M_3,U(N)]$, we find the 1-form symmetry and anomaly
\be \Gamma = H^2(M_3,\Z), \qquad \CA(\gamma_1,\gamma_2) = e^{2 \pi i N \ell(\gamma_1,\gamma_2)}\,,
\ee
where $\ell:TH^2(M_3,\Z) \times TH^2(M_3,\Z) \rightarrow \mathbb{Q}/\Z$ is the linking form. 
As a check, we may verify the set of vacua found above falls into representations of the 1-form symmetry operators.  For example, for $L(p,q)$, $\Gamma=\Z_p$.  When $p$ is odd, the center, $Z$, of the 1-form symmetry is trivial, and so the irreducible representations have dimension $p$, while when $p$ is even, the center is $\Z_2$, and so they have dimension $p/2$.  In both cases, we find the number of vacua, given by $\frac{p(p+1)}{2}$, is indeed divisible by the size of this irreducible representation, which is compatible with this decomposition.   For an example involving a non-linear quiver, in the case $M_3=S^3/\Gamma_{D_{n}}$, the symmetry is $\Z_2 \times \Z_2$ for $n$ even, and $\Z_4$ for $n$ odd.  In the former case, this symmetry is anomaly-free, so the irreducible representations have size 1, while in the latter, it has an anomaly with coefficient 2, and so the irreducible representations have size 2.  Thus the number of vacua must be even in the latter case, and one can verify from \eqref{u2s3dn} that this is always true.

\subsubsection{$T[M_3,U(N)]$}
\label{sec:unBVsubsec}

In principle, the considerations above carry over to the case of general $U(N)$.  That is, we may build up quivers for $T[M_3,U(N)]$ by taking decoupled copies of the $\widehat{T}[\Graph,SU(N)]$ and $\widehat{T}[\Graph,U(1)_N]$ theories together, and gauging appropriate $\Z_N$ 1-form symmetries.  However, this typically leads to much more complicated theories than in the $N=2$ case, for which the Witten index calculations above are often intractable.  Moreover, unlike for $N=2$, these theories will typically be non-Lagrangian.  Specifically, although the $T(U(N))$ theory is a Lagrangian theory, for $N>2$ we do not have a Lagrangian description of it in which the $U(N) \times U(N)$ symmetry is manifest in the UV description, and so when we gauge these symmetries we are led to theories without  Lagrangians.

One case where we can write a Lagrangian is for the $L(k,1)$ theory.  Here the theory $T[L(k,1),U(N)]$ is simply the 3d $\cN=2$ $U(N)$ theory with level $k$ Chern-Simons term and an adjoint chiral multiplet.  As shown in appendix \ref{app:ComputationsU(N)}, the number of solutions to  the Bethe equations for this theory is given by
\be\label{IuN}
I(T[L(k, 1), U(N)])= \binom{N+k-1}{N}\,. \ee
From the discussion above, the 1-form symmetry of this theory is $\Z_k$, with an anomaly with coefficient $N \;( \text{mod} \; k)$.  The center of this symmetry therefore is $\Z_{(N,k)}$, and so the irreducible representations have size $k/(N,k)$.  One may check that the index (\ref{IuN}) is indeed divisible by this integer, giving another check that the 1-form symmetry acts as expected.


\subsection{Bethe Vacua for $T[M_3,\su(2),H]$}
\label{sec:BVSU2SO3}

In this section we construct the physical theories $T[M_3,\su(2),H]$ from $\widehat{T}[\Graph,SU(2)]$ and count the number of vacua. Recall that the unphysical theory $\widehat{T}$ is obtained by gauging all nodes of a graph $\Graph$ corresponding to the manifold $M_3$ at gauge group $SU(2)$. However, as discussed in section \ref{sec:Examples1form} the theory has anomalous $\Z_2$ 1-form symmetries associated to decoupled topological sectors, which need to be eliminated to obtain the physical theories.

\subsubsection{The $\Omega_{[k]}$ Theory}

First, consider the case $M_3=L(k,1)$ using the minimal graph with a single node. The Bethe equation is
\be
\Pi=x^{2k}\left(\frac{x^2t^2-1}{x^2-t^2}\right)^2\,,
\ee
where $x=e^{2\pi i u}$, which has $k+1$ solutions, after accounting for the Weyl symmetry $u\to -u$. These form the Hilbert space of $\widehat{T}[\Graph_{[k]},SU(2)]$.
To obtain the physical Witten index we need to decouple the topological sector. From the prescription in section \ref{sec:ZNAnomalies} we can deduce that
\be
T[L(k,1),\su(2)]=\begin{cases} \widehat{T}[\Graph_{[k]},SU(2)] & \text{$k$ even} \\ \left(\widehat{T}[\Graph_{[k]},SU(2)] \otimes U(1)_{-2} \right)/\Z_2  & \text{$k$ odd} \end{cases}\,.
\ee
For $k$ even, the counting of vacua is therefore the same as above, while for $k$ odd, the decoupling procedure removes a tensor factor of dimension 2, so that
\be\label{IndexLk1}
I(T[L(k,1),\su(2)])=\begin{cases} k+1 & \text{$k$ even} \\ \frac{k+1}{2}  & \text{$k$ odd} \end{cases}\,.
\ee

The above correspond to the $T[L(k,1),\su(2),H=1]$ theories. When $k$ is even, we may also consider the case $H=\Z_2$, which is obtained by gauging the non-anomalous $\Z_2$ 1-form symmetry of this theory.  Thus, it will be important to understand how the 1-form symmetries act on the Bethe vacua of these 3d $\cN=2$ theories.  As discussed in section \ref{sec:ht2reps}, the space of $T^2$ vacua is naturally acted on by two 1-form charge operators, $U_{A,B}$, wrapping the two cycles of the torus.  The action of these charge operators on the Bethe vacua is discussed in \cite{BWHF}, and reviewed in appendix \ref{app:1formBV}.  In the present case, we find these operators act on the Bethe vacua as\footnote{The overall sign of $U^A$ can be fixed by modular invariance.}
\be \label{UiTk}
U^A\ket{\widehat{u}}=\widehat{x}^k\frac{\widehat{x}^2t^2-1}{\widehat{x}^2-t^2}\ket{\widehat{u}}\,, \qquad U^B\ket{\widehat{u}}=\ket{\widehat{u}+\half}\,.
\ee
Note that in general, $U^A U^B=(-1)^k U^B U^A$, so for $k$ even this symmetry is indeed non-anomalous.  The irreducible representations are then 1-dimensional, and the Hilbert space decomposes into the eigenspaces, $\mathcal{H}_{\epsilon_A \epsilon_B}$, introduced in (\ref{Ht2factor}). We will be interested in the space of Bethe vacua $V$, which is a subspace of the full Hilbert space, that also decomposes as $V_{\epsilon_A,\epsilon_B}$ under the 1-form charge operators wrapping the $A$ and $B$ cycles of the torus
\be \label{DecompositionV}
V \to V_{++} \oplus V_{+-} \oplus   V_{-+} \oplus   V_{--}\,.
\ee  
We define
\be
v_{\epsilon_A \epsilon_B} = \text{Tr}_{\mathcal{H}_{\epsilon_A \epsilon_B}} (-1)^F = \text{dim} V_{\epsilon_A \epsilon_B}  \,,
\ee
which we refer to as the {\it refined Witten index}. 
Gauging the $\Z_2$ symmetry then amounts to projecting onto the $V_{++}$ subspace, and including the twisted sectors, as in \eqref{hgauged}.
To analyze the decomposition \eqref{DecompositionV} it is useful to split the common eigenbasis of the $U$ into
\be
\ba
\ket{\widehat{v}_j^\pm}=\frac{1}{\sqrt{2}}\left(\ket{\widehat{u}_j}\pm\ket{\widehat{u}_j+\half}\right),\ j=1,\dots \frac{k}{2}\,, \qquad
\ket{\widehat{u}^{\text{$B$-inv.}}}=\ket{\frac{1}{4}}\,,
\ea
\ee
where $\ket{\widehat{u}^{\text{$B$-inv.}}}$ is invariant under $U^B$ after using the Weyl-symmetry. Clearly
\be \label{ActionUiTk}
U^B\ket{\widehat{v}_j^\pm}=\pm\ket{\widehat{v}_j^\pm}\,, \qquad U^A\ket{\widehat{u}^{\text{$B$-inv.}}}=(-1)^\frac{k}{2}\ket{\widehat{u}^{\text{$B$-inv.}}}\,.
\ee
To determine $v_{++}$ we still need to find the action of $U^A$ on the $\ket{\widehat{v}_j^\pm}$. 
Alternatively, to count the solutions we can directly solve the equation $U^A=\pm1$, and one finds the result always satisfies
\be \label{ModularInv}
v_{+-}=v_{-+}\,.
\ee
Physically, this corresponds to modular invariance along the $T^2$, obtained by exchanging $A$- and $B$-cycle.\footnote{In the following we will always assume that this is guaranteed to give analytic arguments. However, in all examples modular invariance can also be checked explicitly by evaluating $U^A$ on the vacua.}
Combining \eqref{ActionUiTk} and \eqref{ModularInv} we can show that the contribution to the Witten index from the untwisted sectors is
\be \label{IndexLk1SU2Z2}
I_{\text{untwisted}}(T[L(k,1),\su(2),\Z_2])=v_{++}=\left\lfloor \frac{k}{4} \right\rfloor +1\,.
\ee
To account for the twisted sectors, we note, as reviewed in appendix \ref{app:1formBV} that these are associated to fixed points of the $U_B$ operators.  In the present case, there is a single such state\footnote{Note, that the sign of $U^A$ in the twisted sector is again determined by modular invariance, and may differ from the one in the untwisted sector.}, at $\widehat{u}=\frac{1}{4}$, and so we find
	\be \label{IndexLk1SU2Z2full}
I(T[L(k,1),\su(2),\Z_2])=\left\lfloor  \frac{k}{4} \right\rfloor +2\,.
\ee

\subsubsection{General Linear Quivers}
\label{sec:AnomaliesLinQuiver}

Next, consider the general linear quiver $\Graph_{[k_1, \cdots, k_n]}$ of length $n$, corresponding to $M_3=L(p,q)$. The theory $\widehat{T}[\Graph_{[k_1, \cdots, k_n]},SU(2)]$ has $n$ 1-form symmetries, with anomaly matrix given by \eqref{AnomalyMatrix}, which we reproduce here
\be \label{AnomalyMatrix2}
\mathcal{A}= \begin{pmatrix}
	k_1 & 1 & 0 & \cdots & & 0 \\
	1 & k_2 & 1 &  & &  \\
	0 &  1 & k_3 &  &&  \\
	\vdots & & &\ddots &  & \vdots \\
	& & &  & k_{n-1} & 1 \\
	0 & & & \cdots &1 & k_n 
\end{pmatrix} \mod 2 \,.
\ee
This matrix has $\det(\cA) = p$, and admits a single null vector if and only if $p$ is even.  As described in appendix \ref{app:z21formreps}, the set of Bethe vacua on $T^2$ for a theory acted on by such a 1-form symmetry has the form, see (\ref{VRr}),
\be \label{Hthat} 
\widehat{V} = 
V \otimes R^{\otimes m}\,,
\qquad  m = \left\{ \begin{array}{cc} n-1 &p \; \text{even} \\ n &p \; \text{odd} \end{array} \right.\,,
\ee
where $R$ is a two dimensional representation of an anomalous $\Z_2$ 1-form symmetry. 
As noted in \eqref{NvacTSTST}, we find
\be I(\widehat{T}[\Graph_{[k_1, \cdots, k_n]},SU(2)]) = 2^{n-1} (p+1)\,, \ee
and so we deduce that
\be I(T[L(p,q),\su(2)])= \text{dim}(V) = \left\{ \begin{array}{cc} p+1 &p \; \text{even} \\ \frac{p+1}{2} &p \; \text{odd} \end{array} \right.\,, \ee
agreeing with the expected $q$-independent result \eqref{IndexLk1}. 

When $p$ is even we may gauge an additional $\Z_2^{(\delta)}$ 1-form symmetry generated by $\delta$, the central element in \eqref{deltaN}. This amounts to projecting onto the $V_{++}$ eigenspace and by studying a few low $n$ cases explicitly we find
\be 
I_{\text{untwisted}}(T[L(p,q),\su(2),\Z_2^{(\delta)}]) = \left\lfloor \frac{p}{4} \right\rfloor +1\,,
\ee
generalizing \eqref{IndexLk1SU2Z2}. Additionally, we find that there is always a single twisted sector state, and so the answer for the full Witten index is
\be 
I(T[L(p,q),\su(2),\Z_2^{(\delta)}]) = \left\lfloor \frac{p}{4} \right\rfloor +2\,.
\ee

\subsubsection{Gauged Trinion}
\label{sec:VacuaGaugedTrinion}

As a final example consider the manifolds $S^3/\Gamma_{ADE}$ also discussed in section \ref{sec:1-FormGeneral} and appendix \ref{app:Seife}. Recall that the Seifert data can be written as
\be
S^3/\Gamma_{ADE}\cong[0;0;(k_1,1),(k_2,1),(k_3,1)]\,,
\ee
with the $k_i$ given by:
\begin{enumerate}
\item
$(k_1,k_2,1)$ for the Lens space with $\frac{p}{q}=[k_1+1,k_2+1]$
\item
$(-2,2,n-2)$ for $S^3/\Gamma_{D_n}$
\item
$(-2,3,m-3)$ for $S^3/\Gamma_{E_m}$.
\end{enumerate}
The $\widehat{T}[\Omega_{(k_i)},SU(2)]$ theory is the  {\it gauged trinion} theory: a trifundamental chiral multiplet of $SU(2)$ with the three $SU(2)$ flavor symmetries gauged at levels $k_1,k_2$, and $k_3$ and the three Bethe equations are given by
\be \label{BetheGaugedTrinion}
\Pi_i=
x_i^{2k_i}\frac{(x_i x_j x_k-t)(x_i x_j-x_k t)(x_i x_k-x_j t)(x_i-x_j x_k t)}{(x_i x_j x_k t-1)(x_i x_j t-x_k )(x_i x_k t-x_j)(x_i t-x_j x_k)}\,,\qquad i=1, 2, 3 \,.
\ee
This set of equations is discussed in appendix \ref{app:ExplicitComp} and the number of solutions is given by\footnote{All other possibilities can be obtained by permutation of the $k_i$ and parity symmetry.}
\be \label{solGaugedTrinion}
\ba
I(\widehat{T}[\Omega_{(k_i)},SU(2)])&=(k_1+1)(k_2+1)(k_3+1)  && \text{if}\ {k_i\geq 0}\\
I(\widehat{T}[\Omega_{(k_i)},SU(2)])&=(k_1+1)(k_2+1)(-k_3+1)-4k_1k_2 && \text{if}\ {k_{1,2}\geq 0,\ k_3\leq -2}\,,
\ea
\ee
after accounting for the Weyl symmetry.

To find the Witten index of $T[M_3,\su(2)]$ we need to decouple the topological sector.
If not all $k_i$ are even, we can specify $k_3$ odd. Since only $k_i \mod 2$ is relevant for the anomalies, the decoupling of the topological sector  is analogous to the Lens space, using the duality in figure \ref{fig:tststduality}.
This case was discussed in detail in section \ref{sec:AnomaliesLinQuiver}. If all the $k_i$ are even no decoupling is necessary.
Putting this together we find
\be
\ba
I(T[L(p,q),\su(2)])&=\begin{cases} p+1 & \text{$p$ even} \\ \frac{p+1}{2}  & \text{$p$ odd} \end{cases}\\
I(T[S^3/\Gamma_{D_{n}},\su(2)])&=\begin{cases} n+7 & \text{$n$ even} \\ \frac{n+7}{2}  & \text{$n$ odd} \end{cases}\\
I(T[S^3/\Gamma_{E_{m}},\su(2)])&=\begin{cases} 3\ \ \ \ \ & \text{$m$ even} \\ 6  & \text{$m$ odd} \end{cases}\,.
\ea
\ee

The Hilbert space spanned by these vacua splits into the eigenspaces of the charge operators $U^{A,B}_{\delta_j}$
\be
V\to \bigoplus_{\epsilon^{A,B}_j=\pm 1} V_{\epsilon^A_1\epsilon^B_1\cdots\epsilon^A_r\epsilon^B_r}\,, \qquad r=\text{rk}\left(H^1(M_3,\Z_2)\right)\,,
\ee
where the $\delta_j$ generate the $\Z_2^r$ 1-form symmetry.
Let us now study this explicitly.
If two of the $k_i$, say $k_1$ and $k_2$, are even then the theory has a physical non-anomalous $\Z_2$ 1-form symmetry, as argued for in section \ref{sec:1-FormGeneral}. Thus, the theory has a 1-form symmetry generated by $\delta=(1,1,0)$. The charge operators act as, see \eqref{1formgenerators},
\be
\ba
U^A_\delta\ket{\widehat{u}_1,\widehat{u}_2,\widehat{u}_3}&=\left(\Pi_1\Pi_2\right)^{1/2}\ket{\widehat{u}_1,\widehat{u}_2,\widehat{u}_3}=\widehat{x}_1^{k_1}\widehat{x}_2^{k_2} \frac{(\widehat{x}_1\widehat{x}_2\widehat{x}_3-t)(\widehat{x}_1\widehat{x}_2-\widehat{x}_3 t)}{(\widehat{x}_3- \widehat{x}_1 \widehat{x}_2 t)(1-\widehat{x}_1 \widehat{x}_2\widehat{x}_3 t)}\ket{\widehat{u}_1,\widehat{u}_2,\widehat{u}_3}\\ U^B_\delta\ket{\widehat{u}_1,\widehat{u}_2,\widehat{u}_3}&=\ket{\widehat{u}_1+\half,\widehat{u}_2+\half,\widehat{u}_3}\,.
\ea
\ee
We proceed as for the $T^k$ theory and determine the solutions fixed under $U^B_\delta$, which are given by
\be
\ket{\widehat{u}^{\text{$B$-inv.}}}^j=\ket{\frac{1}{4},\frac{1}{4},\widehat{u}_3^j}\,.
\ee
By plugging in these solutions into the Bethe equations \eqref{BetheGaugedTrinion} we find that they reduce to
\be
\Pi_{1,2}=1\,, \qquad\Pi_3=x_3^{2k_3}\left(\frac{x_3^2-t^2}{x_3^2t^2-1}\right)^2\,.
\ee
This is nothing but the $T^{k_3}$ theory with an adjoint and thus has $|k_3|+1$ vacua.\footnote{This is the number of vacua before the decoupling so only half of them are physical if $k_3$ is odd.} These $B$-invariant vacua obey
\be\label{gtfixed}
U^A_\delta \ket{\widehat{u}^{\text{$B$-inv.}}}^j=(-1)^{\frac{k_1+k_2}{2}}\ket{\widehat{u}^{\text{$B$-inv.}}}^j\,, \qquad j=1,\dots \begin{cases} |k_3|+1 & \text{$k_3$ even} \\ \frac{|k_3|+1}{2}  & \text{$k_3$ odd} \end{cases}\,.
\ee
As before, the remaining vacua $\ket{\widehat{v}^\pm}$ split into pairs with eigenvalues $\pm1$ under $U^B$. Modular invariance fixes the dimension of the subspaces with $\epsilon^{A,B}=\pm 1$. We can repeat this for all choices of $\delta_j$, $j=1,\cdots,r$ to determine the refined Witten index
\be
v_{\epsilon^A_1\epsilon^B_1\cdots\epsilon^A_r\epsilon^B_r}=\dim V_{\epsilon^A_1\epsilon^B_1\cdots\epsilon^A_r\epsilon^B_r}\,.
\ee
The following discussion depends heavily on the rank of the maximal 1-form symmetry, $\Upsilon=\Z_2^r$:
\begin{enumerate}
\item Lens spaces $L(p,q)$, with $p$ odd, and $S^3/\Gamma_{E_{m\text{ even}}}$ have no 1-form symmetry, \ie, $r=0$. This means there is no further decomposition of $V$.

\item The manifolds $L(p,q)$, with $p$ even, $S^3/\Gamma_{D_{n\text{ odd}}}$ and $S^3/\Gamma_{E_7}$ have $r=1$. We can study the dimension of the eigenspaces $V_{\epsilon^A \epsilon^B}$ which are summarized in table \ref{tab:dimVr=1}. \footnote{\label{foot:chung}In \cite{Chung:2014qpa}, the number of Bethe vacua for the $\CN=2$ $SU(2)$ $\text{CS}_k$ theory with one adjoint chiral multiplet was computed and compared to the number of $SU(2)$ flat connections on the lens space. In our notation, their counting corresponds to the number of Bethe vacua for the $T[L(p,1), \su(2),1]$ theory with $\epsilon^A=1$. We will see in section \ref{sec:3d3dcorr} that this prescription indeed agrees with the refined 3d-3d correspondence we propose.}  Recall that modular invariance implies $v_{+-}=v_{-+}$.
\begin{table}[h!]
\begin{center}
\begin{tabular}{c||c||c|c|c}
$S^3/\Gamma_{ADE}$, $r=1$ & $\dim V$ & $v_{++}$ & $v_{+-}$ & $v_{--}$ \\
\hline
\hline
$A_{p-1}$, $p=0\mod 4$ & $p+1$ & $\frac{p}{4}+1$ & $\frac{p}{4}$ & $\frac{p}{4}$\\
\hline
$A_{p-1}$, $p=2\mod 4$ & $p+1$ & $\frac{p+2}{4}$ & $\frac{p+2}{4}$ & $\frac{p-2}{4}$\\
\hline
$D_n$, $n$ odd & $\frac{n+7}{2}$ & $\frac{n+1}{2}$ & 1 & 1\\
\hline
$E_7$ & 6 & 2 & 2 & 0
\end{tabular}
\caption{Dimensions of the eigenspaces $V_{\epsilon^A \epsilon^B}$ for $\Upsilon=\Z_2$.}
\label{tab:dimVr=1}
\end{center}
\end{table}

Gauging the 1-form symmetry projects onto the subspace $V_{++}$ and thus
\be
\ba
I{_{\text{untwisted}}}(T[L(p\text{ even},q),\su(2),\Z_2^{(\delta)}])&=\left\lfloor \frac{p}{4} \right\rfloor +1\\
I{_{\text{untwisted}}}(T[S^3/\Gamma_{D_{n\text{ odd}}},\su(2),\Z_2^{(\delta)}])&=\frac{n+1}{2}\\
I{_{\text{untwisted}}}(T[S^3/\Gamma_{E_7},\su(2),\Z_2^{(\delta)}])&=2\,.
\ea
\ee
In addition, one finds the following contribution from the twisted sectors, which we recall are associated to the fixed points, \eqref{gtfixed}
	\be
	\ba
	I_{\text{twisted}}(T[L(p\text{ even},q),\su(2),\Z_2^{(\delta)}])&=1\\
	I_{\text{twisted}}(T[S^3/\Gamma_{D_{n\text{ odd}}},\su(2),\Z_2^{(\delta)}])&=\frac{n-1}{2}\\
	I_{\text{twisted}}(T[S^3/\Gamma_{E_7},\su(2),\Z_2^{(\delta)}])&=2\,.
	\ea
	\ee
	so that the full Witten index is
		\be
	\ba
	I(T[L(p\text{ even},q),\su(2),\Z_2^{(\delta)}])&=\left\lfloor \frac{p}{4} \right\rfloor +2\\
	I(T[S^3/\Gamma_{D_{n\text{ odd}}},\su(2),\Z_2^{(\delta)}])&=n\\
	I(T[S^3/\Gamma_{E_7},\su(2),\Z_2^{(\delta)}])&=4\,.
	\ea
	\ee
\item The remaining case is $S^3/\Gamma_{D_{n\text{ even}}}$ with $r=2$. As discussed in section \ref{sec:1-FormGeneral} there are three choices of non-anomalous 1-form symmetries
\be
\delta_1=(0,1,1)\,, \quad \delta_2=(1,0,1)\,, \quad \delta_3=(1,1,0)\,,
\ee
where we take the first cohomology to be generated by $\delta_1$ and $\delta_2$.\footnote{Note that we need to specify this choice because the charge operators act in a distinct way.}  We note the quiver has a symmetry which means that we can exchange the two $\Z_2$s, \ie,
\be
v_{\epsilon^A_1\epsilon^B_1\epsilon^A_2\epsilon^B_2}=v_{\epsilon^A_2\epsilon^B_2\epsilon^A_1\epsilon^B_1}\,.
\ee
Together with modular invariance we only need to specify the dimensions in table \ref{tab:dimVr=2}.
\begin{table}
\centering
\begin{tabular}{c||c||c|c|c|c|c|c|c}
$S^3/\Gamma_{ADE}$, $r=2$ & $\dim V$ & $v_{++++}$ & $v_{+++-}$ & $v_{+-+-}$ & $v_{+--+}$ & $v_{++--}$ & $v_{+---}$ & $v_{----}$ \\
\hline
\hline
$D_{n}$, $n=0\mod 4$ & $n+7$ & $\frac{n}{4}+1$ & 1 & $\frac{n}{4}$ & 0 & 1 & 0 & $\frac{n}{4}$\\
\hline
$D_{n}$, $n=2\mod 4$ & $n+7$ & $\frac{n+2}{4}$ & 1 & $\frac{n+2}{4}$ & 1 & 0 & 0 & $\frac{n-2}{4}$
\end{tabular}
\caption{Dimensions of the eigenspaces $V_{\epsilon^A_1 \epsilon^B_1\epsilon^A_2 \epsilon^B_2}$ for $\Upsilon=\Z_2^{(1)}\times \Z_2^{(2)}$.\label{tab:dimVr=2}}
\end{table}

To gauge, say, $\Z_2^{(1)}$ we need to project onto the subspace with $\epsilon^{A,B}_1=+1$, \ie,
\be
I_{\text{untwisted}}(T[S^3/\Gamma_{D_{n\text{ even}}},\su(2),\Z_2^{(1,2)}])=\left\{ \begin{array}{c c}
\frac{n}{4}+4 & n=0 \mod 4\\
\frac{n+2}{4}+2 & n=2 \mod 4
\end{array}
\right.\,.
\ee
The twisted sectors can be counted similarly, and contribute $3$ states in both cases, leading to
\be
I(T[S^3/\Gamma_{D_{n\text{ even}}},\su(2),\Z_2^{(1,2)}])=\left\{ \begin{array}{c c}
	\frac{n}{4}+7 & n=0 \mod 4\\
	\frac{n+2}{4}+5 & n=2 \mod 4
\end{array}
\right.\,.
\ee
On the other hand, to gauge $\Z_2^{(3)}$, generated by $\delta_3=\delta_1+\delta_2$, we project onto $\epsilon^A_1\epsilon^A_2=+1=\epsilon^B_1\epsilon^B_2$ which yields
\be
I_{\text{untwisted}}(T[S^3/\Gamma_{D_{n\text{ even}}},\su(2),\Z_2^{(3)}])=n+1\,.
\ee
We also find $n-1$ twisted sector states in this case, and so
\be I(T[S^3/\Gamma_{D_{n\text{ even}}},\su(2),\Z_2^{(3)}])=2n\,.
\ee

Finally, we can gauge the full $\Z_2 \times \Z_2$ 1-form symmetry, projecting onto $V_{++++}$, giving
\be
I_{\text{untwisted}}(T[S^3/\Gamma_{D_{n\text{ even}}},\su(2),\Z_2\times \Z_2])= \begin{cases} \frac{n}{4} + 1 & n =0 \; (\text{mod} \; 4), \\
	\frac{n+2}{4}  & n =2 \; (\text{mod} \; 4). \end{cases}
\ee
Including also the twisted sectors, we find in  both cases
\be
I(T[S^3/\Gamma_{D_{n\text{ even}}},\su(2),\Z_2\times \Z_2])= \frac{n}{2} + 3 \;. 
	\ee

\end{enumerate}


\section{Flat Connections on $M_3$ and  the 3d-3d Correspondence}
\label{sec:3d3dcorr}
In this section we use the results for the refined Witten index of the previous section to test the 3d-3d correspondence \cite{Dimofte:2011ju}.  This correspondence predicts that the Witten index of the theory $T_{\cN=2}[M_3,\g]$ counts the flat $\g^\C$ connections on the 3-manifold $M_3$, which has been extensively studied in the literature \cite{Chung:2014qpa,Pei:2015jsa,Gukov:2016gkn,Gukov:2017kmk,Cheng:2018vpl,Benini:2019dyp}.  However, a more careful consideration of the global structure of the gauge group leads to a refinement of this statement, which we derive below.

The Witten index, or $T^3$ partition function, counts the  number of BPS ground states in the Hilbert space of the $T[M_3,\g]$ theory on $T^2 \times \R$.  This space is equivalently the Hilbert space of the 6d theory of type $\g$ on $M_3 \times T^2 \times \R$.  Compactifying in the opposite order, we see it is also the Hilbert space of a 4d $\cN=4$ SYM theory with Lie algebra $\g$ on $M_3 \times \R$, where we take the GL topological twist along $M_3$ \cite{Gukov:2007ck,Gadde:2013sca}.  Now, the vacua for the $\cN=4$ theory with gauge group $G$ on this space can be identified with flat $G^\C$ connections on $M_3$ \cite{Gukov:2007ck}, and so one expects a relation between these flat connections and the $T^2$ vacua of the $T[M_3,\g]$ theory.  However, as we have seen above, the global form of the gauge group of the SYM theory depends on the choice of polarization of the 6d theory, and this will determine the precise observable in $T[M_3,\g]$ we must consider.

In the case of self-dual $G$, the Witten index of $T[M_3,G]$ simply counts the number of flat $G^\C$ connections on $M_3$.  However, in the general case, we find the following correspondence.  Recall that the $T^2$ Hilbert space of the $T[M_3,\g]$ theory is acted on by an anomaly-free 1-form symmetry with group $\Upsilon=H^1(M_3,\ZtG)$.  Then we may refine the Hilbert space into eigenspaces of the two 1-form operators, $U_A^\gamma$ and $U_B^\gamma$, wrapping the two cycles of the torus
\be \label{DecompV}
V \to \bigoplus_{\chi_{A,B}\in H^2(M_3,\ZtG)} V_{\chi_A,\chi_B}\,, \qquad v_{\chi_A,\chi_B} \equiv \text{dim} \; V_{\chi_A,\chi_B} \;.
\ee

On the other hand, we consider flat $\tG^\C$ connections on $M_3$.  More precisely, in addition to ordinary $\tG^\C$ connections, we may consider connections which are not strictly $\tG^\C$ connections, but rather connections for the quotient group, $\tG^\C/\ZtG$.  These may be classified according to their second Stiefel-Whitney class, $\omega \in H^2(M_3,\ZtG)$, which measures the obstruction to lifting these to well-defined $\tG^\C$ connections.  In addition, the connections are acted on by large gauge transformations (LGTs), which are determined by elements of $H^1(M_3,\ZtG)$.  We may refine the flat connections into eigenspaces of these LGTs,\footnote{More formally, we may consider the vector space with basis given by the flat connections, on which the LGTs act linearly, and refine this space into eigenspaces of this action.} which are labeled by $\chi_{\text{LGT}} \in H^2(M_3,\ZtG)$.  Then we claim\footnote{More precisely, this formula holds as stated when the moduli space of flat connections is  zero-dimensional, which is the case for the homology spheres we consider in this section.  In general, we expect a generalization in terms of a suitably defined Euler characteristic on the moduli space of flat connections on $M_3$.}
\be
\label{3d3dflat}  \# \{ \text{flat $\tG^\C/\ZtG$ connections $A$ with $w_2(A) = \omega$ and $\chi_{\text{LGT}}$} \} = v_{\omega,\chi_{\text{LGT}}} \;.
\ee 
We will derive this formula by reduction from the 6d $\cN=(2,0)$ theory, and verify it in explicit examples below. One interesting feature of this formula is that modular invariance of the torus exchanges the two arguments on the RHS, and so we expect a symmetry under exchange of these arguments.  This implies the geometric quantity on the LHS also exhibits this symmetry under exchange of $\omega$ and $\chi_{\text{LGT}}$, which is not obvious, but which we will verify in examples.  This can also be understood  as a consequence of $S$-duality of the 4d $\cN=4$ SYM theory.

\subsection{Self-dual Case}

Let us first consider the choice of polarization of the 6d theory associated to a self-dual group $G$.  As described in section \ref{sec:selfdualtm3}, this gives rise to an ordinary (rather than relative) QFT in six dimensions.  In this case, the compactification to 3d gives the $T[M_3,G]$ theory, while the compactification on $T^2$ to 4d gives the $\cN=4$ SYM theory with gauge group $G$.  Thus we expect that the Witten index of this theory counts the flat $G^\C$ connections on $M_3$.  This correspondence in the case of $G$ self-dual has been observed and checked in various examples, \eg, in \cite{Gadde:2013sca,Chung:2014qpa,Pei:2015jsa}.

First, we observe that the in case $G=U(1)$ on a general rational homology sphere $M_3$, the number of vacua of $T[M_3,U(1)]$ is the order of $H^1(M_3,\Z)$.  This is the same as the number of $U(1)^\C= \C^\times$ flat connections on $M_3$, as also mentioned  in section \ref{sec:Abelian}, demonstrating the formula in this simple example.

Next consider the case of $G=U(N)$, so that $G^\C = GL(N,\C)$.  The theory $T[L(k,1),U(N)]$ is the 3d $\cN=2$ $U(N)$ theory with level $k$ Chern-Simons term and an adjoint chiral multiplet.   We saw above in \eqref{IuN} that the number of vacua is
\be I(T[L(k, 1), U(N)])= \binom{N+k-1}{N}\,. \ee
On the other hand, as shown in appendix \ref{app:flatGLNCconnections}, this is also the number of flat $U(N)^\C = GL(N,C)$ flat connections on $L(k,1)$.

For more general Seifert $M_3$, we may use the quiver Lagrangians and computations of the previous section in the case of $G=U(2)$. In section \ref{sec:GeneralQuivers}, we found
\be\ba
 \ds I(T[S^3/\Gamma_{D_n},U(2)]) &=n+7 \cr 
 \ds I(T[S^3/\Gamma_{E_m},U(2)])& =3(9-m)\;.
\ea\ee
In appendix \ref{app:flatGLNCconnections} we find the same result for the number of flat $GL(2,\C)$ connections on these spaces.

\subsection{General Case}

Now let us consider the case where $G$ is not necessarily self-dual.  For concreteness, we will focus on $\g=\frak{su}(N)$, but similar arguments apply in general.  Consider the 6d $A_{N-1}$ theory on $M_6 = M_3 \times T^2 \times S^1_t$, where we have singled out one of the circles of $T^3$ as the ``time'' circle.  The partition vector of the 6d theory on this space depends first of all on a choice of polarization,
\be \Lambda \subset H^3(M_6,\Z_N) \;. \ee
Associated to the two compactification orders on this geometry, as discussed in  the introduction to this section, there are two natural polarizations, which we consider in turn.

\subsubsection*{$\mathbf{4d}$ SYM on $\mathbf{M_3 \times S^1_t}$}

We first take a choice which is natural in the context of the reduction to 4d SYM.  We can write (suppressing the coefficients, which are all $\Z_N$)
\be H^3(M_6) \cong H^1(M_3 \times S^1_t)^{AB} \oplus H^2(M_3 \times S^1_t)^A \oplus H^2(M_3 \times S^1_t)^B \oplus H^3(M_3 \times S^1_t) \;, \ee
where the superscripts denote that we take the cup product with the $A$ and/or $B$ cycles of the $T^2$.  Then we take the polarization
\be \label{4dpol} \Lambda_{4d} =  H^1(M_3 \times S^1_t)^{AB} \oplus H^2(M_3 \times S^1_t)^A \;.\ee
As discussed in \cite{Tachikawa:2013hya}, this choice corresponds to considering the $\cN=4$ $SU(N)$ SYM theory on $M_3 \times S^1_t$.  The first factor corresponds to a refinement by a $\Z_N$ $0$-form symmetry of this theory, which we will ignore in what follows, and the second by its 1-form $\Z_N$ electric symmetry.  Specifically, we have, as in \eqref{elecform}
\be
\CZ^{M_6}_\lambda = \left\{\text{contribution from $SU(N)/\Z_N$ bundles $P$ over $M_3 \times S^1$ with $w_2(P)=\lambda \in H^2(M_3 \times S^1_t)$} \right\} \;.\ee
Here the contribution $\CZ^{M_6}_0$ is the undeformed partition function of the $SU(N)$ SYM theory on $M_3 \times S^1_t$.  Then since the moduli space of the SYM theory with this twist is given by flat connections on $M_3$, we expect that this partition function simply counts these flat connections
\be \label{flatPSLconnectionslambda0} \CZ^{M_6}_0 =\# \left\{ \text{flat $SL(N,\C)$ connections $A$ on $M_3$} \right\}  \;, \ee
modulo $SL(N,\C)$ gauge transformations.

Now let us consider the effect of specifying non-zero $\lambda \in H^2(M_3 \times S^1_t)$.  First, if we take $\lambda$ to be contained in $M_3$, then we expect the flat connections on $M_3$ to not be genuine $SL(N,\C)$ connections, but rather to be a $PSL(N,\C)$ connection with second Stiefel-Whitney class equal to $\lambda$.   Thus we find
\be \label{flatPSLconnectionslambdaM3} \CZ^{M_6}_\lambda =\# \left\{ \text{flat $PSL'(N,\C)$ connections $A$ on $M_3$ with $w_2(A)= \lambda$} \right\} , \;\;\;\; \lambda \in H^2(M_3) \;. \ee
Here we must be careful with how we treat large gauge transformations.  Namely, although these connections are $PSL(N,\C)$ connections, we consider them only up to gauge transformations in $SL(N,\C)$, since the 4d gauge group is $SU(N)$.  We denote this by writing $PSL'(N,\C)$, with a prime, which indicates we do not divide by large gauge transformations.

Finally, suppose $\lambda$ has both a component, $\omega$ in $H^2(M_3)$, as well as a component with one leg along $H^1(S^1_t)$  and one given by $\gamma \in H^1(M_3)$.  To create a connection with such a Stiefel-Whitney class, we can proceed as follows: we consider the path-integral on $M_3 \times [0,1]$, and then, when gluing the boundaries to form $M_3 \times S^1_t$, we also perform a large gauge transformation in $SU(N)/\Z_N$ associated to the element $\gamma$.  Note that this is a non-trivial operation, since the gauge group is $SU(N)$, and so such large gauge transformations act non-trivially. When counting the flat connections, they are now weighted by their transformation under this large gauge transformation.  Thus we find
\bea \label{flatPSLconnectionslambdagen} \CZ^{M_6}_{\omega,\gamma} & =\# \left\{ \text{flat $PSL'(N,\C)$ connections $A$ on $M_3$ with $w_2(A)= \omega$, weighted by action of $\gamma$} \right\} , \nn \\
&  \omega \in H^2(M_3)\ ,~~ \gamma  \in H^1(M_3) \ . \eea

Suppose we had instead taken the $B$-polarization, corresponding to the $SU(N)/\Z_N$ theory.  In that case we would be taking a sum over all $PSL(N,\C)$ connections weighted according to their Stiefel-Whitney class.  \Eg, for $\widehat{\lambda}$ with a leg along $S^1$ and one along $\gamma \in H^1(M_3)$, we have
\bea \label{flatPSLconnectionslambdaM3Bpol} {\widehat{\CZ}^{M_6}}_\gamma &=\ds \sum_{\omega \in H^2(M_3)} e^{\frac{2 \pi i}{N} \gamma \cup \omega} ~\# \left\{ \text{flat $PSL(N,\C)$ connections $A$ on $M_3$ with $w_2(A)= \omega$} \right\}\ ,\nn \\ & \gamma \in H^1(M_3,\Z_N)\ , \eea
where in this case we do divide by all $PSL(N,\C)$ large gauge transformations.  Then $S$-duality of the $\cN=4$ SYM theory implies this should be equal to the corresponding quantity computed in the $A$-polarization.  We will implicitly verify this statement in examples anon, when we compare to the 3d field theory, where it follows from modular invariance.

\subsubsection*{$\mathbf{T[M_3]}$ on $\mathbf{T^2 \times S^1_t}$}

Next we consider the polarization more natural for studying the theory $T[M_3]$.  Let us first further decompose the cohomology of $M_6$ as
\bea  
H^3(M_6) \; \cong &
H^3(M_3) \oplus H^2(M_3)^A \oplus H^2(M_3)^B \oplus H^2(M_3)^t \nn\\
& \oplus\; H^1(M_3)^{AB} \oplus H^1(M_3)^{At} \oplus H^1(M_3)^{Bt} \oplus H^0(M_3)^{ABt}\ . \eea
Then the polarization which gives the theory $T[M_3,\frak{su}(N)]$ is
\be \label{3dlambda} \Lambda_{3d} = H^0(M_3)^{ABt} \oplus H^1(M_3)^{Bt} \oplus H^1(M_3)^{At} \oplus H^1(M_3)^{AB}\ . \ee
The partition function labeled by a choice of $\lambda \in \Lambda_{3d}$ is just the expectation value of the 1-form charge operators along the corresponding cycles, \ie, writing $\lambda =(0,\alpha,\beta,\tau)$ for $\alpha,\beta,\tau \in H^1(M_3)$
\be \CZ^{M_6}_\lambda = \CZ^{T[M_3]}_{T^3}[\alpha,\beta,\tau] \equiv \langle U_A^\alpha \; U_B^\beta \; U_t^\tau \rangle_{T^3}\ . \ee
Here we recall that a charge operator lying along the time direction corresponds to taking the trace in a twisted sector of the Hilbert space.  Thus we may write this in terms of the decomposition in \eqref{htwistedsector} as
\be \label{ZTM3decomp} \CZ^{T[M_3]}_{T^3}[\alpha,\beta,\tau] =  \sum_{\chi_A,\chi_B \in H^2(M_3)} \chi_A(\alpha) \chi_B(\beta) \; \text{dim} \; V^\tau_{\chi_A,\chi_B}\ , \;\;\;\; \alpha,\beta,\tau \in H^1(M_3)\ .\ee

\subsubsection*{Comparison}

Now to compare these, let us first rewrite the 4d polarization \eqref{4dpol} as
\be \Lambda_{4d} =   H^0(M_3)^{ABt} \oplus H^1(M_3)^{AB}  \oplus H^1(M_3)^{At} \oplus  H^2(M_3)^A\; .\ee
Comparing to \eqref{3dlambda}, we see they differ in the factor $H^2(M_3)^A$, which  appears here, and the dual factor, $H^1(M_3)^{Bt}$, which appears in \eqref{3dlambda}.  Thus to relate the 3d result to the 4d SYM partition function, and so to the flat connection  counting, we should perform a Fourier transform over the variable $\alpha$ in \eqref{ZTM3decomp}.  For example, taking $\lambda \in \Lambda_{4d}$ to have component $\omega \in H^2(M_3)^A$ and $\beta \in H^1(M_3)^{At}$, we find the 3d observable
\be \frac{1}{|H^1(M_3)|} \sum_{\alpha \in H^1(M_3)} e^{2 \pi i \omega \alpha} \CZ^{T[M_3]}_{T^3}[\alpha,\beta,0] =  \sum_{\chi_B \in H^2(M_3)} \ \chi_B(\beta) v_{\omega,\chi_B}\,.
\ee
On the other hand, from \eqref{flatPSLconnectionslambdagen}, we see this is equal to
\be \CZ^{M_6}_{\omega,\beta} =\# \left\{ \text{flat $PSL'(N,\C)$ connections $A$ on $M_3$ with $w_2(A)= \omega$, weighted by action of $\beta$ } \right\} . \ee
Comparing these, we see we may identify the action of the large gauge transformations on the flat connections with that of the $U_B$ operator on the vacua of $T[M_3]$.  In particular, we may decompose both into their respective eigenspaces, which must agree, and arrive at the relation \eqref{3d3dflat}.  Let us now test this relation in some examples.

\subsection{Example: $T[L(p,1),\frak{su}(N)]$}

\label{sec:PSLLens}

Let us consider the theory associated to the lens space, $L(p,1)$, which is a theory with $\frak{su}(N)$ Lie algebra and level $p$ CS term and an adjoint chiral multiplet.  We focus on the case of $N$ prime for simplicity.  Then there are two cases to consider.

\subsubsection*{$\mathbf{GCD}(N,p)=1$}

Let us first consider the theory $\widehat{T}[\Omega_{[p]},SU(N)]$, which is the $SU(N)_p$ theory with an adjoint chiral multiplet.  As argued in appendix \ref{app:ComputationsU(N)}, this has 
\be
I(\widehat{T}[\Omega_{[p]},SU(N)]) = \binom{N+p-1}{N-1}\,.
\ee
Now, this theory has an anomalous $\Z_N$ 1-form symmetry, and so, as discussed in section \ref{sec:tm3decoupling}, we conjecture that it contains a decoupled TQFT on which the 1-form symmetry acts exclusively.  This TQFT has $N$ states, and after decoupling this theory as in \eqref{decoupconj}, we find
\be \label{nvaclpqgcd1}
I(T[L(p,1),\frak{su}(N)]) = \frac{1}{N} \binom{N+p-1}{N-1}\,.
\ee
Note that in this case, since $H^2(L(p,1),\Z_N)=1$, the formula \eqref{3d3dflat} predicts that the Witten index of this theory should directly count the flat $PSL(N,\C)$ connections on $L(p,1)$, and in this case, these are the same as $SL(N,\C)$ connections.  The latter are given by representations
\be
\phi: \quad \pi_1(L(p,1)) \cong \Z_p \ \rightarrow\  SL(N,\C)\,,
\ee
modulo  $SL(N,\C)$ transformation.
Without loss of generality, we may take these representations to be diagonal, with generator given by
\be
\text{diag}\left(e^{\frac{2 \pi i  m_1}{p}},e^{\frac{2 \pi i  m_2}{p}},\cdots, e^{\frac{2 \pi i  m_N}{p}}\right) \,, \qquad \sum_{i=1}^N m_i=0\mod p\,,
\ee
considered up to permutations.  An arbitrary choice of $\{m_i\}$ may have a non-zero sum modulo $p$, however shifting $m_i \rightarrow  m_i+1$ shifts this sum by $N$, and so precisely one element in the orbit under  this $\Z_p$ action will have sum zero. Thus we have
\be \# \{ \text{flat $SL(N,\C)$ connections on $L(p,1)$} \} \; = \; \frac{1}{p} \binom{N+p-1}{N} \,,
\ee
which equals \eqref{nvaclpqgcd1}.  Note that this agreement required the decoupling of the TQFT, giving some further evidence for this conjecture.

\subsubsection*{$\mathbf{GCD}(N,p)=N$}

Now we find $T[L(p,1),\frak{su}(N)]=\widehat{T}[\Omega_{[p]},SU(N)]$ is the $SU(N)_p$ theory.  This has an anomaly-free 1-form symmetry, and so the Hilbert space decomposes as
\be
V= \bigoplus_{\chi_A,\chi_B \in \Z_N} V_{\chi_A\chi_B} \,,
 \ee
where $\chi_A,\chi_B$ run over the characters of $\Z_N$.  This is a generalization of \eqref{DecompositionV} with the prescription that
\be
\epsilon_{A,B}=\exp\left(\frac{2\pi i \chi_{A,B}}{N}\right)\,.
\ee
Let us consider $N=2$ for simplicity. Then the dimensions of the $V_{\chi_A\chi_B}$ are summarized in table \ref{tab:dimVr=1}, which we reproduce here for convenience:
\be \label{pred}
v_{\chi_A,\chi_B}(T[L(p,1),\frak{su}(2)]) = \begin{cases} \frac{p}{4}+1 & p = 0 \;(\text{mod}\;4) \; \text{and} \; (\chi_A,\chi_B) =(0,0) \\
	\frac{p}{4} & p = 0 \;(\text{mod}\;4) \; \text{and} \; (\chi_A,\chi_B) \in \{(0,1),(1,0),(1,1) \} \\
	\frac{p+2}{4} & p = 2 \;(\text{mod}\;4) \; \text{and} \; (\chi_A,\chi_B) \in \{(0,0),(0,1),(1,0) \} \\
	\frac{p-2}{4} & p = 2 \;(\text{mod}\;4) \; \text{and} \; (\chi_A,\chi_B) =(1,1) \end{cases}
\ee
Now let us compare to the flat connections on $L(p,1)$, using the formula \eqref{3d3dflat}.
Let us check this by counting the flat $PSL'(2,\C)$ connections explicitly.  These are given by homomorphisms from $\pi_1(L(p,1))\cong \Z_p$ into $PSL(2,\C)$, but considered only up to conjugation by $SL(2,\C)$.  Without loss we may write the image of the generator of $\Z_p$ as $\text{diag}(\alpha,\alpha^{-1})$, (written as an $SL(2,\C)$ matrix), where
\be
\alpha^p = \pm 1, \;\;\;\; \alpha \sim \alpha^{-1}  \,,
\ee
where we identify by Weyl symmetry .  As described in appendix \ref{app:flatSL2Cconnections}, we can classify the connections according to their class in $H^2(L(p,1),\Z_2) = \Z_2$, which in in this case is determined by $\alpha^p \in \{\pm 1\}$.  These solutions are also acted on by LGTs, which act as multiplication  by the $\Z_2$ center of $SL(2,\C)$, and take
\be \alpha \rightarrow - \alpha\,. \ee
All the solutions therefor come in pairs, $\pm \alpha$, except for the solution with $\alpha=i$, which is Weyl-equivalent to $-\alpha$.

Now if we project onto the LGT-invariant solutions, which is the same as considering $PSL(2,\C)$ connections, we find
\be
\# \{ \text{flat $PSL(2,\C)$ connections with $w_2= \omega$} \} = \begin{cases} 
	\frac{p}{4}+1 & p = 0 \mod 4 \,,\ \omega = 0 \\
	\frac{p}{4} & p = 0 \mod 4 \,,\ \omega = 1 \\
	\frac{p+2}{4} & p = 2 \mod 4 \,,\ \omega = 0 \\
	\frac{p+2}{4} & p = 2 \mod 4 \,,\ \omega = 1 \,.
\end{cases}
	\ee
From \eqref{3d3dflat}, map to the sectors in \eqref{pred} with $\chi_B=0$, along with $\chi_A=\omega$, and one verifies that these precisely match.  On  the other hand, for each pair $(\alpha,-\alpha)$ with $\alpha \neq i$, we can also construct an antisymmetric formal linear combination of solutions with $\chi_{LGT}=1$.  Thus in this sector find a similar counting, but now with the contribution of $\alpha=i$ removed, namely
\be
\# \{ \text{flat $PSL'(2,\C)$ connections with $w_2= \omega$, $\chi_{LGT}=1$} \} = \begin{cases} 
	\frac{p}{4} & p = 0 \mod 4 \,,\ \omega = 0 \\
	\frac{p}{4} & p = 0 \mod 4 \,,\ \omega = 1 \\
	\frac{p+2}{4} & p = 2 \mod 4 \,,\ \omega = 0 \\
	\frac{p-2}{4} & p = 2 \mod 4 \,,\ \omega = 1 \,.
\end{cases}
\ee
One can check this matches the remaining entries in \eqref{pred}, with $\chi_B=1$.

One interesting feature of this calculation is that modular invariance of the field  theory implies  a relation between the Stiefel-Whitney class and LGT behavior of the flat $PSL'(2,\C)$ connections.  We do not know of a simple derivation of this fact.

\subsection{Example: Gauged Trinion}

As a final example, we consider the theory associated to the $S^3/{\Gamma_{DE}}$ discussed in section \ref{sec:VacuaGaugedTrinion}. Note that the analysis for $L(p,q)$ is independent of $q$ and is thus covered by section \ref{sec:PSLLens}. We again focus on $N=2$, where the results for the vacuum counting, refined by 1-form charges, are summarized in tables \ref{tab:dimVr=1} and \ref{tab:dimVr=2}.

Let us compare this to the flat connection counting, which is reviewed in appendix \ref{app:flatSL2Cconnections}. The Seifert data of a Seifert manifold with three exceptional fibers fixes its fundamental group \eqref{pi1(M3)}. In the general case the number of $PSL(2,\C)$ representations is given by \eqref{PSLConnGeneral} and their Stiefel-Whitney classes can be computed explicitly, see appendix \ref{app:flatSL2Cconnections}. We find for the flat $PSL(2,\C)$ connections with $w_2=\omega\in H^2(M_3,\Z_2)$
\be
\ba
D_n \,, \ n=0\mod 4 :\quad &\begin{cases} \frac{n}{4}+1 & \omega=(0,0) \\ 1 & \omega=(1,0) \\ 1 & \omega=(0,1) \\ \frac{n}{4} & \omega=(1,1) \end{cases}\,,  \quad D_n \,, \ n=2\mod 4  :\quad &&\begin{cases} \frac{n+2}{4} & \omega=(0,0) \\ 1 & \omega=(1,0) \\ 1 & \omega=(0,1) \\ \frac{n+2}{4} & \omega=(1,1) \end{cases}\\
D_{n}\,, \ n \text{ odd} :\quad &\begin{cases} \frac{n+1}{2} & \omega=0 \\ 1 & \omega=1  \end{cases}\,,  \ \, \qquad\qquad\qquad\qquad\qquad E_7 :\quad  &&\begin{cases} 2 & \omega=0 \\ 2 & \omega=1 \\  \end{cases}\,.
\ea
\ee
For $E_6$ and $E_8$ there is no distinction of the connections in terms of SW classes.  

To compare this to the field theory, we recall that the $PSL(2,\C)$ connections correspond to the $PSL'(2,\C)$ connections with $\chi_{LGT}=0$.  Then we may compare the entries above to the relevant entries in tables \ref{tab:dimVr=1} and \ref{tab:dimVr=2} with $\chi_B=0$.  For example, for $D_n$ with  $n=0 \; (\text{mod} \; 4)$, we see $v_{++++}=\frac{n}{4}$, agreeing with the $\omega=(0,0)$ entry above, and one may similarly compare the remaining entries.

\begin{table}
	\centering
	\begin{tabular}{|c||c|c|c|c|}
		\hline
		\backslashbox{$\chi_{LGT}$}{$\omega$} & $(0,0)$ & $(1,0)$ & $(0,1)$ & $(1,1)$ \\
		\hline
		$(0,0)$ & \textcolor{red}{$\frac{n}{4}$} + \textcolor{blue}{1} & \textcolor{darkgreen}{1} & \textcolor{darkgreen}{1} & \textcolor{red}{$\frac{n-4}{4}$}  + \textcolor{darkgreen}{1} \\
		$(1,0)$ & \textcolor{blue}{1} & \textcolor{darkgreen}{1} & 0 & 0 \\
		$(0,1)$ & \textcolor{blue}{1} & 0 & \textcolor{darkgreen}{1} & 0 \\
		$(1,1)$ &  \textcolor{red}{$\frac{n-4}{4}$}  +  \textcolor{blue}{1} & 0 & 0 & \textcolor{red}{$\frac{n-4}{4}$} + \textcolor{darkgreen}{1} \\
		\hline
	\end{tabular}
	
	\begin{centering}
		\smallskip
		$n=0 \; (\text{mod} \; 4)$ 
		
	\end{centering}
	
	\
	
	\begin{tabular}{|c||c|c|c|c|}
		\hline
		\backslashbox{$\chi_{LGT}$}{$\omega$} & $(0,0)$ & $(1,0)$ & $(0,1)$ & $(1,1)$ \\
		\hline
		$(0,0)$ & \textcolor{red}{$\frac{n-2}{4}$} + \textcolor{blue}{1} & \textcolor{darkgreen}{1} & \textcolor{darkgreen}{1} &  \textcolor{red}{$\frac{n-2}{4}$} +\textcolor{darkgreen}{1} \\
		$(1,0)$ & \textcolor{blue}{1} & 0 & \textcolor{darkgreen}{1} & 0 \\
		$(0,1)$ & \textcolor{blue}{1} & \textcolor{darkgreen}{1} &  0 &  0 \\
		$(1,1)$ &  \textcolor{red}{$\frac{n-2}{4}$} + \textcolor{blue}{1} & 0 & 0 & \textcolor{red}{$\frac{n-6}{4}$} + \textcolor{darkgreen}{1} \\
		\hline
	\end{tabular}
	\smallskip
	\begin{centering}
		
		$n=2 \; (\text{mod} \; 4)$ 
		
	\end{centering}
	
	\
	
	\caption{Counting of flat $PSL'(2,\C)$ connections on $S^3/\Gamma_{D_n}$ refined by Stiefel-Whitney class, $\omega$, and $\chi_{LGT}$, for $n=0\;(\text{mod} \;4)$ on the top and $n=2\;(\text{mod} \;4)$ below.  Here we show the contribution from the trivial connection in blue, the other reducible connections in green, and the irreducible connections in red.\label{tab:Dnflatconnections}}
\end{table}
We can also check the flat connections with non-trivial $\chi_{LGT}$.  Let us analyze the case of $S^3/\Gamma_{D_{n}}$ for $n$ even.  Then using the explicit form for the $PSL(2,\C)$ connections in section \ref{app:flatSL2Cconnections}, one can classify them under Stiefely-Whitney class, $\omega$, and $\chi_{LGT}$.  Let us describe this explicitly in the case $n=4$.  Here there is one trivial connection, three other irreducible connections, and one irreducible connection.  The trivial connection is acted on freely by large gauge transformations (LGTs), and so contributes states with $\omega=(0,0)$ and all four values of $\chi_{LGT}$.  Each of the other reducible connections is fixed by by one LGT.  \Eg, the connection with $\omega=(1,0)$ is fixed by the LGT corresponding to $(0,1)$, and so this contributes states with $\chi_{LGT}=(0,0)$ and $(1,0)$, and similarly for the others.  Finally, the irreducible connection is fixed by all LGTs, and so only contributes a state with $\chi_{LGT}=1$.  We summarize the results for general even  $n$ in table \ref{tab:Dnflatconnections}.  One can verify this precisely matches with the refinement of the supersymmetric vacua of $T[S^3/\Gamma_{D_n},\frak{su}(2)]$ in table \ref{tab:dimVr=2}.  Moreover, one can see explicitly the symmetry under reflection in the diagonal, which is implied by modular invariance on the field  theory side.


\section{3d $\mathcal{N}=1$}
\label{sec:3dN=1}

In this final section we extend our discussion to 3d $\mathcal{N}=1$ theories. The analysis of higher-form symmetries will generalize to this case naturally. We focus on a special class of $\CN=1$ theories which have an $\CN=2$ enhancement point in the space of mass parameters. We study the action of higher-form symmetries in the space of Bethe vacua on these supersymmetry enhancement loci.
In particular, starting with a known $\mathcal{N}=2$ enhancement point of the $T_{\mathcal{N}=1}[L(p,1)]$ theory \cite{Acharya:2001dz}, we propose a quiver description for the $T_{\mathcal{N}=1}[L(p,q)]$ theory. We show that this description reproduces the expected number of vacua for $\mathfrak{g}=\mathfrak{u}(2)$ and $\su(2)$ at the supersymmetry enhancement point.

\subsection{$T_{\mathcal{N}=1}[M_3]$ and Boundary Conditions of $\mathcal{N}=1$ Class $\mathcal{S}$}

A natural generalization of the 3d $\mathcal{N}=2$ theories associated to $M_3$ is to consider instead a topological twist that preserves $\mathcal{N}=1$ supersymmetry \cite{Eckhard:2018raj}. From a 6d $(2,0)$ point of view, this is obtained by starting with the decomposition $Sp(4)_R\to SU(2)_\ell \times SU(2)_r$ and twisting $SU(2)_{\text{twist}}=\diag(SO(3)_M,SU(2)_r)$. The supercharges decompose as 
\be \label{TwistMin}
\ba
SO(1,5)_L\times Sp(4)_R  \quad &\to \quad SO(1,2)_L \times SU(2)_{\text{twist}} \times SU(2)_\ell\\
 (\bold{4},\bold{4}) \quad &\mapsto \quad \underline{(\bold{2},\bold{1},\bold{1})} \oplus (\bold{2},\bold{3},\bold{1})\oplus(\bold{2},\bold{2},\bold{2})\,,
\ea
\ee
preserving 3d $\mathcal{N}=1$. This 3d $\mathcal{N}=1$ twist, and the associated 3d-3d correspondence, was studied in \cite{Eckhard:2018raj} and the resulting theories were referred to as $T_{\mathcal{N}=1} [M_3]$. It has a geometric realization, where $M_3$ is an associative three-cycle with normal bundle
\be\label{NAssoc}
N_{M_3}\cong \mathbb{S}\otimes V\,,
\ee
where $\mathbb{S}$ is the spin bundle of $M_3$ and $V$ is a rank two $SU(2)_\ell$-bundle. 
The 3d theory depends on the choice of this bundle -- in the local $G_2$, the existence of sections of this bundle will depend on the local metric that is induced from the ambient $G_2$. The simplest choice is when 
$V$ is the trivial bundle, which will be referred to as the {\it minimal twist}.

The 3d-3d correspondence for this twist was discussed in \cite{Eckhard:2018raj}, where the BPS equations 
along $M_3$ were determined to be the generalized Seiberg-Witten equations (gSW)
\be\label{gSW}
\ba
0&=\varepsilon_{abc}F^{bc}-
{i \over 2}[\phi_{\alpha\widehat{\alpha}},\phi^{\beta\widehat{\alpha}}](\sigma_a)^\alpha_{\ \beta}\\
0&=(\slashed{\mathcal{D}}\phi)^{\alpha\widehat{\alpha}}\,.
\ea
\ee
Here $\phi^{\alpha\widehat{\alpha}}$ are sections of the normal bundle \eqref{NAssoc}, where $\alpha,\widehat{\alpha}$ are indices labeling the spinor representations. It was conjectured in \cite{Eckhard:2018raj}, that the number of ``abelian" flat (real) $G$-connections reproduces the Witten index, whenever the three-manifold does not admit any twisted harmonic spinors (in the above equations, this corresponds to trivial $\phi^{\alpha \widehat\alpha}$). Here the ``abelian" flat connection\footnote{These are referred to as ``inequivalent" flat connections in \cite{Acharya:2001dz}.  They can be equivalently characterized as the connections whose stabilizer group is abelian.} is defined as a representation $\pi_1(\CM_3)\rightarrow G$ such that all the irreducible representations contribute at most once. We will confirm this for $U(2)$ by counting the Bethe vacua for these theories anon.

The minimal twist can also be thought of as arising from 4d $\mathcal{N}=1$ class $\mathcal{S}$ theories on an interval, \ie, $T_{\mathcal{N}=1}[\Sigma_{g}]$, where $\Sigma_{g}$ is a genus $g$ curve, which now is embedded inside a local Calabi-Yau three-fold  \cite{Bah:2012dg}. This is given by the total space of the rank two bundle
\be
N_{\Sigma_{g}}=L_1 \oplus L_2\,,
\ee
where the $L_i$ are line bundles with $\deg{L_1}+\deg{L_2}=2g-2 $. The BPS equations of these theories are given by a generalized Hitchin system \cite{Xie:2013gma}
\be \label{GenHitchin}
\ba
0&=F_{z\bar{z}}+[\Phi_1,\bar{\Phi}_1]+[\Phi_2,\bar{\Phi}_2]\\
0&=[\Phi_1,\Phi_2]\\
0&=D_z\Phi_i\,,
\ea
\ee
where the $\Phi_i$ are adjoint-valued sections of the $L_i$.

In order to study 3d $\mathcal{N}=1$ theories we can put this 4d $\mathcal{N}=1$ class $\mathcal{S}$ theory on an interval. The most general boundary conditions for $T_{\mathcal{N}=1}[\Sigma]$ have not been studied, however a nice observation is that the BPS equations of  $T_{\mathcal{N}=1}[M_3]$ can be related to those of the Hitchin system \eqref{GenHitchin}. Geometrically, we can view the setup as a consequence of the Heegaard splitting of $M_3$, see \eg \cite{MR1886684,MR1712769}. Each three-manifold can be obtained by gluing two manifolds $H_{\pm}^g$ along its boundaries $\p H_{\pm}^g=\Sigma_g$. The gluing map is given by an element of the mapping class group of $\Sigma_g$.

This also means that the Hitchin system \eqref{GenHitchin} should be related to the generalized SW equations (\ref{gSW}). Indeed, locally, the geometry of $M_3$ is given by $\Sigma_g \times \R$ with coordinates $(z,\bar{z},x_3)$ and the gSW equations become
\be
\ba
0&=F_{z\bar{z}}+[\Phi_1,\bar{\Phi}_1]+[\Phi_2,\bar{\Phi}_2]\\
0&=F_{\bar{z}3}-[\Phi_1,\Phi_2]\\
0&=D_3\bar{\Phi}_1-D_z \Phi_2\\
0&=D_3\bar{\Phi}_2+D_z \Phi_1\,,
\ea
\ee
where we identify $\Phi_1=\phi^{11}=(\phi^{22})^*$ and $\Phi_2=\phi^{12}=-(\phi^{21})^*$. We see that this agrees with the generalized Hitchin system in \cite{Xie:2013gma} after imposing Neumann boundary conditions. Clearly this relation between generalized Hitchin and generalized Seiberg-Witten equations deserves further study.

\subsection{Higher-Form Symmetries for  $T_{\cN=1} [M_3]$ }

The analysis of sections \ref{sec:6dO} and \ref{sec:tm3ghdef}, \ie, the higher-form symmetries of 3d theories obtained by dimensional reduction of the 6d $(2,0)$ theory, carries over directly to the $\mathcal{N}=1$ case, as it does not explicitly depend on the choice of topological twist.  In particular, we expect that for any subgroup,
\be
H  \leq \widehat{\Upsilon} = H^2 (M_3, \ZGtilde) \,,
\ee
we may define a theory
\be
T_{\mathcal{N}=1}[M_3,\g,H]  \,.
\ee
These theories have 1-form symmetry given by $\Upsilon_H$ and a 0-form symmetry $H$, as defined in section \ref{sec:tm3ghdef}.  In particular, the versions of the theory with general $H$ may be obtained from the $H=1$ theory by gauging a suitable subgroup of its 1-form symmetry.

In the case of $M_3$ a graph manifold, to define these theories explicitly, we must first define analogues of the theories $\widehat{T} [\Graph, G]$ associated to a graph, which were defined in section \ref{sec:THatDef}.  These relied on a description of the theories as 3d $\mathcal{N}=2^*$ preserving boundary conditions of 4d $\mathcal{N}=4$ SYM. A first principle derivation of the $\mathcal{N}=1$ versions should exist as boundary conditions on $\mathcal{N}=1$ theories of class $\mathcal{S}$, as discussed in the last subsection, or in  terms of $\frac{1}{4}$ BPS boundary conditions of the $\cN=4$ theory.  Unfortunately, a full classification of such boundary conditions does not exist. However, we expect that a qualitatively similar quiver gauge theory description can be assigned to a graph, $\Graph$, of $M_3$ as in the $\cN=2$ case, with suitable modifications.  For example, we expect that these will involve $\cN=1$ rather than $\cN=2$ Chern-Simons terms, and the coupling of the nodes to the $S$-walls must also be suitably modified.  However, one important point is that the higher-form symmetry structure of this theory should be the same as their $\cN=2$ cousins.  This is because these theories have the same gauge groups and Chern-Simons couplings, and differ only in the couplings of additional adjoint-valued matter fields, which do not affect the higher-form symmetry structure.

It would be interesting to develop a detailed Lagrangian description of these theories for general graph manifolds.  For the remainder of this section, we consider this description in the special case of $M_3=L(p,q)$, where we will argue that a deformation of $T[M_3]$ leads to a theory with $\cN=2$ supersymmetry, which we can study more explicitly.

\subsection{3d $\mathcal{N}=1$ Lens Space Quivers}
\label{sec:N=1Seifert}

Let us consider the theories $T_{\cN=1}[M_3,\g]$ for $M_3$ a lens space.  We first consider the case $M_3=L(k,1)$, which can be associated to the quiver $\Graph_{[k]}$, corresponding to $T^k \in SL (2, \mathbb{Z})$.  Extending our notation in the natural way, we first define the $\widehat{T}_{\mathcal{N}=1}[\Graph_{[k]},G]$ theory.  
Following \cite{Acharya:2001dz,Eckhard:2018raj}, we find that this consists of an $\mathcal{N}=1$ $G$ vector multiplet at Chern-Simons level $k$ and a massless adjoint scalar multiplet.\footnote{There are no fields associated to the zero-sections of the normal bundle as the Lens space has positive scalar curvature \cite{Eckhard:2018raj,MR0156292}.}  Then it is known that the associated 3d $\mathcal{N}=1$ theory  has an $\mathcal{N}=2$ enhancement point in the space of mass parameter. 
Tuning on the scalar mass to
\be
m_{\mathcal{N}=2}=-\frac{kg^2}{4\pi}\,,
\ee
where $g$ is the 3d gauge coupling, yields a pure $\mathcal{N}=2$ $G_k$ CS-theory. The mass deformation from $m=0$ to $m_{\CN=2}$ preserves the number of vacua \cite{Bashmakov:2018wts}. At the enhancement point, we can define and count Bethe vacua of $\widehat{T}_{\mathcal{N}=1}[\Graph_{[k]},G]$, which can be described as an $\cN=2$ quiver consisting of a single gauged node at Chern-Simons level $k$, but  now without an adjoint chiral multiplet.  We will return to this computation below.

\begin{figure}
\centering	
		\includegraphics[width=13cm]{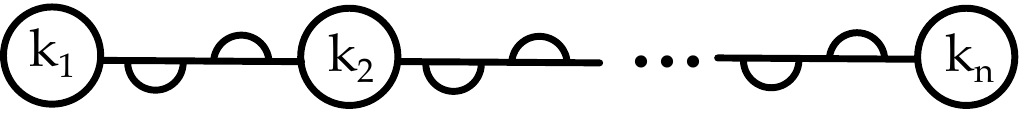}
	\caption{Quiver corresponding to the $\mathcal{N}=2$ enhancement point of $\widehat{T}_{\mathcal{N}=1}[\Graph_{[k_1, \cdots, k_n]},G]$
		\label{fig:tstst}}		
\end{figure}

We conjecture that we can similarly define a more general $\mathcal{N}=2$ quiver which computes the number of vacua of  the $\mathcal{N}=1$ theory for general $[k_1, \cdots, k_n]$, which are the Lens spaces $L(p,q)$. This can be done by taking each $T^{k_i}$ theory, enhancing it to $\mathcal{N}=2$ supersymmetry and coupling them together with the $T(G)$ theory. This quiver description can also be obtained from a procedure similar to section \ref{sec:BounCond}, with the adjoint multiplets $\text{adj}(\CV_\pm)$ removed from all the basic building blocks and gluing rules. The quiver for $\widehat{T}_{\mathcal{N}=1}[\Graph_{[k_1, \cdots, k_n]},G]$ is then given in figure \ref{fig:tstst}. It is obtained from the quiver in figure \ref{fig:tststdgg} by removing the adjoint attached to the left-most node. 
It would be interesting to show that this is indeed the same as the prescription that is obtained from the direct dimensional reduction on $L(p,q)$ with the $\mathcal{N}=1$ twist. This prescription is supported by the vacuum count and the compatibility with the dualities of section \ref{sec:QuiverSymmetries}, which will also apply for these $\cN=2$ quivers. 

The physical theory $T_{\mathcal{N}=1}[M_3, \mathfrak{g}]$ is then obtained by decoupling various topological sectors, precisely as in the earlier discussions for the $\mathcal{N}=2$ case. 

\subsection{Bethe Vacua for 3d $\mathcal{N}=1$}

To check this proposed description of the quivers we now compute the number of vacua.  Let us first consider the case $M_3=L(k,1)$ and $G=U(N)$, where, since $U(N)$ is self-dual, we may identify $T_{\cN=1}[L(k,1),U(N)] = \widehat{T}_{\mathcal{N}=1}[\Graph_{[k]},U(N)]$.   The Bethe equations for this theory are given simply by
\be z {x_i}^k = 1, \;\;\; i=1,...,N \;,\ee
where $z$ is the fugacity for the $U(1)_J$ symmetry.  Accounting  for Weyl symmetry, we find this has
\be I(T_{\cN=1}[L(k,1),U(N)]) = \binom{k}{N} \ee
From section \ref{sec:selfdualtm3} we expect this to be acted on by a $\Z_k$ 1-form symmetry with anomaly coefficient $(N,k)$, as in the $\cN=2$ case, and this is indeed true of this CS theory.  For example, recall from section \ref{sec:unBVsubsec} that the representations of this 1-form symmetry have size $\frac{k}{(N,k)}$, and one verifies that the number of vacua is indeed divisible by this integer.

Next, for $M_3=L(p,q)$, we consider the gauge algebra $\mathfrak{g}=\frak{(s)u}(2)$. As in section \ref{sec:THatCount} we first determine the number of vacua of $\widehat{T}_{\mathcal{N}=1}[\Graph_{[k_1, \cdots, k_n]},SU(2)]$.
The simplest case is the single node $T^k$, where the single Bethe equation is
\be
\Pi(x)=x^{2k}\,.
\ee
The solutions are $\widehat{u}_j=\frac{j}{2k}$, with $j=0,\dots,2k-1$, with the $\Z_2$ Weyl symmetry acting as $\widehat{u}_j\to \widehat{u}_{-j}$. Except for the two degenerate solutions $\widehat{u}_0,\widehat{u}_k$ they appear in Weyl pairs, so
\be
I({\widehat{T}_{\mathcal{N}=1}[\Graph_{[k]},SU(2)]})= k-1\,.
\ee
The anomaly is exactly as for the $\mathcal{N}=2$ case in section \ref{sec:BVSU2SO3}, so 
\be
I(T_{\mathcal{N}=1}[L(k,1),\su(2)])=\begin{cases} k-1 & \text{$k$ even} \\ \frac{k-1}{2}  & \text{$k$ odd} \end{cases}\,.
\ee

If $k$ is even we can gauge the $\Z_2$ 1-form symmetry of this theory. Since the Bethe vacua are known explicitly we can do this by hand. The Hilbert space is spanned by the physical vacua $\ket{\widehat{u}_j}$, $j=1,\dots,k-1$ and the two generators of the 1-form symmetry \eqref{1formgenerators} act as\footnote{The overall sign of $U^A$ is fixed by modular invariance.}
\be
U^A\ket{\widehat{u}_j}= (-1)^{j+1} \ket{\widehat{u}_j}\,,\qquad U^B\ket{\widehat{u}_j}=\ket{\widehat{u}_{k-j}}\,.
\ee
Thus, the common eigenvectors of the $U$ are $\ket{\widehat{u}_{k/2}}$ and $\ket{\widehat{v}_j^\pm}=\frac{1}{\sqrt{2}}\left(\ket{\widehat{u}_j}\pm \ket{\widehat{u}_{k-j}}\right)$, $j=1,\dots,k/2-1$.
The eigenvalues of the $U$ are given by
\be
\left(U^A,U^B\right)\ket{\widehat{u}_{k/2}}=\left((-1)^{\frac{k}{2}+1},+1\right)\ket{\widehat{u}_{k/2}}\,, \qquad \left(U^A,U^B\right)\ket{\widehat{v}^\pm_j}=\left((-1)^{j+1},\pm 1\right)\ket{\widehat{v}^\pm_j}\,.
\ee
From this we easily see that the subspace with eigenvalues $(+1,+1)$ has dimension $\left\lceil \frac{k}{4} \right\rceil$.

We can explicitly check this structure for general linear quivers. This is analogous to the $\mathcal{N}=2$ theories discussed in section \ref{sec:ResultsU(N)}. We find
\be
I({\widehat{T}_{\mathcal{N}=1}[\Graph_{[k_1, \cdots, k_n]},SU(2)]})= 2^{n-1} p-1\,,\\
\ee
from which we can deduce that
\be
\ba
I(T_{\mathcal{N}=1}[L(p,q),SU(2)])&=\begin{cases} p-1 & \text{$p$ even} \\ \frac{p-1}{2}  & \text{$p$ odd} \end{cases}\,,\\
I(T_{\mathcal{N}=1}[L(p,q),U(2)])&={p \choose 2}\,.
\ea
\ee
Furthermore, we can gauge the $\Z_2$ 1-form symmetry if $p$ is even. Doing this explicitly for small values of $n$ and $p$ we find
\be
I_{\text{untwisted}}(T_{\mathcal{N}=1}[L(p,q),\su(2),\Z_2])=\left\lceil \frac{p}{4}\right\rceil\,.
\ee
as for the $\cN=2$ twist, there is also a single twisted sector, leading to
\be
I(T_{\mathcal{N}=1}[L(p,q),\su(2),\Z_2])=\left\lceil \frac{p}{4}\right\rceil + 1 \,.
\ee
The interesting point to note here is that the Bethe vacua counting is -- as expected, although it is not manifest -- independent of $q$. 
It would be interesting to extend this to higher rank and more general, non-linear quivers and to test the $\mathcal{N}=1$ 3d-3d correspondence for these theories.


\subsection*{Acknowledgments}
We thank C.~Closset, T.~Dimofte, D.~Gang, S.~Giacomelli, S.~Gukov, M.~H{\"u}bner, D.~Morrison, S.~Razamat, N.~Seiberg, Y.~Tanaka, and K.~Yonekura for discussions.
We also thank the Simons Collaboration "Special Holonomy in Geometry, Analysis and Physics", where this collaboration was started.
JE, HK, and SSN are supported by the ERC Consolidator Grant number 682608 
``Higgs bundles: Supersymmetric Gauge Theories and Geometry (HIGGSBNDL)'',
SSN is grateful to the CERN Theory Group and KIAS for hospitality.  BW is supported by the Simons Foundation Grant \#488629 (Morrison) and the Simons Collaboration on ``Special Holonomy in Geometry, Analysis and Physics.''


\appendix

\section{Notations and Nomenclature}
\label{app:NotNom}

As there are several notational intricacies in this paper, we summarize these in the following table.

\begin{tabular}{r|l}
Label & Meaning \cr \hline
$G$, $\mathfrak{g}$ & Lie group (not necessarily simply connected) and associated Lie algebra \cr 
$\widetilde{G}$ & Simply-connected Lie group associated to $\mathfrak{g}$\cr 
$\Omega$ & Graph \cr 
$T(G)$ & $S$-duality wall theory of \cite{Gaiotto:2008ak} \cr 
$FT(G)$ & T(G) theory with one flavor symmetry flipped, as defined in \cite{Aprile:2018oau} \cr 
$\widehat{T}[\Omega, G ]$ & Unphysical theory associated to graph manifold $\Omega$. Defined in section \ref{sec:THatDef}. \cr 
$Z_{G}$ & Center of $G$ \cr 
$\Upsilon$ & $H^1(M_3,\ZGtilde)$ \cr 
$\widehat{\Upsilon}$ & $H^2(M_3,\ZGtilde)$  \cr 
$H$ & Subgroup of $\widehat{\Upsilon}$\cr 
$\Upsilon_H$ & $\{ x \in \Upsilon \; | \; \chi(x) = 1, \; \; \forall \chi \in H \} \subset \Upsilon$ \cr 
$T[M_3, \mathfrak{g}, H]$& Theory with $0$-form symmetry $H$, and 1-form symmetry $\Upsilon_H$. \cr 
& Defined in section \ref{sec:Tm3gH}. \cr 
$T[M_3, \mathfrak{g}]$  & $T[M_3, \mathfrak{g}, H=1]$ \cr 
$T[M_3, G]$ & The theory $T[M_3,\g]$ with the natural polarization for self-dual $G$\cr
$I(T[\cdots])$ & Witten index of theory $T[\cdots]$
\end{tabular}

\section{Seifert Manifolds}
\label{app:Seife}

In this appendix we provide concrete specializations of the Seifert data to three-manifolds that will be studied in the paper.  For concreteness, let us specialize to Seifert manifolds over a genus zero base, and with three special fibers
\be M_3 \cong [d=0;~g=0;~(p_1,q_1), \;(p_2,q_2), \;(p_3,q_3) \; ]\,. \ee
We are interested in Seifert manifolds with positive scalar curvature which is equivalent to 
\be
\left(\frac{1}{p_1}+\frac{1}{p_2}+\frac{1}{p_3}-1\right)>0\,.
\ee
Such spaces can always be written as quotients
\be
M_3\cong S^3/(\Gamma_1\times\Gamma_2)\,,
\ee
where $\Gamma_1\times\Gamma_2$ is a finite subgroups of $SO(4)$ of which at least one is cyclic.\footnote{Additionally, the orders of the two groups have to be coprime.}
We can classify the possibilities, which come in three series:
\begin{enumerate}
\item $M_3\cong[0;0;(1,1),(p_2,q_2),(p_3,q_3)]$ are the Lens spaces $L(p,q)$ introduces above. They are given by $\Z_p$ quotients of $S^3$ where the group action is given by
\be
(z_1,z_2)\to(e^{2\pi i/p}z_1,e^{2\pi i q/p}z_2)\,,
\ee
where $(z_1,z_2)\in \C^2$. The first homology group is
\be
H_1(L(p,q),\Z)\cong \Z_p\,,
\ee
with $p=|p_2p_3+p_2q_3+p_3q_2|$.

\item $M_3\cong[0;0;(2,-1),(2,1),(p_3,q_3)]$ are the prism manifolds $M^{\text{Prism}}_{p_3,q_3}$. The fundamental group is a product of a cyclic and a binary dihedral group $\mathbb{D}_m$. Taking $(p_3,q_3)=(n-2,1)$ the cyclic group becomes trivial. Then, the manifolds are $S^3/{\mathbb{D}_{4(n-2)}}$ with fundamental group
\be \label{HomologyDn}
H_1(S^3/{\mathbb{D}_{4(n-2)},\Z)}\cong \left\{
\begin{array}{l l}
\Z_2 \times \Z_2 & n\text{ even}\\
\Z_4 & n\text{ odd}
\end{array}\right. \,.
\ee
In the more general case $q_3\neq 1$ the $\Gamma_i$ depend non-trivially on $q_3$.

\item $M_3\cong[0;0;(2,-1),(3,q_2),(p_3,q_3)]$, where $p_3=3,4,5$. The fundamental group is a product of a cyclic and one of the groups $\mathbb{T}_k,\mathbb{O},\mathbb{I}$,\footnote{These are the binary tetrahedral, octahedral and icosahedral groups of order $(8\cdot 3^k,48,120)$ respectively.} depending on $p_3$. Choosing $(p_3,q_3,q_2)=(m-3,1,1)$ leaves the cyclic group trivial. Then, the manifolds are $S^3/\{\mathbb{T}_1,\mathbb{O},\mathbb{I}\}$ respectively with fundamental groups
\be \label{HomologyEm}
\ba
H_1(S^3/\mathbb{T}_1)&=\Z_3\,, \qquad
H_1(S^3/\mathbb{O})&=\Z_2\,, \qquad
H_1(S^3/\mathbb{I})&=1\,.
\ea
\ee
The last manifold is also called the Poincar\'e homology sphere, and is a famous example of an integer homology sphere, \ie, with the same homology as $S^3$ but different fundamental group. For simplicity we will not consider the general case with non-trivial $q_i$.
\end{enumerate}
These three series are connected to the semi-simple Lie algebras by the McKay correspondence. In particular, if all the $q_i=1$ they are known as the $S^3/\Gamma_{ADE}$
\be\label{S3/Gamma}
\ba
S^3/\Gamma_{A_{k}}&\cong[0;0;(\ell,1),(k-\ell+1,1)]\ \forall \ell\\
S^3/\Gamma_{D_{n}}&\cong[0;0;(2,-1),(2,1),(n-2,1)]\\
S^3/\Gamma_{E_{m}}&\cong[0;0;(2,-1),(3,1),(m-3,1)]\,.
\ea
\ee

There is no simple classification of Seifert manifolds with negative curvature. However, if all the $q_i=1$ the spaces are known as the Brieskorn manifolds \cite{MR0418127}. They are given by
\be
M^{\text{Brieskorn}}_{p_1,p_2,p_3}=[1;0;(p_i,1)]\,,
\ee
and also include the positive curvature manifolds in \eqref{S3/Gamma}. They are defined as the intersection of the complex three-sphere $\{z_i\in \C^3\,,\ \sum_i|z_i|^2=1\}$ and the hypersurface $\sum_i z_i^{p_i}=0$.

\section{Twisted Superpotentials and Bethe Equations}
\label{app:BetheEquations}

In this appendix we review some aspects of the twisted superpotential and associated Bethe equations for 3d $\cN=2$ theories, focusing on applications to the theories $T[M_3]$.  We give arguments for the number of Bethe vacua for several examples of these theories.

\subsection{Twisted Superpotential for $\widehat{T}[\Omega, SU(2)]$}
\label{app:SU2Supo}

Let $M_3$ be a Seifert manifold, with associated graph $\Graph$ and linking matrix $Q_{ij}$. Consider the theory $\widehat{T}[\Graph, SU(2)]$. To compute the twisted superpotential for this theory, we first need to compute the contribution from the building blocks: $\varphi=T^k$ and $\varphi=S$. 

Let us start with the $T(SU(2))$ theory, \ie, $\varphi =S$.
It has a UV Lagrangian description as a $U(1)$ gauge theory with two hypermultiplets. This theory has a manifest $SU(2)_m$ flavor symmetry acting on the hypers, and the $U(1)_J$ topological symmetry enhances in the IR to another $SU(2)_\zeta$ flavor symmetry.  We consider the $\CN=2^*$ version of the theory by turning on a coupling to the diagonal combination of the R-symmetry
\be
U(1)_\tau = 2[U(1)_C - U(1)_H]\ ,
\ee
where $U(1)_C$ and $U(1)_H$ are a maximal torus of the $SU(2)_C$ and $SU(2)_H$ factors of the R-symmetry. The global symmetry of the theory in the infrared is then
\be \label{SymmmetryTSU(2)}
SU(2)_\zeta \times SU(2)_m \times U(1)_\tau\,.
\ee
The twisted superpotential can be written as
\be
\CW_{S}(u,\zeta,m;\tau) = \CW_{\text{FI}}(u,\zeta) + \CW_{\text{matter}}(u,m;\tau) \,,
\ee
where the first term is the contribution from the FI parameter
\be
\CW_{\text{FI}}(u,\zeta)=2u\zeta\,,
\ee
and the contribution from the matter can be written as
\be
\CW_{\text{matter}}(u,m;\tau)=\CW_{\Phi}( \tau  \pm u \pm m)+\CW_{\Phi}(-2\tau)\,,
\ee
where $\CW_\Phi$ is defined in \eqref{wphi}. Here and below  we use the convention that each choice of sign in `$\pm$' is  summed over.

As mentioned in section \ref{sec:ExSU2} this theory has a dual description as an $SU(2)$ theory with a $k=1$ CS term and $N_f=2$ fundamental flavors, see figure \ref{fig:tststduality}. This dual theory has twisted superpotential
\be \label{dualtsu2}
\widetilde{\CW}_{S}(u,\zeta,m;\tau) =  \CW_\Phi(\tau \pm u \pm m \pm \zeta) + u^2 - \CW_{\text{adj}}(\zeta;\tau) - \zeta^2 - m^2\,, \ee
where the first term comes from the trinion coupling. We have added an adjoint chiral multiplet (the ``flip'' field) charged under the $SU(2)_\zeta$ symmetry, as well as background CS terms (``contact terms'') for the $SU(2)_m \times SU(2)_\zeta$ symmetries.  These are both necessary for a precise duality to the usual description of  $T(SU(2))$ above.
Here, and in general, coupling an adjoint chiral multiplet of positive $U(1)_t$ charge corresponds to
\be
\mathcal{W}_{\text{adj}}(\zeta;\tau)=\mathcal{W}_\Phi(-2\tau\pm 2\zeta)+\mathcal{W}_\Phi(-2\tau)\,,
\ee
whereas an adjoint of negative charge contributes with $-\mathcal{W}_{\text{adj}}$ (up to $\tau$-dependent contact terms, which we ignore).
In addition, we have equipped the flavor groups in $T(SU(2))$ with a background Chern-Simons term. The corresponding contribution to the twisted superpotential is
\be
\mathcal{W}_{T^k}(\zeta)=k\zeta^2\,,
\ee
where $k$ is the background level.

We may use these ingredients to construct the twisted superpotential of a general quiver theory, $\widehat{T}[\Omega,SU(2)]$.  Including a copy of the twisted superpotential, $\CW_S$, of the $T(SU(2))$ theory for each edge, and an appropriate Chern-Simons term for each node, we may write this conveniently in terms of the linking matrix $Q_{ij}$ of \eqref{LinkingMatrix}, as
\be
\mathcal{W}_{\Graph}(u_i;\tau)=\sum_{i=1}^n \left(Q_{ii} u_i^2 + N^{\text{adj}}_i \mathcal{W}_{\text{adj}}(u_i;\tau)\right) - \sum_{i< j} Q_{ij} \mathcal{W}_S(u_i,u_j;\tau)\,,
\ee
where $N^{\text{adj}}_i$ is the number of adjoints attached to the $i$th node, which can be determined following the prescription in section  \ref{sec:SeifertQuivers}, and $I$ runs over the connecting $T(SU(2))$s. Here $u_i$ is the $SU(2)$ fugacity for the $i$th node.

The number of vacua is given by the number of solutions to the vacuum equations using the definition (\ref{FluxOperator})
\be
\Pi_i = e^{2\pi i \partial_{u_i} \CW_\Omega}=1\,, \quad i=1,...,n\,,
\ee
modulo the action of $\mathbb{Z}_2$ Weyl group
\be \label{WeylZ2}
x_i \to x_i^{-1}\,,
\ee
where $x_i=e^{2 \pi i u_i}$.

\subsection{Vacuum Count for  $\widehat{T}_{\mathcal{N}=2}[\Graph_{[k_1, k_2]},SU(2)]$}
\label{app:ExplicitComp}

As a concrete example we will go through the explicit computation of the index for the Lens space with $\frac{p}{q}=k_1-\frac{1}{k_2}$. The quiver for this theory is shown in figure \ref{fig:QuiverTST}.

We first compute the vacua for $\widehat{T}[\Graph_{[k_1,  k_2]},SU(2)]$.  The linear quiver with $\varphi=  T^{k_1} S T^{k_2}$  has twisted superpotential
\be
\ba
\mathcal{W}(\zeta,u_1,u_2;\tau)=&\mathcal{W}_S(\zeta,u_1,u_2;\tau)+\mathcal{W}_{T^{k_1}}(u_1)+\mathcal{W}_{T^{k_2}}(u_2)+\mathcal{W}_{\text{adj}}(u_1;\tau)\\
=&2\zeta u_1+\log(-z)^2+\log(-x_2)^2\\
&+\text{Li}_2(zx_2t)+\text{Li}_2(zx_2^{-1}t)+\text{Li}_2(z^{-1}x_2t)+\text{Li}_2(z^{-1}x_2^{-1}t)\\
&+k_1 u_1^2+k_2 u_2^2+\half\log(-x_1^2)^2+\text{Li}_2(x_2^2t)+\text{Li}_2(x_2^{-2}t)+\dots\,,
\ea
\ee
where the $\dots$ only depend on $\tau$.
However, we can also use the dual trinion description, discussed in section \ref{sec:ExSU2}. Then, the theory is described by a $T_2$ trinion with the three flavor groups gauged at levels $(\widetilde{k}_1, \widetilde{k}_2, \widetilde{k}_3) = (k_1-1,k_2-1,1)$ respectively and no additional adjoints. The superpotential for general $\widetilde{k}_i$ is
\be
\ba
\widetilde{\mathcal{W}}(u_1,u_2,u_3;\tau)=&\mathcal{W}_{\Phi}(\tau\pm u_1\pm u_2 \pm u_3)
+\mathcal{W}_{T^{\widetilde{k}_1}}(u_1)+\mathcal{W}_{T^{\widetilde{k}_2}}(u_2)+\mathcal{W}_{T^{\widetilde{k}_3}}(u_3)\\
=&\text{Li}_2(x_1x_2x_3t)+\text{Li}_2(x_1x_2x_3^{-1}t)+\text{Li}_2(x_1x_2^{-1}x_3t)+\text{Li}_2(x_1^{-1}x_2x_3t)\\
&+\text{Li}_2(x_1x_2^{-1}x_3^{-1}t)+\text{Li}_2(x_1^{-1}x_2x_3^{-1}t)+\text{Li}_2(x_1x_2^{-1}x_3^{-1}t)+\text{Li}_2(x_1^{-1}x_2^{-1}x_3^{-1}t)\\
&+2\log(-x_1)^2+2\log(-x_2)^2+2\log(-x_3)^2 +\widetilde{k}_1 u_1^2+\widetilde{k}_2 u_2^2+\widetilde{k}_3 u_3^2+\dots\,.
\ea
\ee
Using the description in terms of the trinion with $SU(2)^3$ gauge group,  the Bethe equations are given by
\be \label{tstbe} 
\Pi_i = {x_i}^{2 \widetilde{k}_i} \frac{(x_i - t x_j x_k)(x_i x_j- t x_k)(x_i x_k- t x_j)(x_i x_j x_k- t)}{(1- t x_i x_j x_k)(x_j- t x_i x_k)(x_k- t x_i x_j)(x_j x_k- t x_i)} = 1  \,,
\ee
where $(i,j,k) \in \{(1,2,3),(2,3,1),(3,1,2) \}$. 
This gives a system of three couple polynomial equations in the $x_i$.  

Next we must count the solutions to these equations, modulo Weyl invariance.  We first factor out the Weyl invariant solutions at $x_i =\pm1$, which are unphysical, and we may then change to Weyl invariant variables, $y_i=x_i + x_i^{-1}$.  Note that an arbitrary Weyl-invariant polynomial in $x_i$ can be written as a linear combination of the polynomials
\be p_n(y_i) \equiv  \frac{{x_i}^{n+1} - {x_i}^{-n-1}}{{x_i}-{x_i}^{-1}} \,.\ee
For $n$ non-negative, this is a degree $n$ polynomial in $y_i$, which has only even or odd degree terms.  This definition extends also to negative $n$, and satisfies
\be \label{pref} p_n(y_i) = - p_{-2-n}(y_i) \,.
\ee
Then one finds the Bethe equations, \eqref{tstbe}, lead to the following equations for $y_i$
\be \label{Pdef} 
\ba
&P_{1}(y_1,y_2,y_3) \cr 
\equiv &p_{\widetilde{k}_1+1}(y_1) + t^4 p_{\widetilde{k}_1-3}(y_1) - t y_2 y_3(p_{\widetilde{k}_1}(y_1) + t^2 p_{\widetilde{k}_1-2}(y_1))+ t^2 ({y_2}^2 + {y_3}^2 -2) p_{\widetilde{k}_1-1}(y_1) = 0\,, \ea\ee
and similarly for $P_2$ and $P_3$.\footnote{Note, that by rewriting the Bethe equations as polynomial equations can add spurious solutions, that are not solutions to the original Bethe equations. These are removed in the counting. }  This represents a system of three coupled polynomial equations, which is difficult to solve analytically.  However, we may attempt to count the solutions by using the fact that their number does not jump as we vary the parameter $t$ continuously.  Thus, let us consider the behavior in the $t \rightarrow 0$ limit.  We first assume all $\widetilde{k}_i \geq 0$.   Then this limit is well-behaved because the leading term of each $P_{i}$ survives in the limit.  Then the equations simplify to
\be \label{anargpeqs} p_{\widetilde{k}_1+1}(y_1)  \approx 0 , \;\;\;\; p_{\widetilde{k}_2+1}(y_2)  \approx 0 , \;\;\;\; p_{\widetilde{k}_3+1}(y_3)  \approx 0\,, \ee
and we find that the number of vacua is
\be \label{nvackpos} N_{\text{vac}} = (\widetilde{k}_1+1)(\widetilde{k}_2+1)(\widetilde{k}_3+1)\,,  \;\;\;\;\; \tk_i\geq 0\,. \ee
One can check that the equations are invariant under taking all $\widetilde{k}_i \rightarrow -\widetilde{k}_i$ and also $t \rightarrow t^{-1}$, so one finds a similar formula when all $\widetilde{k}_i \leq 0$.  

When two of the $\widetilde{k}_i$'s have different signs, we may not use this argument, as the leading terms of some of the polynomials vanish as we take $t \rightarrow 0$, so some solutions run off to infinity in a way that is difficult to control analytically.  In this case, the equation \eqref{Pdef} can be solved using Gr\"obner basis methods in Mathematica or {\sc Singular}.  These methods are similar to solving systems of linear equations using Gaussian  elimination:  the system of polynomial equations is mapped to the Gr\"obner basis, retaining the same set of solutions. The system of equations can however now be successively solved. For small values of $\tk_i$ this can be implemented, and the results are consistent with the following formula
\be N_{\text{vac}} = (\widetilde{k}_1 + 1)(\widetilde{k}_2+1)(-\widetilde{k}_3+1)-4 \widetilde{k}_1 \widetilde{k}_2, \;\;\;\; \widetilde{k}_1,\widetilde{k}_2 \geq 0,\ \widetilde{k}_3 \leq -2\,. 
\ee
If we now take the values 
\be 
\widetilde{k}_1 = k_1 - 1, \;\;\; \widetilde{k}_2 = k_2 - 1 , \;\;\; \widetilde{k}_3 = 1  \,,
\ee
these reduce to the counting of vacua in terms of $p$ to be 
\be 
\label{nvactsta} 
I(\widehat{T}[\Graph_{[k1,k2]},SU(2)]) = 2 \left(|k_1 k_2-1|+1\right) = 2(p+1)  \,.
\ee

For larger quivers, the procedure to obtain twisted superpotential and Bethe equations is completely analogous, but the resulting systems of polynomial equations quickly become difficult to solve, even using computer algebra programs like Mathematica and {\sc Singular}.  We have studied some further quivers cases with up to seven gauge group factors (leading to seven coupled equations), and some of the results are quoted and utilized in the main text.

\subsection{Vacuum Count for $T[L(k, 1), (S)U(N)]$ from Bethe Equations}
\label{app:ComputationsU(N)}

Next we consider higher rank examples, with $G=U(N)$ and $SU(N)$.
 
First, for the $\mathcal{N}=1$ twist, after passing to the $\cN=2$ enhancement point, the $T^k$ theory with $G= U(N)$ is the topological $U(N)_k$ CS theory and the number of vacua can be computed by standard techniques, but let us illustrate the counting using the Bethe equations, to exemplify the method. 
These are given here by
\be \Pi_i =  z {x_i}^k = 1 , \;\;\;\; i=1,...,N \,.\ee
The solutions are then $x_i=e^{2 \pi i n_i/k}$, $n_i \in \{0,...,k-1\}$.  In addition, we must impose the Weyl-invariance condition, \ie, that no two $n_i$ are equal, and that we count the solutions up to permutations.  Thus we are led to
\be I(T_{\mathcal{N}=1}[L(k,1),U(N)])= \binom{k}{N} \,.\ee
For the $\mathcal{N}=2$ twist, the Bethe equations are instead
\be \label{dggun}z {x_i}^k \prod_{j \neq i} \frac{t^2 x_i - x_j}{t^2 x_j - x_i} = 1 ,\;\;\;\; i=1,...,N \,. \ee
This is now a set of coupled equations.  To make progress, let us first consider a closely related theory, where we have instead a CS level $\frac{k}{2}$ and $k$ anti-fundamental chiral multiplets, which are acted on by a $U(k)$ flavor symmetry which we assign fugacities $w_a$, $a=1,...,k$.  Then the Bethe equations become
\be \label{dggun2} z \prod_{a=1}^k (x_i-w_a) \prod_{j \neq i} \frac{t^2 x_i - x_j}{t^2 x_j - x_i} = 1 \,. \ee
Notice that if we take all $w_a \rightarrow 0$, we recover the equations in \eqref{dggun}.  This reflects the fact that giving all the anti-fundamental chirals large real masses and integrating them out, the CS level is shifted from $\frac{k}{2}$ to $k$.  Thus we may instead do the vacuum counting  in this theory with anti-fundamentals.

The equations \eqref{dggun2} are still coupled, so this is is not obviously an improvement.  However, let us now consider these equations in the limit $z \rightarrow \infty$.  Then we can see that the remaining factor in the Bethe equations must be approximately zero, which implies
\be x_i \approx w_a, \;\;\; \text{or} \;\;\; x_i \approx t^{-2} x_j, \;\;\;\;\; \forall i \,.\ee
Thus the approximate solutions can be organized into towers over each  $w_a$
\bea \label{watowers} x_i \in \{ & w_1, \; w_1 t^{-2}, ..., \;w_1 t^{-2(\ell_1-1)}, \nonumber \\
& w_2, \;w_2 t^{-2} ,..., \;w_2 t^{-2(\ell_2-1)} \nonumber \\
& ...\nonumber \\
&w_k, \;w_k t^{-2} ,... ,\;w_k t^{-2(\ell_k-1)} ~~\} \,,\eea
where there are $\ell_a \geq 0$ \footnote{If $\ell_a=0$, $w_a$ and its tower is absent from \eqref{watowers}.} solutions in the tower above $w_a$, and $\sum_a \ell_a=N$.  The number of such solutions, up to permutations, is just the number of ways of distributing $N$ elements among $k$ boxes, which gives
\be \label{UNcount} I(T_{\mathcal{N}=2}[L(k,1),U(N)])= \binom{k+N-1}{N}\,. \ee

Next we consider the counting for $G=SU(N)$.   For this, we recall that $U(N) \cong (SU(N) \times U(1))/\Z_N$.  Then each $U(N)$ CS theory above is a 1-form gauging of a tensor product of the corresponding $SU(N)$ theory with a $U(1)$ theory.\footnote{Note this argument would not hold if there were fundamental matter.}  This $U(1)$ theory has a level $Nk$ Chern-Simons term, and so $Nk$ vacua, and we must gauge the diagonal of its $\Z_N$ 1-form symmetry with that of the $SU(N)$ factor.  Then we expect this gauging to simply remove a factor of $N^2$ from the number of vacua of the tensor product.\footnote{More precisely, if $(N,k)=1$, these two tensor factors have a 1-form symmetry with a maximal anomaly (in the sense that the center is trivial).  In this situation, we have seen that gauging the $\Z_N$ symmetry removes a factor of $N^2$ from the vacuum counting. We have checked for some low values of $N$ and $k$ that this holds also when $(N,k)>1$.}  Thus we find
\be I(T_{\mathcal{N}=1,2}[L(k,1),U(N)]) = \frac{k}{N} I(T_{\mathcal{N}=1,2}[L(k,1),\su(N)]) \,. \ee
For the $\cN=1$ twist, this gives
\be I(T_{\mathcal{N}=1}[L(k,1),\su(N)]) = \binom{k-1}{N-1}\,, \ee
which is the known result for the pure CS theory.
For the $\CN=2$ twist, we find
\be \label{SUNcount} I(T_{\mathcal{N}=2}[L(k,1),\su(N)])= \binom{k+N-1}{N-1}\,. \ee
We note that the results \eqref{UNcount} and \eqref{SUNcount} agree with \cite{Gadde:2013sca}, which were computed by different means.

\section{One-form Symmetries and Anomalies for Bethe Vacua}

\label{app:SU(2)Anomalies}

In this appendix we discuss the action of 1-form symmetries on the supersymmetric Bethe vacua of a 3d $\cN=2$ theory.  We mostly focus on $\Z_2$ 1-form symmetries in the context of $SU(2)^m$ gauge theories, but much of the discussion below can be easily generalized.  

\subsection{1-form symmetries and Bethe vacua}
\label{app:1formBV}

Let us first describe how 1-form symmetries act on the space of supersymmetric, or Bethe vacua on $T^2$.  We refer to \cite{BWHF} for a derivation of these statements, and more details.  

First recall that the supersymmetric vacua are given by solutions to the Bethe equations
\be
\Pi_i = e^{2 \pi i \partial_{u_i}\CW} = 1, \quad i = 1, ... , r_G\,,
\ee
where $r_G$ is the rank of the gauge group.  These are rational functions of the gauge variables, $u_i$, $i=1,...,r_G$, as well as the other flavor symmetry parameters, which we suppress from the notation.  We may associate states in the Hilbert space of vacua to the solutions, $\widehat{u}$, to these equations (modulo Weyl symmetry)
\be \label{appcbv} \widehat{V} 
= \text{span}\{ \; \ket{\widehat{u}_i} \; | \; \widehat{u}_i \in \CS_{BE} \; \}\,. 
\ee
Recall that 1-form electric symmetries in a gauge theory are associated to a subgroup $\Gamma$ of the center of the gauge group which acts trivially on the matter content.  To such a 1-form symmetry, we expect that we may define operators acting on the vacua on $T^2$, $U^{A,B}_\gamma$, for each element $\gamma \in \Gamma$, and associated to the two cycles of the torus, as described in section \ref{sec:ht2reps}.

To see how these operators act on the Bethe vacua of \eqref{appcbv}, note that we may identify the center of the gauge group with a set of transformations acting on the gauge variables, $u_i \rightarrow u_i + \gamma_i$, which are  defined up to large gauge transformations.  Now suppose an element $\gamma = (\gamma_1,...,\gamma_{r_G})$ in the center acts trivially on the matter.  Then we may define the 1-form operators acting on $\widehat{V}$ as above, which act on the basis of Bethe vacua as
\be \label{1formgenerators}
\ba
&U^A_\gamma \; : \; \ket{\widehat{u}_i} \rightarrow \Pi_\gamma(\widehat{u}_i) \ket{\widehat{u}_i}, \qquad \Pi_\gamma \equiv \prod_i {\Pi_i}^{\gamma_i} =e^{2 \pi i \gamma_i \partial_{u_i} \CW} \\
&U^B_\gamma \; : \;| \widehat{u}_i \rangle \rightarrow | \widehat{u}_i + \gamma_i \rangle\,.
\ea
\ee
We note both operators are well-defined when acting on the space of Bethe vacua (one can show $\Pi_\gamma$ is a rational function, given the assumption that $\gamma$ acts trivially), and satisfy the appropriate group law.  One also finds their commutation relations correctly encode the anomalies of the theory
\be U^A_{\gamma_1} \; U^B_{\gamma_2}  = e^{2 \pi i \gamma_1 \cA \gamma_2} \; U^B_{\gamma_2} \; U^A_{\gamma_1} \,. \ee
Thus in addition to the space of vacua, we can consider the extra data of how the vacua are acted on by these symmetries.  This data will be important when gauging the symmetries.

To illustrate this with a simple example, consider the $\widehat{T}[\Omega_{[k]},SU(2)]$ theory, which is an $SU(2)$ gauge theory with level $k$ CS term and an adjoint.  The Bethe equation is
\be \Pi = x^{2k} \bigg(\frac{t^2 x^2-1}{t^2-x^2} \bigg)^2 = 1\,. \ee
Then the center acts as $u \rightarrow u + \frac{1}{2}$, or $x=e^{2 \pi i u} \rightarrow -x$, and so we may take $\gamma=(\frac{1}{2})$ above, and find from \eqref{1formgenerators}
\be \label{1formgenerators_ex}
\ba
&U^A_\gamma \; : \; \ket{\widehat{u}} \rightarrow - x^{k} \bigg(\frac{t^2 x^2-1}{t^2-x^2} \bigg) \ket{\widehat{u}}, \qquad 
U^B_\gamma \; : \;| \widehat{u} \rangle \rightarrow | \widehat{u}+ \frac{1}{2} \rangle\,.
\ea
\ee
These satisfy $U^A_\gamma U^B_\gamma  = (-1)^k U^B_\gamma U^A_\gamma$, which is the expected anomaly coefficient.
The sign for $U_A$ is fixed by modular invariance. 

Finally, we must describe the twisted sectors, as in \eqref{htwistedsector}.  As shown in \cite{BWHF}, for each state that is fixed by the operator $U^B_\gamma$, this state also contributes to the twisted sector $\CH^\gamma$.  Conversely, if a solution, $\widehat{u}$, is fixed by $U^B_\gamma$ for all $\gamma$ in a subgroup, $\Gamma_{\widehat{u}}$, of $\Gamma$, then this solution contributes $|\Gamma_{\widehat{u}}|$ states, one untwisted sector and a twisted sector for each non-trivial element in $\Gamma_{\widehat{u}}$.  
	
These twisted sector states become physical states in the theory obtained by gauging $\Gamma$.  Then we can understood the counting above physically from the fact that, in this situation, $\Gamma_{\widehat{u}}$ corresponds to an unbroken subgroup of the gauge group acting at $\widehat{u}$, and so there remains a low energy $\Gamma_{\widehat{u}}$-gauge theory at this solution, which is known to have $|\Gamma_{\widehat{u}}|$ states.  In the example above, we see that the non-trivial twisted sector has a single state, corresponding to  the $\widehat{u}=\frac{1}{4}$ fixed point.

\subsection{Representations of $\Z_2$ 1-form symmetries}
\label{app:z21formreps}

Now let us specialize to a gauge theory with $SU(2)^n$ gauge group, as all the $\widehat{T}[\Graph,SU(2)]$ theories may be taken to have this form.  The 1-form symmetry group, $\Gamma \cong {\Z_2}^n$, can be identified with the subgroup of the center of the gauge group, $Z_G={\Z_2}^m$, that acts trivially on the matter.  This 1-form symmetry may have 't Hooft anomalies, which are obstructions to gauging, as discussed in section \ref{sec:HigherFormSymm}. Namely, if we pick a basis for $\Gamma$ as a $\Z_2$ vector space, then we may define the ``anomaly matrix,'' $\cA$, and the mutual anomaly of two 1-form symmetries, $\gamma,\gamma' \in \Gamma$, is given by $\gamma^T \cA \gamma'$.  Then in order to gauge a subgroup, $\Lambda$, of $\Gamma$, we require
\be \gamma^T \cA \gamma'=  0, \;\;\; \gamma,\gamma' \in \Lambda\,. \ee

The anomaly matrix defines a symmetric bilinear form on the $\Z_2$ vector space, $\Gamma$.  Such a bilinear form, $B$, is classified according to its dimension $d$ its rank $d-r$ and whether or not it is ``even'', meaning that every element has vanishing inner product with itself \cite{hopkinsnotes}.  For a non-even bilinear form, we have
\be B \cong E^{d-r} \oplus Z^{r}\,, \ee
where $E$ is the unique non-degenerate bilinear form on $\Z_2$, and $Z$ is the trivial bilinear form on $\Z_2$.  For an even bilinear form, the rank must be even, and we find
\be B \cong H^{(d-r)/2} \oplus Z^{r}\,, \ee
where $H$ is the bilinear form on ${\Z_2}^2$ associated to the matrix
\be \begin{pmatrix} 0 & 1 \\ 1 & 0  \end{pmatrix}\,. \ee

The space of vacua forms a representation of the group of $U^A_\gamma$ and $U^B_\gamma$, defined above, which can give strong constraints on the structure of this space, especially in the presence of anomalies.  To illustrate this, let us consider a single $\Z_2$ symmetry with a non-vanishing self-anomaly, \ie, corresponding to the space $E$ above.  Then the $U^A$ and $U^B$ operators each form a $\Z_2$ group acting on the vacua, but the anomaly implies the full group acting on the vacua is a central extension of $\Z_2 \times \Z_2$, which is given by the dihedral group, $D_4$.  Then the only non-trivial\footnote{By non-trivial, we mean that the central element must be represented by $-1$, as this is how the algebra acts on the vacua above.} representation of this group is two dimensional, and we may take $U^A$ and $U^B$ to act as
\be
U^A \rightarrow \begin{pmatrix} 1 & 0 \\ 0 & -1 \end{pmatrix}\,, \qquad U^B \rightarrow \begin{pmatrix} 0 & 1 \\ 1 & 0 \end{pmatrix}\,.
\ee
We will denote this basic two dimensional representation as $R$.  More generally, if the 1-form group is isomorphic to $E^m$, then the irreducible representations are tensor products $R^{\otimes m}$, each of dimension $2^m$.  Similarly, for $H$, the basic irrep is the tensor product of two copies of $R$, associated to the pairs $U^A_1,U^B_2$ and $U^A_2, U^B_1$.  Explicitly, we may represent these as
\be \label{Hrep} U^A_1 \rightarrow  \begin{pmatrix} 
	1 & 0 & 0 & 0 \\
	0 & -1 & 0 & 0 \\
	0 & 0 & 1& 0 \\
	0 & 0 & 0 & -1 \end{pmatrix}, \;\;\;\; 
U^A_2 \rightarrow \begin{pmatrix} 
	1 & 0 & 0 & 0 \\
	0 & 1 & 0 & 0 \\
	0 & 0 & -1& 0 \\
	0 & 0 & 0 & -1 \end{pmatrix}, \nn \ee

\be U^B_1 \rightarrow \begin{pmatrix} 
	0 & 0 & 1 & 0 \\
	0 & 0 & 0 & 1 \\
	1 & 0 & 0 & 0 \\
	0 & 1 & 0 & 0 \end{pmatrix}, \;\;\;\; 
U^B_2  \rightarrow \begin{pmatrix} 
	0 & 1 & 0 & 0 \\
	1 & 0 & 0 & 0 \\
	0 & 0 & 0& 1 \\
	0 & 0 & 1 & 0 \end{pmatrix}. \ee
Finally, the case of $Z$, with no anomaly, is the least constraining, as the group acting on the vacua is $\Z_2 \times \Z_2$ and so the representations are one-dimensional, determined by a pair of signs, $(\epsilon_A,\epsilon_B)$ indicating the eigenvalues of $U^A$ and $U^B$.  

The above implies the space of vacua of a theory can be written as a tensor product
\be \label{VRr}
\widehat{V} = V \otimes R^{\otimes (d-r)}\,, \ee
where we observe that each non-degenerate dimension of $B$ contributes a tensor product of an $R$ representation, for a total of $2^{(d-r)}$ in each irrep, and $\text{dim}(V)$ is the number of such irreps.  In particular, we see that the number of vacua must be a multiple of $2^{(d-r)}$.  Finally, $V$ itself may be decomposed into the eigenspaces associated to the various $Z$ factors
\be \label{vdecomp} V\to \bigoplus_{\epsilon^{A,B}_j=\pm 1} V_{\epsilon^A_1\epsilon^B_1\cdots\epsilon^A_r\epsilon^B_r}\,. \ee
Note that \eqref{VRr} and \eqref{vdecomp} are a special case of the more general factorization and decomposition of the Hilbert space discussed around \eqref{Ht2factor}.

\paragraph{Gauging}

To gauge a non-anomalous symmetry group, $\Lambda \subset \Gamma$, as noted above, we project onto the subspace
\be 
V^{\text{$\Lambda$ gauged, untwisted}}  = \{ \ket{\widehat{u}} \in V  \; | \; U^A_\gamma \ket{\widehat{u}} =\ket{\widehat{u}}, \;\; U^B_\gamma \ket{\widehat{u}} =\ket{\widehat{u}}, \;\; \gamma \in \Lambda \}\,. \ee
Here the superscript ``untwisted'' indicates that in general this is only a subset of states of the gauged theory, as there may also be ``twisted sector'' states, described above.

There are two basic examples of non-anomalous subgroups, from which we can construct general examples.
\begin{itemize}
\item $\Lambda \cong Z$:  The Hilbert space before gauging decomposes into $(\epsilon_A,\epsilon_B)$ representations as
\be
 V = V_{++} \oplus V_{+-} \oplus V_{-+} \oplus V_{--}\,. \ee
Then we simply have
\be
V^{\text{$Z$ gauged, untwisted}}  = V_{++}\,.
\ee
\item $\Lambda = \langle (1,0) \rangle \subset H$ - Now the Hilbert space has the structure
\be
\label{secondgauging}
\widehat{V}= V \otimes R^{\otimes 2}\,,
\ee
where the 1-form operators act as in \eqref{Hrep}.  We can see from this representation that the simultaneous 1-eigenspace of $U^A_1$ and $U^B_1$ is generated by the state $(1,0,1,0)$, and so gauging effectively cuts the dimension by a factor of 4.  In particular, we find simply
\be
\widehat{V}^{\text{$ \langle (1,0) \rangle$ gauged, untwisted}} = V\,.
\ee
The case where $\Lambda = \langle(1,1)\rangle \subset E^2$, which also has no self-anomaly, is similar.
\end{itemize}
Iterating the second example, we see that each time we gauge a non-anomalous subgroup of the non-degenerate part of $B$, we simply remove two tensor powers of $R$ from the Hilbert space.

\subsection{$T[M_3,U(2)]$}
\label{app:SU(2)Anomalies:u2gauging}

As an example, let us consider the theory $T[M_3,U(2)]$.  As described in the main text, this  consists of two decoupled sectors, the $\widehat{T}[\Graph,SU(2)]$ theory, and the $\widehat{T}[\Graph,U(1)_2]$ theory  One can check that both theories admit a $\Z_2$ 1-form symmetry acting on each node, and the anomaly matrices are identical, given in both cases by, \eg, \eqref{AnomalyMatrix} in the case of $M_3=L(p,q)$.  Thus if we take the diagonal sum of these two $\Z_2$ symmetries at each node, this combination is non-anomalous, and the Hilbert space takes the form of \eqref{secondgauging}. We may then gauge these symmetries to form the gauge group $(U(1) \times SU(2))/\Z_2 \cong U(2)$, which simply removes a tensor factor of $R^{\otimes 2}$ from the Hilbert space.  Thus each such gauging reduces the dimension of the Hilbert space by a factor of four, as claimed in the main text.

%

\section{Flat Connections on $S^3/\Gamma_{ADE}$}
\label{app:flatconnections}

According to the 3d-3d correspondence we expect that the index of the theory $T[M_3,\mathfrak{g}]$ counts the number of flat connections on $M_3$.  The precise statement is discussed in section \ref{sec:3d3dcorr}.  In this section we will investigate this correspondence by computing the $G_{\C}$-connections on $M_3=S^3/\Gamma_{ADE}$ for $G_{\C}=GL(N,\C)$ and $PSL(2,\C)$, which are given by the corresponding representations of the fundamental group
\be
\ba
\left\{A|F_A=0\right\}/G_{\C}&=\text{Hom}\left(\pi_1(M_3),G_{\C}\right)/G_{\C}\\
\pi_1(S^3/\Gamma_{ADE})&=\Gamma_{ADE}\,.
\ea
\ee

\subsection{$GL(N,\C)$}
\label{app:flatGLNCconnections}

\label{app:FlatConn}

We start by computing the number of flat $GL(N,\C)$-connections on $S^3/\Gamma_{ADE}$ in order to compare them with the Witten index of the $T[M_3,U(2)]$ determined in section \ref{sec:THatCount}.
In principle we expect
\be
I(M_3, U(N))=\#( \text{$GL(N,\C)$ reps of $\pi_1(M_3)$})
\ee
In the case of $M_3 = S^3/\Gamma_{ADE}$, we can compute the number of $GL(N,\C)$-representations of $\pi_1(M_3)=\Gamma_{ADE}$ by the McKay correspondence \cite{mackey1983graphs}. It states that the irreducible representation of $\Gamma_{ADE}$ are in one-to-one correspondence with the nodes of the $ADE$ affine Dynkin diagram. Each node of a Dynkin diagram is assigned a Dynkin number $d_{i}$, which is the dimension of the $i$-th irreducible representation. The extended node represents the trivial representation with $d_0=1$.

The dimension $GL(N,\C)$-representations of $\Gamma_{ADE}$ can be constructed from the linear combinations of these irreducible representations, which are labeled by a set of integers $(n_0,\cdots, n_r)$
\be
\#( \text{$GL(N,\C)$ reps of $\Gamma_{ADE}$})=\left\{(n_0,\cdots, n_r)~\Big|~ N = \sum_{i=0}^{r} d_i n_i \right\}\,.
\ee

For $A_{p-1}$ all the $p$ Dynkin numbers are one, so $\Z_p$ has $p$ irreducible $\C$-connections of dimension one. The number of $N$-dimensional representations is thus given by the number of ways to assign $N$ indistinguishable objects to $p$ distinct boxes
\be\ba\label{flat connection 2}
\#( \text{$GL(N,\C)$ reps of $\Z_p$}) =  \binom{p+N-1}{N}\,,
\ea
\ee
with the special case
\be \label{FlatConnU2N=2}
\#( \text{$GL(2,\C)$ reps of $\Z_p$}) = \binom{p+1}{2}\,.
\ee

For $\Gamma_{D_n}$, the Dynkin numbers are $d_0=d_1=d_2=d_3=1$ for the two left-most nodes and the two right-most nodes, and $d_i=2$ for the remaining $(n-3)$ nodes. We can easily find
\be
\#( \text{$GL(2,\C)$ reps of $\Gamma_{D_n}$}) =\binom{5}{2} + (n-3)=n+7\,,
\ee
where the first and second term denote the reducible and irreducible solutions respectively.
Similarly, for $\Gamma_{E_m}$
\be
\#( \text{$GL(2,\C)$ reps of $\Gamma_{E_m}$})=3(9-m)\,.
\ee
All these results match the number of vacua of the corresponding theories computed in section \ref{sec:GeneralQuivers}, see table \ref{tab:ResultsTrinions}.

In the $\CN=1$ twist we only count the flat connections with $n_i=0 \text{ or } 1,\forall i$, which behave like fermionic states. In this case we find
\be \label{FlatConnU2N=1}
\#( \text{``abelian" $GL(N,\C)$ reps of $L(k,1)$})=  \binom{k}{N} \,.
\ee

\subsection{$PSL(2,\C)$}
\label{app:flatSL2Cconnections}

Next we consider the $PSL(2,\C)$ connections for the Seifert manifolds $M_3=[0;0;(p_i,q_i)]$ with three exceptional fibers. The fundamental group is given by
\be \label{pi1(M3)}
\pi_1(M_3)=\braket{x,y,h|x^{p_1}=h^{q_1},y^{p_2}=h^{q_2},(xy)^{-p_3}=h^{q_3}}\,,
\ee
where $h$ is a central element \cite{MR565450}.\footnote{Note that this group reduces to a cyclic group if any of the $p_i=1$.}

The $PSL(2,\C)$ connections of the groups in \eqref{pi1(M3)} are discussed in \cite{MR1829566}. The strategy is to first consider the reducible representations which factor through the abelianization of $\pi_1(M_3)$
\be
\pi_1^{\text{ab}}(M_3)=\Z_{c_1} \oplus \Z_{c_2}\,, \quad c_1=\frac{|H_1(M_3,\Z)|}{(p_1,p_2,p_3)}\,, \quad c_2=(p_1,p_2,p_3)\,,
\ee
generated by $z_{1,2}$.
The reducible $PSL(2,\C)$-representations of the cyclic groups are given by
\be
\rho_{j_1,j_2}(z_i)=\begin{pmatrix} e^{\pi i j_i/c_i} & 0 \\ 0 & e^{-\pi i j_i/c_i}
\end{pmatrix}\,, \qquad j_i\in\{0,\cdots, c_i-1\}\,.
\ee
After accounting for Weyl symmetry we find
\be \label{reduciblerepspi1}
\#( \text{reducible $SL(2,\C)$ reps of $\pi_1(M_3)$})  = \left\lfloor \frac{|H_1(M_3,\Z)|}{2} \right\rfloor + \begin{cases} 1 & (p,q,r) \; \text{odd} \\ 2 & (p,q,r) \; \text{even} \end{cases} \,. \ee

To find the number of irreducible representations we use the related group
\be
\Delta_{p_i}=\braket{x,y|x^{p_1}=y^{p_2}=(xy)^{p_3}=1}\,,
\ee
as $\rho(h)$ needs to be trivial if $\rho$ is irreducible. The irreducible representations of these groups are counted by three integers
\be
\ba
&\rho_{j}(x)=\begin{pmatrix} \alpha_{j} & 0 \\ 0 & \alpha^{-1}_j \end{pmatrix}\,, \qquad \rho_{j,k,\ell}(y)=\begin{pmatrix} \gamma_{j,k\ell} & 1 \\ \gamma_{j,k,\ell} \left(\beta_k-\gamma_{j,k,\ell}\right)-1 & \beta_k-\gamma_{j,k,\ell} \end{pmatrix}\,,\\
&\alpha_j=e^{\frac{\pi i j}{p_1}}\,, \quad \beta_k=2\cos\left(\frac{\pi k}{p_2}\right)\,, \quad \gamma_{j,k,\ell}=\frac{2\cos\left(\frac{\pi \ell}{p_3}\right)-\alpha^{-1}_j \beta_k}{2i \Im \lambda_k}\,,\\
&j=1,\cdots \left\lfloor\frac{p_1}{2}\right\rfloor\,, \quad k=1,\cdots \left\lfloor\frac{p_2}{2}\right\rfloor\,,\quad \ell=1,\cdots p_3-1\,.
\ea
\ee
After accounting for the Weyl symmetry, and including the reducible connections in \eqref{reduciblerepspi1} we obtain
\be \label{PSLConnGeneral}
\ba
&\#( \text{$PSL(2,\C)$ reps of $\pi_1(M_3)$})=\left\lfloor \frac{p_1}{2} \right\rfloor \left\lfloor \frac{p_2}{2} \right\rfloor \left\lfloor \frac{p_3}{2} \right\rfloor +  \left\lfloor \frac{p_1-1}{2} \right\rfloor \left\lfloor \frac{p_2-1}{2} \right\rfloor \left\lfloor \frac{p_3-2}{2} \right\rfloor  \\
&  + \left\lfloor \frac{(p_1,p_2)}{2} \right\rfloor + \left\lfloor \frac{(p_1,p_3)}{2} \right\rfloor + \left\lfloor \frac{(p_2,p_3)}{2} \right\rfloor  + 1  + \left\lfloor \frac{|H_1(M_3,\Z)|}{2} \right\rfloor -  \left\lfloor \frac{(p_1p_2,p_1p_3,p_2p_3)}{2} \right\rfloor\,.
\ea
\ee
We can plug in the choices of $p_i$ for the $S^3/\Gamma_{ADE}$ to obtain
\be
\ba
\#( \text{$PSL(2,\C)$ reps of $\Z_p$})&=\left\lfloor\frac{p}{2}\right\rfloor+1\\
\#( \text{$PSL(2,\C)$ reps of $\Gamma_{D_n}$})&=\begin{cases} \frac{n}{2}+3 & \text{$n$ even} \\ \frac{n+3}{2}  & \text{$n$ odd} \end{cases}\\
\#( \text{$PSL(2,\C)$ reps of $\Gamma_{E_m}$})&=\begin{cases} 3 & \text{$m$ even} \\ 4  & \text{$m$ odd} \end{cases}
\,.
\ea
\ee

Next we need to compute the Stiefel-Whitney (SW) class of these connections.  For this, we note that a $PSL(2,\C)$ representations of a group $\Gamma$ determines an $SL(2,\C)$ representation of some $\Z_2$ central extension of $\Gamma$.  Such extensions are classified by the group $H^2(\Gamma,\Z_2)$, and this determines the Stiefel-Whitney class of the connection. To compute this we note that for the ADE groups we are interested in
\be H^2(\Gamma,\Z_2) \cong \text{Ext}^1(\Gamma^{\text{ab}},\Z_2)\,, \ee
where $\Gamma^{\text{ab}}$ is the abelianization of $\Gamma$.  In other words, all central extensions of $\Gamma$ are determined by an abelian extension of the group $\Gamma^{\text{ab}}$.  

For example, consider the $A_{p-1}$ case.  Then $\Gamma = \Gamma^{\text{ab}} = \Z_p$, and we have
\be \text{Ext}(\Z_p,\Z_2) = \Z_{(2,p)}\,. \ee
For $p$ odd, all extensions are trivial, and all $PSL(2,\C)$ representations lift to $SL(2,\C)$ representations.  For $p$ even, the two extensions are $\Z_p \times \Z_2$ (trivial) and $\Z_{2p}$ (non-trivial).  Recall the representations are labeled by $j \in \{0,...,\frac{p}{2} \}$, and map the generator, $z$, of $\Z_p$ to
\be \rho(z) = \pm \begin{pmatrix} e^{\frac{2\pi i j }{2p}} & 0 \\ 0 & e^{-\frac{2\pi i j}{2p} } \end{pmatrix}\,. \ee
Then we note
\be \rho(z)^{p} = (-1)^{j}\,, \ee
which implies this is an $SL(2,\C)$ representation of $\Z_p \times \Z_2$ for $j$ even, and $\Z_{2p}$ for $j$ odd.  Thus we find the counting of flat $PSL(2,\C)$ connections with $w_2= \omega$ is given by
\be
p=0 \mod 4 :\quad \begin{cases} \frac{n}{4}+1 & \omega=(0,0) \\ 1 & \omega=(1,0)  \end{cases}\,,  \qquad p=2 \mod 4 :\quad  \begin{cases} \frac{n+2}{4} & \omega=(0,0) \\ 1 & \omega=(1,0) \end{cases}\,.
\ee

For an example involving irreducible connections, we consider $D_{n}$ with even $n$.  Here $\Gamma_{D_{n}}$ is the binary dihedral, or dicyclic group, $\text{Dic}_{n-2}$, with presentation
\be \text{Dic}_{n-2} = \braket{ x,y,h \;| \; x^2 = y^2 = (xy)^{n-2} = h , \;\; h \; \text{central} }\,.\ee
The commutator subgroup is $\Z_{n-2}$, generated by $(xy)^2$, and the quotient is
\be {\Gamma_{D_{n}}}^{\text{ab}} = \Z_2 \oplus \Z_2\,, \ee
generated by $x$ and $y$. Thus
\be  
\label{d2nh2} H^2(\Gamma_{D_{n}},\Z_2) = \Z_2 \oplus \Z_2 \,.
\ee
There are four reducible representations for all $n$, and by a similar argument as above, one finds these take values in the four elements of \eqref{d2nh2}.   The irreducible representations in this case can be written as
\be \rho_\ell(x) = \pm \begin{pmatrix} i & 0 \\ 0 & -i \end{pmatrix}, \;\;\;\; \rho_\ell(y) = \pm \begin{pmatrix} - i \cos \frac{\ell \pi}{n-2} & 1 \\ -\sin^2 \frac{\ell \pi}{n-2} & i \cos \frac{\ell \pi}{n-2} \end{pmatrix}, \;\;\; \rho_\ell(x y) = \pm \begin{pmatrix}\cos \frac{\ell \pi}{n-2} & i \\ i \sin^2 \frac{\ell \pi}{n-2} & \cos \frac{\ell \pi}{n-2} \end{pmatrix}\,, \ee
where $\ell \in \{1,...,\frac{n-2}{2}\}$.  From this one can compute
\be \rho_\ell(x)^2 = \rho_\ell(y)^2 = -1, \;\;\; \rho_\ell(xy)^{n-2} = (-1)^\ell \,.\ee
Thus $\rho_\ell(x)^2$ agrees with $\rho_\ell(xy)^{n-2}$ only when $\ell$ is odd, which means when $\ell$ is odd (respectively, even) this is a representation of an extension with trivial (respectively, non-trivial) component in the first factor in \eqref{d2nh2}.  The same holds for $y$, and so we find
\be \omega_{\rho_\ell} = \begin{cases} (0,0) & \ell \; \text{odd} \\  (1,1) & \ell \; \text{even} \end{cases}\,. \ee
Thus we expect the following total distribution of flat connections over the four classes in $H^2$
\be 
n=0 \mod 4 :\quad \begin{cases} \frac{n}{4}+1 & \omega=(0,0) \\ 1 & \omega=(1,0) \\ 1 & \omega=(0,1) \\ \frac{n}{4} & \omega=(1,1) \end{cases}\,,  \qquad n=2 \mod 4 :\quad  \begin{cases} \frac{n+2}{4} & \omega=(0,0) \\ 1 & \omega=(1,0) \\ 1 & \omega=(0,1) \\ \frac{n+2}{4} & \omega=(1,1) \end{cases}\,.
\ee
Similar computations can be performed in the remaining cases and the results are given in the main text.

\bibliography{M5}{}
\bibliographystyle{JHEP}

\end{document}